\documentclass[showpacs,amsmath,amsfonts,amssymb,aps,superscriptaddress]{revtex4}
\usepackage{amsmath,amssymb}
\usepackage[utf8]{inputenc}
\usepackage[T1]{fontenc}
\usepackage{microtype}
\usepackage{caption}
\usepackage{parskip}
\usepackage{physics}
\usepackage{xcolor}
\definecolor{RED}{named}{red}
\usepackage{url} 
\newtheorem{theorem}{Theorem}[section]

\newtheorem{proposition}[theorem]{Proposition}
\newtheorem{corollary}[theorem]{Corollary}

\newenvironment{proof}[1][Proof.]{\begin{trivlist}
		\item[\hskip \labelsep {\bfseries #1}]}{\end{trivlist}}

\usepackage{mathtools}
\usepackage{listings}
\usepackage{booktabs}
\usepackage{float}
\usepackage{graphicx}
\usepackage{subcaption}

\usepackage{natbib}
\begin{document}

    \title{Scalar quasinormal modes of the exact spinning Morris--Thorne wormhole}

    \author{Mark Essa Sukaiti}
    \email{100064482@ku.ac.ae}
    \affiliation{Mathematics Department, Khalifa University of Science and Technology, PO Box 127788, Abu Dhabi, United Arab Emirates}

    \author{Davide Batic}
    \email{davide.batic@ku.ac.ae}
    \affiliation{Mathematics Department, Khalifa University of Science and Technology, PO Box 127788, Abu Dhabi, United Arab Emirates}

    \author{Denys Dutykh}
    \email{denys.dutykh@ku.ac.ae}
    \affiliation{Mathematics Department, Khalifa University of Science and Technology, PO Box 127788, Abu Dhabi, United Arab Emirates}

    \date{\today}

\begin{abstract}
We study the quasinormal modes (QNMs) of a minimally coupled massless scalar field on the exact spinning Morris-Thorne wormhole. The problem is posed on the complete two-ended geometry, whose asymptotic rest frames rotate relative to each other. After extracting the axis behaviour and the outgoing phases at both infinities, we compactify the radial coordinate and construct a quadratic eigenvalue problem by two-dimensional Chebyshev collocation. The analytic background yields exact constraints on the spectrum. At fixed azimuthal number $m$, the frequencies $\widehat{\omega}=r_0\omega$ are mirror symmetric about $\operatorname{Re}\widehat{\omega}=3\pi jm/2$, where $r_0$ is the throat radius and $j=J/r_0^2$ is the dimensionless angular momentum. Each multipole family depends on $m$ and $j$ only through $mj$ and three angular Chebyshev polynomials represent exactly the
multipoles $L=|m|,|m|+1,|m|+2$. A conserved flux confines nonthreshold
real quasinormal frequencies to the superradiant window $0<\omega/(m\Omega_-)<1$, where $\Omega_-=3\pi J/r_0^3$. For the sampled sectors $m=1,2,3$ and spins up to $j=19$, every converged frequency is damped and lies outside this window. Independent integration verifies the low-lying modes, while Kerr spectra provide a separate benchmark for the two-dimensional solver. A limiting radial equation explains why the computed mode frequencies approach the upper edge of the superradiant window with frequency offsets proportional to $(mj)^{-1/2}$ at large $mj$. A complete stability assessment requires further analysis of the
unresolved scalar spectrum and of perturbations of the coupled
metric-matter system.
\end{abstract}

\maketitle

\section{Introduction}
QNMs describe the radiative response of relativistic compact objects. Their boundary conditions select waves that escape the system and their complex frequencies determine the oscillation and damping or
growth rates. With the convention $e^{-i\omega t}$ used here, $\operatorname{Im}\omega<0$ corresponds to damping and
$\operatorname{Im}\omega>0$ to exponential growth. Quasinormal spectra
therefore connect ringdown signals with questions of linear stability
\cite{Kokkotas1999LR,Berti2009CQG,Konoplya2011RMP}. Moreover, gravitational-wave observations have made this connection central to strong-field tests of compact-object models \cite{Abbott2016PRL,LVK2023GWTC3}. Wormholes provide a complementary setting in which topology, matter support and radiative propagation can be examined together. The Einstein-Rosen bridge, the Morris-Thorne construction and the Ellis-Bronnikov family illustrate different aspects of wormhole geometry \cite{Einstein1935PR,Morris1988AJP,Ellis1973JMP,Bronnikov1973APP,Visser1996}. For traversable wormholes, the throat also raises the familiar question of compatibility with the standard energy conditions. Studies of static wormholes have explored how the geometry, supporting matter and radiation conditions affect the quasinormal spectrum and late-time response \cite{Konoplya2005PRD,Konoplya2010PRD,Konoplya2016JCAP,
Konoplya2018PLB,Aneesh2018PRD,BlazquezSalcedo2018PRD,Azad2023PRD,
Churilova2020PLB,Gonzalez2022PRD,Batic2024CGG,Batic2025EPJC,Batic2025CQG}. The distinction between a test-field calculation and a perturbation of the supporting system is essential. A growing test-field mode establishes an instability in that sector, whereas the absence of such a mode does not establish stability of the coupled metric-matter equations
\cite{Gonzalez2009CQG,Azad2024PLB}. Rotation introduces frame dragging and may produce an ergoregion, making the relation between the two asymptotic ends particularly important. Teo's stationary, axisymmetric ansatz provides a general framework for rotating traversable wormholes \cite{Teo1998PRD}. Subsequent constructions include slowly rotating configurations and rotating Ellis and phantom-scalar solutions
\cite{Kashargin2008GC,Kashargin2008PRD,Kim2005NCB,Kim2008PTPS,
Kleihaus2014PRD,Chew2019PRD,Cisterna2023PRD,Clement2023PLB,
Batic2026CQGSlow}. Perturbation calculations are also available for
geometrically distinct rotating wormholes and black-hole mimickers. These
include scalar modes and ergoregion-instability diagnostics of the
Kerr-black-bounce family \cite{Franzin2022PRD}, metric perturbations of
rapidly rotating Ellis--Bronnikov wormholes \cite{Khoo2024PRD} and a WKB
analysis of scalar waves on a Teo rotating wormhole \cite{Radhakrishnan2026}.

The closest methodological precedent to our present work is the study by Khoo \textit{et al.} \cite{Khoo2024PRD}. They formulate the perturbation problem on the complete two-ended manifold, compactify the radial coordinate by $x=(2/\pi)\arctan{(r/r_0)}$ and impose the outgoing condition at the second end using its asymptotically shifted frequency. Their Chebyshev--Legendre discretisation yields a quadratic eigenvalue problem on a numerically constructed background. They also identify empirically the relation between co- and counter-rotating modes that we derive here from an exact symmetry. The present work builds on this formulation for an analytic spinning Morris-Thorne background and a scalar test field. Our main contributions are the exact spectral identities available for this geometry, the resulting mode families and their large-spin asymptotics. The background we analyze is the exact spinning Morris-Thorne wormhole of \cite{Batic2026EPJC}. It has unit lapse, shape function $b(r)=r_0^2/r$, round two-spheres at fixed $t$ and $r$ and an analytic frame-dragging profile determined by the angular momentum $J$. Introducing the signed proper radial distance $\ell$ and the dimensionless variables $\eta=\ell/r_0$, $j=J/r_0^2$ and $\widehat{\omega}=r_0\omega$ removes the overall throat scale from the
scalar equation. For $J>0$, frame dragging decreases monotonically
from $\Omega_-=3\pi J/r_0^3$ at one infinity to zero at the other. The two asymptotic rest frames therefore rotate relative to each other. Outgoing radiation must be defined separately in these frames with
frequencies $\omega$ and $\omega-m\Omega_-$. The throat itself is an
ordinary point of the equation in $\ell$, so the global solution passes
through it without an additional boundary condition. Our numerical formulation keeps the radial and polar dependence together. We extract the axis behaviour and an analytic phase that interpolates between the two outgoing factors, compactify the real radial line and apply tensor-product Chebyshev collocation. The resulting quadratic matrix pencil can be treated with standard spectral and eigenvalue methods \cite{Ascher1979MC,Boyd2000,Trefethen2000,Tisseur2001}. For this particular background, the radial dependence of frame dragging also permits separation in associated Legendre functions. We use that structure to prove that the two-dimensional pencil decomposes exactly
into radial blocks labelled by the multipole index $L$. It provides
both an unambiguous mode label and a direct check of the assembled
operator. The same numerical framework can also be adapted to backgrounds that do not admit this angular decomposition. This requires deriving the operator coefficients and boundary factors for the new geometry and verifying convergence as the angular resolution is increased.

The global formulation gives several constraints before any spectrum is
computed. Two discrete symmetries imply mirror symmetry about
$\operatorname{Re}\widehat{\omega}=3\pi jm/2$, while each fixed multipole
family depends on $m$ and $j$ only through $mj$. In particular, the
axisymmetric spectrum is independent of rotation and inherits the modal
stability of the static scalar problem. A conserved flux places every
nonthreshold real quasinormal frequency in the window $0<\omega/(m\Omega_-)<1$. Thus, a continuous mode branch can cross from
damping to growth away from the thresholds only within that window. The global geometry also clarifies the meaning of $j_{\rm{ergo}}=2/(3\pi)$. This is the spin at which the ergoregion of a chosen asymptotic frame reaches the throat. For every nonzero spin, that ergoregion already extends to the opposite infinity, so the hypotheses of the classical horizonless ergoregion-instability argument \cite{Friedman1978CMP,Moschidis2018CMP} do not apply directly. We compute the sectors $m\in\{1,2,3\}$ at eleven spins up to $j=19$. Within the multipoles and frequencies resolved by the convergence tests, all modes are damped and lie outside the superradiant window. An independent evaluation of the radial pencils checks the discrete spectra, while integration along complex radial contours verifies the low-lying outgoing modes and extends the fundamental families beyond the range resolved by the fixed spectral grids. The frequency trends have a
geometric-optics interpretation in terms of an unstable circular null
orbit that moves into the second exterior. At large $mj$, a limiting
radial problem yields the scaling $\widehat{\omega}_-=O((mj)^{-1/2})$ for the computed families and explains their approach to the window edge. We also apply the two-dimensional solver to scalar perturbations of the Kerr black hole and compare the resulting quasinormal frequencies with those computed independently using Leaver's continued-fraction method \cite{refLeaver1985_Kerr,Stein2019}. Unlike the wormhole's exact angular reduction, the Kerr discretisation requires convergence in both
coordinates. For spins up to $a/M=0.99$, the identified converged Kerr
frequencies agree with the reference values to relative deviations
between $10^{-15}$ and approximately $3\times 10^{-7}$. Comparisons
with published continued-fraction, pseudospectral and asymptotic-iteration values \cite{KonoplyaZhidenko2006,BertiCardoso2006,LinEtAl2017,
Mamani2022EPJC,ChoEtAl2012} and with matrix-method and WKB approximations
\cite{LinEtAl2017,IyerWill1987PRD,Konoplya2003PRD} provide further
checks at their respective levels of accuracy.

The paper is organised as follows. Section \ref{metric_eom} develops
the global scalar problem, from the two asymptotic frames to the
compactified equation and its exact spectral constraints. Section \ref{spec_meth} turns this formulation into a Chebyshev matrix pencil and specifies how converged modes are selected. The resulting spectra and their large-spin interpretation are examined in Section \ref{results} before Section \ref{Kerr_valid} tests the same numerical framework against Kerr. The conclusions identify the limits of the present scalar calculation and the analyses needed to extend it.

\section{Global geometry and scalar-field equations}\label{metric_eom}

We first describe the rotating wormhole and its two asymptotic frames, then formulate the massless scalar wave problem on the complete geometry.
Throughout, we use geometrised units $c=G_{N}=1$ and adopt the metric signature $(-,+,+,+)$. In the coordinates $(t,r,\chi,\varphi)$, where
$\chi=\cos\theta\in[-1,1]$ and $\varphi\in[0,2\pi)$, the general Teo-type line element may be written as \cite{Teo1998PRD}
\begin{equation}\label{rotmet}
ds^{2}=-\left[N^{2}-\kappa^{2}\Omega^{2}(1-\chi^{2})\right]dt^{2}+\frac{r}{\Sigma}dr^{2}
-2\Omega\kappa^{2}(1-\chi^{2})dt d\varphi+\kappa^{2}\left[
\frac{d\chi^{2}}{1-\chi^{2}}+(1-\chi^{2})d\varphi^{2}\right],
\end{equation}
where $\Sigma(r,\chi)=r-B(r,\chi)$ and $\kappa(r,\chi)=rK(r,\chi)$. Here, $N$ is the lapse, $\Omega$ is the angular velocity of locally nonrotating frames, $B$ generalises the static shape function and $K$ determines the circumferential scale of the surfaces of constant $t$ and $r$. We require the rotating family to approach a static Morris-Thorne geometry continuously as its angular momentum vanishes. Thus, as $J\rightarrow 0$, we impose $N(r,\chi)\longrightarrow e^{\Phi(r)}$, $B(r,\chi)\longrightarrow b(r)$, $K(r,\chi)\longrightarrow 1$ and $\Omega(r,\chi)\longrightarrow 0$, where $\Phi(r)$ and $b(r)$ denote, respectively, the redshift and shape functions of the associated static wormhole. Under these conditions, \eqref{rotmet} reduces to
\begin{equation}
ds^{2}=-e^{2\Phi(r)}dt^{2}+\frac{dr^{2}}{1-b(r)/r}+r^2\left[
\frac{d\chi^{2}}{1-\chi^{2}}+(1-\chi^{2})d\varphi^{2}\right].
\end{equation}
In the static limit, $r$ is the areal radius. Along a path through the
wormhole, it decreases to the throat radius $r_0$ and then increases towards the other infinity. The same value $r>r_0$ therefore labels one sphere in each exterior, so $r$ alone is not a global radial coordinate. Thereofore, we introduce a signed radial coordinate below. The static throat condition is $b(r_{0})=r_{0}$, while the usual flare-out requirement is $b'(r_{0})<1$. In a chart that is asymptotically nonrotating at the chosen end, the asymptotic-flatness conditions are $b(r)/r\longrightarrow 0$, $N(r,\chi)\longrightarrow 1$, $K(r,\chi)\longrightarrow 1$ and $\Omega(r,\chi)\longrightarrow 0$ as $r\rightarrow\infty$. Notice that the condition on $\Omega$ refers to that end's rest frame and need not hold at the other end in the same coordinates. The supporting matter is an anisotropic fluid. Let $\{U^{\alpha},n_{(r)}^{\alpha},n_{(\chi)}^{\alpha},n_{(\varphi)}^{\alpha}\}$ be a comoving orthonormal tetrad with fluid four-velocity $U^\alpha$ and
spacelike principal-stress directions $n_{(i)}^\alpha$. Its normalisation is $U_{\alpha}U^{\alpha}=-1$, $n_{(i)\alpha}n_{(j)}^{\alpha}=\delta_{ij}$ and $U_{\alpha}n_{(i)}^{\alpha}=0$. In this frame, the mixed-index energy-momentum tensor takes the diagonal principal-pressure form
\begin{equation}
\widetilde{T}^{\alpha}{}_{\beta}=\widetilde{\rho}\,U^{\alpha}U_{\beta}+
P_{r}\,n_{(r)}^{\alpha}n_{(r)\beta}+P_{\chi}\,n_{(\chi)}^{\alpha}n_{(\chi)\beta}
+P_{\varphi}\,n_{(\varphi)}^{\alpha}n_{(\varphi)\beta}.
\end{equation}
The comoving energy density is $\widetilde{\rho}$ and $P_r$, $P_\chi$, $P_\varphi$ are the radial, polar and azimuthal principal pressures. Projecting the Einstein equations onto this tetrad identifies the density and stresses separately. It also gives the principal combinations entering the null energy condition, i.e. $\widetilde{\rho}+P_{r}$, $\widetilde{\rho}+P_{\chi}$ and $\widetilde{\rho}+P_{\varphi}$. The exact solution of \cite{Batic2026EPJC} specialises the Teo ansatz to $N(r,\chi)=1$, $K(r,\chi)=1$, $B(r,\chi)=b(r)$ and $\Omega(r,\chi)=\Omega(r)$. Hence, $\kappa(r,\chi)=r$, $\Sigma(r)=r-b(r)$ and the intrinsic geometry of each constant-$r$ two-surface remains that
of a round sphere of areal radius $r$. Since $r/\Sigma(r)=(1-b(r)/r)^{-1}$, the resulting line element assumes the form \cite{Batic2026EPJC}
\begin{equation}
ds^{2}=-\left[1-r^{2}\Omega^{2}(1-\chi^{2})\right]dt^{2}+\frac{dr^{2}}{1-\frac{b(r)}{r}}-2\Omega r^2(1-\chi^{2})dtd\varphi+r^{2}\left[\frac{d\chi^{2}}{1-\chi^{2}}+(1-\chi^{2})d\varphi^{2}\right].
\end{equation}
Completing the square in the $t$-$\varphi$ sector, we write the metric
in the final form
\begin{equation}\label{shift_form_metric}
ds^{2}=-dt^{2}+\frac{dr^{2}}{1-\frac{b(r)}{r}}+\frac{r^{2}d\chi^{2}}{1-\chi^{2}}
+r^{2}(1-\chi^{2})\left[d\varphi-\Omega dt\right]^{2}.
\end{equation}
For the Ellis-Bronnikov shape function $b(r)=r_0^2/r$, the solution on the
first exterior has frame-dragging function
\begin{equation}\label{frame_dragging_function}
\Omega(r)=\frac{3J}{r_{0}^{3}}\left[\operatorname{arctan}{\left(\frac{r_{0}}{\sqrt{r^{2}-r_{0}^{2}}}\right)}-\frac{r_{0}\sqrt{r^{2}-r_{0}^{2}}}{r^{2}}
\right]
\end{equation}
for $r\geq r_{0}$. The parameters $J$ and $r_0$ are the angular momentum and throat radius, respectively. Moreover, the frame dragging remains finite at the throat where we have
\begin{equation} \label{Omega_throat}
\Omega(r_{0})=\frac{3\pi J}{2r_{0}^{3}}
\end{equation}
and its asymptotic expansion is
\begin{equation}
\Omega(r)=\frac{2J}{r^{3}}+\frac{3Jr_{0}^{2}}{5r^{5}}+\mathcal{O}\left(\frac{1}{r^7}\right)
\end{equation}
as $r\rightarrow\infty$. In this regime, the $t\varphi$-metric coefficient reads
\begin{equation}
g_{t\varphi}=-\frac{2J}{r}(1-\chi^{2})+\mathcal{O}\left(\frac{1}{r^3}\right),
\end{equation}
which is the standard angular-momentum falloff. We underline that \eqref{frame_dragging_function} describes only one exterior \cite{Batic2026EPJC}. To distinguish the two exteriors and continue the metric through the throat, we introduce the signed proper radial distance $\ell\in\mathbb{R}$ defined by
\begin{equation}
r^{2}=\ell^{2}+r_{0}^{2},\qquad
d\ell=\pm\frac{dr}{\sqrt{1-b(r)/r}}
\end{equation}
with $\ell>0$ on the exterior described by \eqref{frame_dragging_function}, $\ell<0$ on the second exterior and $\ell=0$ at the throat. In terms of $\ell$, the smooth two-ended frame-dragging profile of the exact solution is
\begin{equation}
\Omega_{\rm{full}}(\ell)=\frac{3J}{r_0^3}\left[\frac{\pi}{2}\arctan{\left(\frac{\ell}{r_0}\right)}-\frac{r_0\ell}{\ell^2+r_0^2}\right].
\end{equation}
For $\ell>0$, this profile agrees with \eqref{frame_dragging_function} and provides its unique analytic continuation through the throat. Reflecting the positive-$\ell$ profile evenly would instead produce a jump in its derivative, from $+6J/r_0^4$ to $-6J/r_0^4$ at $\ell=0$ for $J\neq 0$, as it can be seen  from \eqref{Omega_derivative}. Such a reflection would not represent the same smooth solution. The one-exterior description in \cite{Batic2026EPJC} does not by itself fix the asymptotic frame on the other side. The analytic continuation does and its second-end limit is the nonzero constant $\Omega_-$ defined below.  Differentiating the global frame-dragging profile gives
\begin{equation}\label{Omega_derivative}
\frac{d\Omega_{\rm{full}}}{d\ell}=-\frac{6J}{(\ell^{2}+r_{0}^{2})^{2}}=-\frac{6J}{r^{4}}.
\end{equation}
For $J>0$, this derivative is strictly negative, so $\Omega_{\rm{full}}$ decreases as $\ell$ increases from the second asymptotic end to the first. Its limiting value at the second end is
\begin{equation}\label{Omega_minus}
\Omega_{-}:=\lim_{\ell\to-\infty}\Omega_{\rm{full}}(\ell)=\frac{3\pi J}{r_{0}^{3}}=2\Omega(r_{0}).
\end{equation}
At the throat where $\ell=0$, the angular velocity has decreased to
$\Omega(r_{0})=\Omega_{-}/2$, as it can be seen from \eqref{Omega_throat}. Moreover, it continues to decrease in the first exterior and tends to zero as $\ell\to+\infty$. Furthermore, the profile obeys the exact reflection identity
\begin{equation}\label{Omega_reflection}
\Omega_{\rm{full}}(-\ell)=\Omega_{-}-\Omega_{\rm{full}}(\ell),
\end{equation}
so that the frame dragging measured relative to the throat, i.e. $\Omega_{\rm{full}}(\ell)-\Omega(r_0)$, is an odd function of $\ell$. In particular, we have $\Omega_{\rm{full}}(-\ell)\neq\Omega_{\rm{full}}(\ell)$ for
$J\neq 0$ and
\begin{equation}\label{Omega_minus_expansion}
\Omega_{\rm{full}}(\ell)=\Omega_{-}-\frac{2J}{|\ell|^{3}}+\frac{12Jr_{0}^{2}}{5|\ell|^{5}}+\mathcal{O}\left(\frac{1}{|\ell|^7}\right)
\end{equation}
as $\ell\rightarrow-\infty$. Since $1-b(r)/r=\ell^{2}/r^{2}$, the line element \eqref{shift_form_metric} takes the global form
\begin{equation}\label{global_metric}
ds^{2}=-dt^{2}+d\ell^{2}+(\ell^{2}+r_{0}^{2})\left\{\frac{d\chi^{2}}{1-\chi^{2}}
+(1-\chi^{2})\left[d\varphi-\Omega_{\rm{full}}(\ell)dt\right]^{2}\right\}
\end{equation}
with $\ell\in\mathbb{R}$. The coefficients of \eqref{global_metric} are analytic in $\ell$, so the throat is regular in this chart. The first end is asymptotically Minkowskian in $(t,\ell,\chi,\varphi)$. At the second end, an asymptotically inertial chart is obtained by the rigid change of the azimuthal angle, that is
\begin{equation}\label{rotating_angle}
\varphi_{-}=\varphi-\Omega_{-}t.
\end{equation}
Indeed, we have $d\varphi-\Omega_{\rm{full}}dt=d\varphi_- +[\Omega_- -\Omega_{\rm{full}}(\ell)]dt$ and the bracket is $\mathcal O(|\ell|^{-3})$ by \eqref{Omega_reflection} and \eqref{Omega_minus_expansion}. Thus, the two asymptotic rest frames differ by the constant angular velocity $\Omega_-$. Notice that this relation between the ends cannot be determined from one exterior alone and it fixes the frequency shift in the outgoing condition at the second infinity, as in the numerical rotating Ellis-Bronnikov calculation of \cite{Khoo2024PRD}.

\subsection{Asymptotic frames and the ergoregion threshold}

The norm of the Killing field $\partial_t$ follows from \eqref{global_metric}, i.e. $g_{tt}=-1+(\ell^{2}+r_{0}^{2})(1-\chi^{2})\Omega_{\rm{full}}^{2}(\ell)$. For $J\neq 0$, the limit $\Omega_{\rm{full}}\to\Omega_-\neq 0$ implies
that $g_{tt}$ grows as $\ell^2$ at the second end, away from the axis. The first-end time translation $\partial_t$ therefore becomes spacelike there. Relative to the second-end inertial chart, its orbits are those of rigid rotation and cross a light-cylinder boundary. This observation makes the choice of time translation essential. Below, the \emph{ergoregion of a frame} means the region where that frame's asymptotic time translation is spacelike. More generally, for constant $\Omega_*$ the stationary Killing field $\xi_*=\partial_t+\Omega_*\partial_\varphi$ has norm $g(\xi_{*},\xi_{*})=-1+(\ell^{2}+r_{0}^{2})(1-\chi^{2})\left[\Omega_{\rm{full}}(\ell)-\Omega_{*}\right]^{2}$. To remain timelike near the first infinity, away from the axis, this field requires $\Omega_*=0$, while, at the second infinity, we must have $\Omega_*=\Omega_-$. For $J\neq 0$, these conditions are incompatible, so no field of this form is timelike throughout the geometry. Notice that the asymptotic time translations are $\partial_t$ at the first end and $\partial_t+\Omega_-\partial_\varphi$ at the second. On the equatorial plane, $\partial_{t}$ is spacelike wherever $F(\ell)>1$ with
\begin{equation}\label{ergo_function}
F(\ell):=(\ell^{2}+r_{0}^{2})\Omega_{\rm{full}}^{2}(\ell),\qquad
F'(\ell)=2\Omega_{\rm{full}}(\ell)\left[\ell\Omega_{\rm{full}}(\ell)-\frac{6J}{\ell^{2}+r_{0}^{2}}\right],
\end{equation}
where \eqref{Omega_derivative} has been used.  For $J>0$, \eqref{Omega_derivative} gives $\Omega_{\rm{full}}(\ell)=6J\int_{\ell}^{\infty}(\sigma^2+r_{0}^{2})^{-2}d\sigma>0$. For $\ell\leq 0$ the bracket in \eqref{ergo_function} is therefore negative, while for $\ell>0$ the estimate $\Omega_{\rm{full}}(\ell)\leq 6J(\ell^{2}+r_{0}^{2})^{-1}r_{0}^{-1}\arctan{(r_{0}/\ell)}<6J/[\ell(\ell^{2}+r_{0}^{2})]$ leads to the same conclusion. The function $F$ is thus strictly decreasing on $\mathbb{R}$, from $+\infty$ at $\ell\to-\infty$ to zero at $\ell\to+\infty$. The equatorial ergoregion of the first asymptotic frame is consequently the connected set $\ell<\ell_{e}(J)$, where $F(\ell_{e})=1$ and it contains the throat if and only if $F(0)=r^2_{0}\Omega^2(r_{0})>1$. In terms of $j=J/r_{0}^{2}$, this condition reads
\begin{equation}\label{j_ergo}
j>j_{\rm{ergo}}:=\frac{2}{3\pi}\simeq 0.2122 .
\end{equation}
Since, away from the equator, the factor $1-\chi^2$ reduces the Killing norm, the ergoregion first reaches the throat in the equatorial plane. Moreover, \eqref{Omega_reflection} maps the first-frame ergoregion to the second-frame ergoregion. On the equator, the latter is $\ell>-\ell_e(J)$. For $0<j<j_{\rm{ergo}}$, a neighbourhood of the throat separates the two regions and both asymptotic time translations are timelike there. For $j>j_{\rm{ergo}}$, the regions overlap near the equatorial throat. Thus, $j_{\rm{ergo}}$ is the threshold for an ergoregion to reach the throat in the chosen asymptotic frame. However, it is not a threshold for its existence
on the complete manifold. For every $j>0$, each frame's ergoregion extends
to the opposite infinity. Hence, the familiar ergoregion-instability argument for horizonless spacetimes assumes an asymptotic time translation with positive
outgoing energy flux at infinity \cite{Friedman1978CMP,CominsSchutz1978PRSA,Cardoso2008PRD,Moschidis2018CMP}.
Negative-energy perturbations then cannot decay by radiating only positive
energy. Here, the energy associated with $\partial_t$ can escape with either
sign through the second end. The usual argument therefore does not apply
directly and the ergoregion alone does not decide whether growing scalar
modes exist. We address that question through the spectral problem below.

\subsection{Scalar wave equation and radiation conditions}

We consider a minimally coupled massless scalar field $\Phi$ on the fixed
wormhole background \eqref{global_metric}. Its backreaction on the metric and
the supporting matter is neglected. The field satisfies the covariant
Klein-Gordon equation
\begin{equation}\label{KG_covariant}
\Box\Phi:=\nabla_{\mu}\nabla^{\mu}\Phi=\frac{1}{\sqrt{-g}}\partial_{\mu}\left(
\sqrt{-g}g^{\mu\nu}\partial_{\nu}\Phi\right)=0,
\end{equation}
where $g=\det(g_{\mu\nu})$. In the global coordinates $(t,\ell,\chi,\varphi)$, the nonzero inverse-metric components are
\begin{equation}
g^{tt}=-1,\quad
g^{t\varphi}=-\Omega_{\rm{full}}(\ell),\quad
g^{\varphi\varphi}=\frac{1}{(\ell^{2}+r_{0}^{2})(1-\chi^{2})}-\Omega_{\rm{full}}^{2}(\ell),\quad
g^{\ell\ell}=1,\quad
g^{\chi\chi}=\frac{1-\chi^{2}}{\ell^{2}+r_{0}^{2}}.
\end{equation}
The determinant and invariant volume density are
\begin{equation}\label{metric_determinant}
g=-(\ell^{2}+r_{0}^{2})^{2},\qquad
\sqrt{-g}=\ell^{2}+r_{0}^{2}.
\end{equation}
The frame-dragging terms cancel from the determinant because of the shift
form of the $t\varphi$-part of \eqref{global_metric}. Unlike the
areal-coordinate density, the volume density \eqref{metric_determinant} is finite and nonzero at the throat. Substitution into \eqref{KG_covariant} gives
\begin{equation}\label{KG_explicit_operator}
-\left[\partial_{t}+\Omega_{\rm{full}}(\ell)\partial_{\varphi}\right]^{2}\Phi
+\frac{1}{r^{2}}\partial_\ell\left(r^2\partial_\ell\Phi\right)+
\frac{1}{r^{2}}\partial_\chi\left[(1-\chi^{2})\partial_\chi\Phi\right]
+\frac{\partial_{\varphi\varphi}\Phi}{r^{2}(1-\chi^{2})}=0
\end{equation}
with $r^{2}=\ell^{2}+r_{0}^{2}$. The $t\varphi$-term combines the two Killing derivatives as $\partial_t+\Omega_{\rm{full}}\partial_\varphi$. This identifies the local frequency shift that enters the wave equation. Stationarity and axisymmetry allow us to construct the following ansatz
\begin{equation}
\Phi(t,\ell,\chi,\varphi)=e^{-i\omega t}e^{im\varphi}\Psi_{\omega m}(\ell,\chi),
\end{equation}
where $\omega\in\mathbb{C}$ and $m\in\mathbb{Z}$ are the mode frequency and
azimuthal number. With the convention $e^{-i\omega t}$, a negative imaginary
part of $\omega$ describes damping. The $t\varphi$-derivative acts on
this mode through $-i[\omega-m\Omega_{\rm{full}}(\ell)]$ and \eqref{KG_explicit_operator} becomes
\begin{equation}\label{scalar_reduced_PDE}
(1-\chi^{2})\left[(\ell^{2}+r_{0}^{2})\partial_{\ell\ell}\Psi_{\omega m}
+2\ell\partial_\ell\Psi_{\omega m}+
(1-\chi^{2})\partial_{\chi\chi}\Psi_{\omega m}-2\chi\partial_\chi\Psi_{\omega m}
+(\ell^{2}+r_{0}^{2})\bigl(\omega-m\Omega_{\rm full}(\ell)\bigr)^{2}
\Psi_{\omega m}\right]=m^{2}\Psi_{\omega m}.
\end{equation}
To remove the overall throat scale, we introduce the dimensionless quantities
$\eta=\ell/r_{0}$, $j=J/r_{0}^2$, $\widehat{\omega}=r_0\omega$ and $\widehat{\Omega}(\eta)=r_{0}\Omega_{\rm{full}}(r_{0}\eta)$. In these variables, \eqref{scalar_reduced_PDE} is independent of $r_0$. Suppressing the subscripts $\omega m$ on the reduced field yields
\begin{equation}\label{dimensionless_scalar_PDE}
(1+\eta^2)(\chi^2-1)\partial_{\eta\eta}\Psi-(1-\chi^2)^2
\partial_{\chi\chi}\Psi+2\eta(\chi^2-1)\partial_\eta\Psi
+2\chi(1-\chi^2)\partial_\chi\Psi+V(\eta,\chi)\Psi=0,
\end{equation}
where the effective frequency-dependent term is
\begin{equation}
V(\eta,\chi)=(1+\eta^2)(\chi^2-1)\left[\widehat{\omega}-m\widehat{\Omega}(\eta)
\right]^{2}+m^{2}.
\end{equation}
On the angular interior $-1<\chi<1$, division by $\chi^2-1$ gives the
useful divergence form
\begin{equation}\label{divergence_form_PDE}
\partial_\eta\left[(1+\eta^2)\partial_\eta\Psi\right]+(1+\eta^2)
\left[\widehat{\omega}-m\widehat{\Omega}(\eta)\right]^{2}\Psi+\mathcal{L}_{m}\Psi=0,
\end{equation}
where the angular differential operator is
\begin{equation}\label{angular_operator_Lm}
\mathcal{L}_{m}=(1-\chi^2)\partial_{\chi\chi}-2\chi\partial_\chi-\frac{m^{2}}{1-\chi^{2}}.
\end{equation}
Moreover, the dimensionless frame-dragging profile and its derivative are
\begin{equation}\label{dimensionless_frame_dragging}
\widehat{\Omega}(\eta)=3j\left(\frac{\pi}{2}-\arctan{\eta}-\frac{\eta}{1+\eta^2}
\right),\qquad
\frac{d\widehat{\Omega}}{d\eta}=-\frac{6j}{(1+\eta^2)^{2}}
\end{equation}
with $\eta\in\mathbb{R}$. Its values at the throat and at the second infinity are
\begin{equation}
\widehat{\Omega}(0):=\widehat{\Omega}_{0}=\frac{3\pi j}{2},\qquad
\widehat{\Omega}_{-}:= r_{0}\Omega_{-}=3\pi j=2\widehat{\Omega}_{0}
\end{equation}
and its asymptotic behaviour is
\begin{equation}\label{Omega_hat_asymptotics}
\widehat{\Omega}(\eta)=\frac{2j}{\eta^{3}}-\frac{12j}{5\eta^{5}}+\mathcal{O}\left(\frac{1}{\eta^7}\right),\quad \eta\rightarrow+\infty,\qquad
\widehat{\Omega}(\eta)=\widehat{\Omega}_{-}-\frac{2j}{|\eta|^{3}}+\frac{12j}{5|\eta|^{5}}+\mathcal{O}\left(\frac{1}{\eta^7}\right),\quad\eta\rightarrow-\infty.
\end{equation}
We shall use two shifted frequencies. The first is measured in the
asymptotic rest frame of the second exterior, while the second in the frame
co-rotating with the throat, that is
\begin{equation}
\widehat{\omega}_{-}:=\widehat{\omega}-m\widehat{\Omega}_{-},\qquad
\varpi:=\widehat{\omega}-m\widehat{\Omega}_{0}=\frac{\widehat{\omega}+\widehat{\omega}_{-}}{2}.
\end{equation}
At fixed $m$, the dimensionless eigenvalue problem depends only on $j$. The throat radius sets the physical frequency through $\omega=\widehat{\omega}/r_0$. Furthermore, the coordinate $\eta$ covers the complete manifold with the throat at the interior point $\eta=0$ and the two infinities at $\eta\to\pm\infty$. QNMs are regular at the two components of the symmetry axis ($\chi=\pm1$) and outgoing at both infinities. At each infinity, the outgoing condition is defined in that end's asymptotic rest frame. No additional boundary condition is imposed at the throat. We first derive these local behaviours and then extract their leading factors from the field. A compact radial coordinate $x$ will place the remaining equation on the square $(x,\chi)\in[-1,1]\times[-1,1]$. This compact domain supports a tensor-product Chebyshev discretisation. The distinction between smoothness of the residual field and selection of the outgoing branch will be important at the irregular endpoints.

The radial principal coefficient in \eqref{dimensionless_scalar_PDE} is $(1+\eta^2)(\chi^2-1)$, which is nonzero for real $\eta$ and $-1<\chi<1$, while all radial coefficients are analytic at $\eta=0$. In particular, projection onto an associated Legendre function gives a radial equation whose
two independent solutions are analytic through the throat. The areal coordinate $z=r/r_0=\sqrt{1+\eta^2}$ obscures this regularity because it covers only one exterior at a time. In that coordinate, the radial equation degenerates at $z=1$, with leading behaviours $(z-1)^0$ and $(z-1)^{1/2}$. Since $z-1=2^{-1}\eta^{2}+\mathcal{O}(\eta^{4})$, an analytic solution
$\Psi=a_0(\chi)+a_1(\chi)\eta+\mathcal{O}(\eta^2)$ restricted to either
exterior $\eta=\pm\sqrt{z^2-1}$, generally contains both behaviours. Its even
part produces the exponent $0$ and its odd part the exponent $1/2$. Thus,
regularity through the throat does not select one of the two areal-coordinate
branches. Their relative amplitude is determined by the global radiation
conditions. At $j=0$, reflection symmetry separates the modes into even solutions with $\partial_\eta\Psi|_{\eta=0}=0$ and odd solutions with
$\Psi|_{\eta=0}=0$. For $mj\neq 0$, the coefficient $[\widehat{\omega}-m\widehat{\Omega}(\eta)]^2$ is not invariant under $\eta\mapsto-\eta$ at fixed $(\widehat{\omega},m)$, so definite radial parity is no longer required. The corresponding transformation of the spectral parameters is discussed in Section \ref{spec_symm}. We determine now the first-infinity asymptotics using the construction in \cite{Batic2024EPJC,Olver1994MAA}. The angular dependence can be kept explicit by treating $\mathcal{L}_m$ as an operator on functions of $\chi$. Dividing \eqref{divergence_form_PDE} by $1+\eta^2$ gives
\begin{equation}\label{PDE_infinity_standard_form}
\partial_{\eta\eta}\Psi+p(\eta)\partial_\eta\Psi+\mathcal{Q}(\eta)\Psi=0,\quad  p(\eta)=\frac{2\eta}{1+\eta^{2}},\quad  
\mathcal{Q}(\eta)=\left[\widehat{\omega}-m\widehat{\Omega}(\eta)\right]^{2}+
\frac{1}{1+\eta^{2}}\mathcal{L}_{m}
\end{equation}
with $\mathcal{L}_m$ defined in \eqref{angular_operator_Lm}. The asymptotic construction below may be applied separately to each associated Legendre component, where $\mathcal{L}_m$ is replaced by $-L(L+1)$. Notice that the operator notation collects those componentwise results without imposing separation in the numerical formulation. To classify the first infinity, we introduce the reciprocal radial coordinate $\tau=1/\eta$. The differential equation \eqref{PDE_infinity_standard_form} then becomes
\begin{equation}\label{PDE_tau_equation}
\partial_{\tau\tau}\Psi+C(\tau)\partial_\tau\Psi+\mathcal{D}(\tau)\Psi=0,\qquad
C(\tau)=\frac{2\tau}{1+\tau^{2}},\qquad
\mathcal{D}(\tau)=\frac{1}{\tau^{4}}\left\{\left[\widehat{\omega}-m\widehat{\Omega}\left(\frac{1}{\tau}\right)\right]^{2}+\frac{\tau^{2}}{1+\tau^{2}}\mathcal{L}_{m}\right\}
\end{equation}
with
\begin{equation}
\widehat{\Omega}\left(\frac{1}{\tau}\right)=2j\tau^{3}-\frac{12j}{5}\tau^{5}
+\mathcal{O}(\tau^{7})
\end{equation}
by means of \eqref{Omega_hat_asymptotics}. The rank-one classification corresponds to $k=1$ \cite{Bender1999}, for which $\tau^{k+1}C(\tau)=\tau^{2}C(\tau)=\mathcal{O}(\tau^{3})$, whereas
$\tau^{2k+2}\mathcal{D}(\tau)=\tau^{4}\mathcal{D}(\tau)=\widehat{\omega}^{2}
+\tau^{2}\mathcal{L}_{m}-4mj\widehat{\omega}\tau^{3}+\mathcal{O}(\tau^{4})$.
Since the leading coefficient $D_{0}=\widehat{\omega}^{2}$ is nonzero when $\widehat{\omega}\neq 0$, this point at infinity is an irregular
singular point of rank one. The threshold $\widehat{\omega}=0$ requires a
separate asymptotic analysis. Away from that threshold, the construction of \cite{Olver1994MAA} starts from the inverse-radius expansions
\begin{equation}
p(\eta)=\sum_{n=0}^{\infty}\frac{\mathfrak{f}_{n}}{\eta^{n}}=\frac{2}{\eta}
-\frac{2}{\eta^{3}}+\mathcal{O}\left(\frac{1}{\eta^5}\right),\quad     \mathcal{Q}(\eta)=\sum_{n=0}^{\infty}\frac{\mathfrak{g}_{n}}{\eta^{n}}= \widehat{\omega}^{2}+\frac{\mathcal{L}_{m}}{\eta^{2}}-\frac{4mj\widehat{\omega}}{\eta^{3}}+\mathcal{O}\left(\frac{1}{\eta^4}\right).
\end{equation}
The coefficients entering the characteristic equations are therefore
\begin{equation}\label{Olver_coefficients}
\mathfrak{f}_{0}=0,\qquad
\mathfrak{f}_{1}=2,\qquad
\mathfrak{g}_{0}=\widehat{\omega}^{2},\qquad
\mathfrak{g}_{1}=0.
\end{equation}
The angular operator first enters at $\mathfrak{g}_2$ and frame dragging
at $\mathfrak{g}_3$. They affect the subleading coefficients but not the
leading exponential and algebraic factors. We therefore seek formal solutions
of the form
\begin{equation}
\Psi^{(s)}(\eta,\chi)=\eta^{\mu_{s}}e^{\lambda_{s}\eta}\sum_{n=0}^{\infty}
\frac{a_{n,s}(\chi)}{\eta^{n}}\qquad
\end{equation}
with $s\in\{+,-\}$ labelling the two asymptotic branches. Their exponential
rates $\lambda_s$ and algebraic exponents $\mu_s$ satisfy
\begin{equation}
\lambda^{2}+\mathfrak{f}_{0}\lambda+\mathfrak{g}_{0}=0,\qquad
\mu_{s}=-\frac{\mathfrak{f}_{1}\lambda_{s}+\mathfrak{g}_{1}}{\mathfrak{f}_{0}
+2\lambda_{s}}.
\end{equation}
Using \eqref{Olver_coefficients}, the first characteristic equation
becomes $\lambda^{2}+\widehat{\omega}^{2}=0$, and hence, we find $\lambda_{\pm}=\pm i\widehat{\omega}$. The corresponding algebraic exponents are $\mu_{\pm}=-1$. Therefore, the two independent asymptotic branches are
\begin{equation}
\Psi^{(\pm)}(\eta,\chi)\underset{\eta\rightarrow+\infty}{\sim}\frac{e^{\pm i\widehat{\omega}\eta}}{\eta}a_{0,\pm}(\chi).
\end{equation}
For real nonzero frequency, the factor $e^{-i\omega t+i\widehat{\omega}\eta}$
describes radiation travelling towards increasing $\eta$. Its continuation to
complex frequency defines the outgoing branch. The first-end radiation
condition is therefore
\begin{equation}\label{PDE_outgoing_boundary_condition}
\Psi(\eta,\chi)\underset{\eta\rightarrow+\infty}{\longrightarrow}\frac{e^{i\widehat{\omega}\eta}}{\eta}a_{+}(\chi),
\end{equation}
where $a_{+}(\chi)$ is required to possess the appropriate regular
behaviour at the symmetry axis. At the second infinity, we set $\zeta=-\eta\to+\infty$. The radial first-derivative term and the angular term keep the same form, but the frame-dragging profile now tends to $\widehat{\Omega}_-$. Moreover, \eqref{Omega_hat_asymptotics} gives
\begin{equation}
\left[\widehat{\omega}-m\widehat{\Omega}(\eta)\right]^{2}=\widehat{\omega}_{-}^{2}+\frac{4mj\widehat{\omega}_{-}}{\zeta^{3}}+\mathcal{O}\left(\frac{1}{\zeta^5}\right)
\end{equation}
for $\zeta\rightarrow+\infty$. Thus, the leading coefficients are the same as at the first infinity with $\widehat{\omega}$ replaced by $\widehat{\omega}_-$. For $\widehat{\omega}_-\neq 0$, we again obtain a rank-one irregular singular point with exponential rates $\lambda_\pm=\pm i\widehat{\omega}_-$ and algebraic exponents $\mu_\pm=-1$. The two branches are
\begin{equation}
\Psi^{(\pm)}(\eta,\chi)\underset{\eta\rightarrow-\infty}{\sim}\frac{e^{\pm i\widehat{\omega}_{-}|\eta|}}{|\eta|} a_{0,\pm}(\chi).
\end{equation}
The outgoing branch is identified in the asymptotically inertial coordinates $(t,\ell,\chi,\varphi_-)$ of \eqref{rotating_angle}. Notice that in the original ansatz for $\Phi$, the time and azimuthal factors must be adjusted accordingly so that $\Phi=e^{-i(\omega-m\Omega_{-})t}e^{im\varphi_{-}}\Psi_{\omega m}(\ell,\chi)$. The local asymptotic frequency is therefore $\omega-m\Omega_-=\widehat{\omega}_-/r_0$. Notice that radiation travelling towards increasing $|\eta|$, and hence towards the second infinity, has the radial factor $e^{i\widehat{\omega}_-|\eta|}$. Its analytic continuation defines the second-end outgoing condition
\begin{equation}\label{PDE_outgoing_boundary_condition_minus}
\Psi(\eta,\chi)\underset{\eta\rightarrow-\infty}{\longrightarrow}
\frac{e^{i\widehat{\omega}_{-}|\eta|}}{|\eta|}a_{-}(\chi).
\end{equation}
For $mj\neq 0$, using $e^{i\widehat{\omega|}\eta|}$ at this end would impose
the radiation condition in the wrong asymptotic frame. The shift by
$m\widehat{\Omega}_-$ is required by the relative rotation of the two exteriors. Let us now discuss the regularity on the symmetry axis. The endpoints $\chi=1$ and $\chi=-1$ are the north and south poles of
each constant-$t$, constant-$\eta$ two-sphere. To obtain their local
behaviour, we write \eqref{dimensionless_scalar_PDE} as
\begin{equation}\label{dimensionless_scalar_PDE_axis_form}
(1+\eta^{2})\partial_{\eta\eta}\Psi+2\eta\partial_\eta\Psi+(1+\eta^{2})
\left[\widehat{\omega}-m\widehat{\Omega}(\eta)\right]^{2}\Psi+
(1-\chi^{2})\partial_{\chi\chi}\Psi-2\chi\partial_\chi\Psi-\frac{m^{2}}{1-\chi^{2}}\Psi=0.
\end{equation}
The radial terms in \eqref{dimensionless_scalar_PDE_axis_form} remain
regular as $\chi\rightarrow\pm 1$. The leading singular behaviour is
therefore determined entirely by the angular differential operator
$\mathcal{L}_{m}$ defined in \eqref{angular_operator_Lm}. We first examine the neighbourhood of $\chi=1$. Introducing the local coordinate $u=1-\chi$ with $u\rightarrow 0^{+}$ yields $1-\chi^{2}=2u+\mathcal{O}(u^{2})$. We seek a local solution of the form
\begin{equation}
\Psi(\eta,\chi)=u^{\sigma}\sum_{n=0}^{\infty}a_{n}(\eta)u^{n}
\end{equation}
with $a_{0}\neq 0$. If we substitute this expansion into \eqref{dimensionless_scalar_PDE_axis_form} and collect the leading terms of order $u^{\sigma-1}$, we end up with the corresponding indicial equation $
4\sigma^{2}-m^{2}=0$ whose characteristic exponents are $\sigma_{\pm}=\pm|m|/2$. For $m\neq 0$, the negative exponent gives a divergent branch and is excluded by axis regularity. When $m=0$, the exponents coincide and the second independent local solution is logarithmic, so it is excluded as well. In both cases,as $\chi\rightarrow 1^{-}$, the regular behaviour is
\begin{equation}
\Psi(\eta,\chi)\sim(1-\chi)^{|m|/2}A_{+}(\eta),
\end{equation}
where $A_{+}(\eta)$ remains finite. An analogous analysis near $\chi=-1$ is obtained by introducing $v=1+\chi$ with $v\rightarrow 0^{+}$. $1-\chi^{2}=2v+\mathcal{O}(v^{2})$, the same indicial equation and the same pair of characteristic exponents are obtained. As $\chi\rightarrow-1^{+}$, regularity at the south pole therefore requires
\begin{equation}
\Psi(\eta,\chi)\sim(1+\chi)^{|m|/2}A_{-}(\eta),
\end{equation}
where $A_{-}(\eta)$ is finite. The local analysis now provides the factors needed to reformulate the problem on a compact domain. We extract the axis powers, the inverse-radius amplitude and an outgoing phase, and define the residual field $\varPhi$ by
\begin{equation}\label{scalar_asymptotic_factorisation}
\Psi(\eta,\chi)=(1-\chi^{2})^{|m|/2}\frac{e^{iS(\eta)}}{\sqrt{1+\eta^{2}}}
\varPhi(\eta,\chi),
\end{equation}
with the phase
\begin{equation}
S(\eta)=\eta\left[\sigma(\eta)\widehat{\omega}
-\left(1-\sigma(\eta)\right)\,\widehat{\omega}_{-}\right],\qquad
\sigma(\eta)=\frac{1}{2}\left(1+\frac{\eta}{\sqrt{1+\eta^{2}}}\right).
\end{equation}
Notice that the function $\sigma(\eta)$ interpolates smoothly between $0$ at the second asymptotic end, $\eta\to-\infty$, and $1$ at the first, $\eta\to+\infty$. Accordingly, $S=\widehat{\omega}\eta+\mathcal{O}(\eta^{-1})$ as $\eta\to+\infty$ and $S=\widehat{\omega}_-|\eta|+\mathcal{O}(|\eta|^{-1})$ as $\eta\to-\infty$. Hence, the exponential and inverse-radius factors reproduce the leading outgoing behaviours at both ends, while $(1-\chi^2)^{|m|/2}$ supplies the regular axis powers. Using the throat-frame frequency $\varpi=\widehat{\omega}-m\widehat{\Omega}_{0}$ defined earlier, we can rewrite the phase as
\begin{equation}\label{phase_alternative}
S(\eta)=\varpi h(\eta)+m\widehat{\Omega}_{0}\eta,\qquad
h(\eta)=\frac{\eta^{2}}{\sqrt{1+\eta^{2}}},
\end{equation}
which makes the analyticity of $S$ on the real radial line explicit. In particular, the factorisation introduces no new throat condition. Its local radial wave number is
\begin{equation}
k(\eta):=S'(\eta)=\varpi\,H(\eta)+m\widehat{\Omega}_{0},\qquad
H(\eta)=h'(\eta)=\frac{\eta(\eta^{2}+2)}{(1+\eta^{2})^{3/2}}
\end{equation}
with $k\to\pm\widehat{\omega}$ as $\eta\to\pm\infty$. If we substitute \eqref{scalar_asymptotic_factorisation} into \eqref{dimensionless_scalar_PDE} and cancel the common prefactor, we get 
\begin{equation}\label{regular_pde}
P_{20}(\eta,\chi)\partial_{\eta\eta}\varPhi+
P_{02}(\eta,\chi)\partial_{\chi\chi}\varPhi+
P_{10}(\eta,\chi)\partial_\eta\varPhi+
P_{01}(\eta,\chi)\partial_\chi\varPhi+
P_{00}(\eta,\chi)\varPhi=0,
\end{equation}
where the coefficient functions are given by
\begin{align}
P_{20}(\eta,\chi)&=1+\eta^{2},\qquad
P_{02}(\eta,\chi)=1-\chi^{2},\qquad
P_{10}(\eta,\chi)=2i(1+\eta^{2})k(\eta),\qquad
P_{01}(\eta,\chi)=-2(|m|+1)\chi,\\
P_{00}(\eta,\chi)&=(1+\eta^{2})\left\{\left[\widehat{\omega}-m\widehat{\Omega}(\eta)\right]^{2}-k^{2}(\eta)\right\}+\frac{i\varpi(2-\eta^{2})}{(1+\eta^{2})^{3/2}}-\frac{1}{1+\eta^{2}}-|m|(|m|+1).
\end{align}
Since $P_{20}>0$ and the radial coefficients are analytic, the throat
remains an ordinary point. To place the two infinities at finite endpoints,
we use the compactification
\begin{equation}\label{compactification}
x=\frac{2}{\pi}\arctan{\eta},\qquad
\eta=\tan{\left(\frac{\pi x}{2}\right)},
\end{equation}
The throat maps to $x=0$, while the two asymptotic ends are sent to $x=\pm 1$. For brevity, we define the following functions $c_{x}:=\cos{(\pi x/2)}$ and $s_{x}:=\sin{(\pi x/2)}$. Then, $1+\eta^2=c_x^{-2}$ and $\partial_\eta=(2/\pi)c_x^2\partial_x$. We use the same function symbols after composition with $\eta=\tan{(\pi x/2)}$. In particular, the compact frame-dragging profile is the entire function
\begin{equation}
\widehat{\Omega}(x)=\frac{3j}{2}\left[\pi(1-x)-s_{2x}\right],\qquad
\frac{d\widehat{\Omega}}{dx}=-3\pi jc_{x}^{2},
\end{equation}
while
\begin{equation}
H(x)=s_{x}\left(1+c_{x}^{2}\right),\qquad
k(x)=\varpi H(x)+m\widehat{\Omega}_{0}.
\end{equation}
Changing variables in \eqref{regular_pde} and multiplying by $-1$ to
match the implementation gives
\begin{equation}
M_{20}(x,\chi)\partial_{xx}\varPhi+
M_{02}(x,\chi)\partial_{\chi\chi}\varPhi+
M_{10}(x,\chi)\partial_x\varPhi+
M_{01}(x,\chi)\partial_\chi\varPhi+
M_{00}(x,\chi)\varPhi=0,
\end{equation}
where the coefficients are given by
\begin{align}
M_{20}(x,\chi)&=-\frac{4c_{x}^{2}}{\pi^{2}}\qquad
M_{02}(x,\chi)=\chi^2 -1,\qquad
M_{10}(x,\chi)=\frac{2s_{2x}}{\pi}-\frac{4i}{\pi}k(x),\qquad
M_{01}(x,\chi)=2(|m|+1)\chi,\label{M20}\\
M_{00}(x,\chi)&=\frac{k^{2}(x)-\left[\widehat{\omega}-m\widehat{\Omega}(x)\right]^{2}}{c_{x}^{2}}+i\varpi c_{x}\left(1-3c_{x}^{2}\right)+c_{x}^{2}+|m|(|m|+1).\label{M00}
\end{align}
All coefficients extend analytically to the closed square. Only the first
term of $M_{00}$ requires checking because it contains the term $c_x^{-2}$. At $x=1$, $k=\widehat{\omega}$ and $\widehat{\Omega}=0$, while at $x=-1$,
$k=-\widehat{\omega}_-$ and $\widehat{\Omega}=\widehat{\Omega}_-$. Moreover, the derivatives $d\widehat{\Omega}/dx$ and $dH/dx=(\pi/2)c_x(3c_x^2-1)$ are zero at both endpoints. The numerator therefore vanishes to at least second order, thus cancelling the zeros of $c_x^2$. The limiting values are
\begin{equation}
M_{00}\big|_{x=1}=\widehat{\omega}\varpi+|m|(|m|+1),\qquad
M_{00}\big|_{x=-1}=\widehat{\omega}_{-}\varpi+|m|(|m|+1).
\end{equation}
Because the coefficient of $\partial_x^2\varPhi$ vanishes at $x=\pm 1$, the endpoint equations reduce to the first-order relations
\begin{equation}\label{endpoint_relations}
-\frac{4i}{\pi}\widehat{\omega}\partial_{x}\varPhi(1,\chi) +\mathcal{A}_{m}\varPhi(1,\chi)+\widehat{\omega}\varpi\varPhi(1,\chi)=0, \qquad
\frac{4i}{\pi}\widehat{\omega}_{-}\partial_{x}\varPhi(-1,\chi) +\mathcal{A}_{m}\varPhi(-1,\chi)+\widehat{\omega}_{-}\varpi\varPhi(-1,\chi)=0,
\end{equation}
where
\begin{equation}\label{angular_operator_Am}
\mathcal{A}_{m}=(\chi^{2}-1)\partial_{\chi\chi}+2(|m|+1)\chi\partial_\chi+|m|(|m|+1).
\end{equation}
We underline that analyticity of these coefficients does not make the endpoints ordinary points, nor does it imply analyticity of the residual eigenfunctions. When $\widehat{\omega}$ and $\widehat{\omega}_-$ are nonzero, the radial endpoints remain irregular singular points. The outgoing solutions admit formal expansions in integer powers of $1\mp x$ and satisfy \eqref{endpoint_relations}. The incoming branches instead contain the
factors
\begin{equation}\label{excluded_branches}
\varPhi\propto\exp\left[-\frac{4i\widehat{\omega}}{\pi(1-x)}\right],\quad x\rightarrow 1,\qquad
\varPhi\propto\exp\left[-\frac{4i\widehat{\omega}_{-}}{\pi(1+x)}\right],\quad x\rightarrow-1.
\end{equation}
As functions of the complexified radial coordinate, these factors have
essential singularities at the endpoints. They may nevertheless remain
bounded as the endpoints are approached along the real interval. For
a damped mode, the incoming contribution at $x=1$ is exponentially small and
extends smoothly with all derivatives zero at that endpoint. The same holds
at $x=-1$, since the real frequency shift leaves the imaginary part
unchanged. Thus, $\varPhi_{\rm{out}}$ and $\varPhi_{\rm{out}}+c\varPhi_{\rm{in}}$ can have the same power asymptotics. The outgoing series itself is generally divergent, so the outgoing residual is smooth but need not be analytic at the endpoints. We therefore define the radiation conditions by analytic continuation. The outgoing solution is continued from the frequency region where it decays on the real radial line. For a damped mode, it can instead be selected as the subdominant solution along a suitable complex radial ray on which $e^{i\widehat{\omega}\eta}$ or $e^{-i\widehat{\omega}_-\eta}$ at the second end, decays. However, smoothness on the real compact interval alone does not make this selection. The Chebyshev pencil below approximates the residual equation without additional endpoint rows, as in spectral formulations that encode radiation
conditions through asymptotic factorisation \cite{Jansen2017}. Its convergent
roots must nevertheless be checked against the outgoing problem just defined.
Section \ref{suscheck} provides this independent check for the low-lying
modes by integration along complex rays and Section \ref{Kerr_valid} tests the same numerical framework against the Kerr spectrum. We finally bring the present problem in quadratic operator form. We start by observing that since the phase depends linearly on $\widehat{\omega}$, the transformed wave equation is quadratic in the spectral parameter. Collecting its powers
gives
\begin{equation}\label{quadratic_differential_eigenproblem}
\mathcal{L}_0 [\varPhi]+\mathcal{L}_1 [\varPhi]\widehat{\omega}+\mathcal{L}_2[\varPhi]\widehat{\omega}^2=0
\end{equation}
with
\begin{align}
\mathcal{L}_0[\varPhi]&=L_{000}(x,\chi)\varPhi+L_{020}(x,\chi)\partial_{xx}\varPhi+L_{010}(x,\chi)\partial_x\varPhi+L_{002}(x,\chi)\partial_{\chi\chi}\varPhi+L_{001}(x,\chi)\partial_\chi\varPhi,\label{L0}\\
\mathcal{L}_1[\varPhi]&=L_{100}(x,\chi)\varPhi+L_{110}(x,\chi)\partial_x\varPhi,\qquad
\mathcal{L}_2[\varPhi]=L_{200}(x,\chi)\varPhi. \label{L2}
\end{align}
The nonvanishing coefficient functions are
\begin{align}
L_{200}&=1-c_{x}^{2}-c_{x}^{4},\qquad
L_{110}=-\frac{4is_{x}}{\pi}\left(1+c_{x}^{2}\right),\qquad
L_{100}=\frac{2m}{c_{x}^{2}}\left[\widehat{\Omega}(x)+H(x)\left(1-H(x)\right)\widehat{\Omega}_{0}\right]+ic_{x}\left(1-3c_{x}^{2}\right), \label{L100}\\
L_{020}&=-\frac{4c_{x}^{2}}{\pi^{2}},\qquad
L_{010}=\frac{2s_{2x}}{\pi}-\frac{4im\widehat{\Omega}_{0}}{\pi}\left(1-H(x)\right),\qquad
L_{002}=\chi^{2}-1,\qquad
L_{001}=2(|m|+1)\chi,\\
L_{000}&=\frac{m^{2}}{c_{x}^{2}}\left[\widehat{\Omega}_{0}^{2}\left(1-H(x)\right)^{2}-\widehat{\Omega}^{2}(x)\right]-im\widehat{\Omega}_{0}c_{x}\left(1-3c_{x}^{2}\right)+c_{x}^{2}+|m|(|m|+1). \label{L000}
\end{align}
Notice that angular derivatives enter only through $\mathcal{A}_m$ in $\mathcal{L}_0$, whereas $\mathcal{L}_1$ contains only radial derivatives and $\mathcal{L}_2$ is a multiplication operator. This structure will allow the angular discretisation discussed in Section \ref{Subsandexact}. Finally, the apparent endpoint singularities in $L_{100}$ and $L_{000}$ are removable
for the same reason as in $M_{00}$. For example, we have $L_{100}(1)=-m\widehat{\Omega}_0$, $L_{100}(-1)=-3m\widehat{\Omega}_0$, $L_{000}(1)=|m|(|m|+1)$ and $L_{000}(-1)=2m^2\widehat{\Omega}_0^2+|m|(|m|+1)$.

\subsection{Spectral symmetries and real-frequency constraints} \label{spec_symm}

The global equation and its outgoing conditions imply exact relations
between mode families. We first derive two discrete symmetries and the
reduction of the rotation dependence to $mj$. We then use flux conservation
to constrain real quasinormal frequencies and possible nonthreshold crossings
of the real axis. These results provide algebraic checks of the numerical
implementation in addition to the resolution tests of Section \ref{spec_meth}. The next result shows that complex conjugation relates QNMs with opposite azimuthal numbers.

\begin{proposition}\label{prop_conj}
If $\widehat{\omega}$ is a quasinormal frequency with azimuthal number $m$ and eigenfunction $\Psi$, then $-\widehat{\omega}^{*}$ is a QNM with azimuthal number $-m$ and eigenfunction $\Psi^{*}$.
\end{proposition}

\begin{proof}
Since the Klein-Gordon operator has real coefficients, the conjugate field
$\Phi^*=e^{-i(-\omega^*)t}e^{-im\varphi}\Psi^*$ also solves the equation.
Under $(\widehat{\omega},m)\mapsto(-\widehat{\omega}^*,-m)$, the second-end
frequency transforms as $\widehat{\omega}_-\mapsto-\widehat{\omega}_-^*$. The conjugated factors in \eqref{PDE_outgoing_boundary_condition} and
\eqref{PDE_outgoing_boundary_condition_minus} therefore satisfy the outgoing
conditions of the transformed problem. Axis regularity is preserved because
it depends only on $|m|$. Finally, the residual field transforms as
$\varPhi\mapsto\varPhi^*$.\hfill$\square$
\end{proof}

Reflection through the throat exchanges the two asymptotic ends, giving a second spectral symmetry whose frequency shift accounts for the relative rotation of their rest frames.

\begin{proposition}\label{prop_refl}
If $\widehat{\omega}$ is a QNM with azimuthal number $m$ and eigenfunction $\Psi(\eta,\chi$, then $\widehat{\omega}_{-}=\widehat{\omega}-m\widehat{\Omega}_{-}$ is a quasinormal frequency with azimuthal number $-m$ and eigenfunction $\Psi(-\eta,\chi)$.
\end{proposition}

\begin{proof}
Equation \eqref{Omega_reflection} gives $\widehat{\omega}-m\widehat{\Omega}(-\eta)=\widehat{\omega}_- -(-m)\widehat{\Omega}(\eta)$. The remaining terms of \eqref{dimensionless_scalar_PDE} are invariant under $\eta\mapsto-\eta$ and depend on $m$ only through $m^2$. Moreover, reflection exchanges the two infinities. In the transformed problem, the first-end frequency is $\widehat{\omega}_-$ and the second-end frequency is $\widehat{\omega}_- -(-m)\widehat{\Omega}_-=\widehat{\omega}$, so the outgoing conditions are exchanged as well and the throat-frame frequency $\varpi$ is unchanged. Finally, \eqref{phase_alternative} gives $S\mapsto S(-\eta)$ and
$\varPhi(\eta,\chi)\mapsto\varPhi(-\eta,\chi)$.\hfill$\square$
\end{proof}

Combining complex conjugation with reflection through the throat restores the original azimuthal number and yields a mirror symmetry of the quasinormal spectrum at fixed $m$.

\begin{corollary}\label{cor_mirror}
For fixed $m$, the quasinormal spectrum is invariant under the mirror symmetry
\begin{equation}\label{mirror_symmetry}
\widehat{\omega}\longmapsto m\widehat{\Omega}_{-}-\widehat{\omega}^{*},
\end{equation}
or equivalently, $\varpi\longmapsto-\varpi^{*}$, i.e., it is symmetric with respect to the line $\operatorname{Re}\widehat{\omega}=m\widehat{\Omega}_{0}=3\pi jm/2$. The
corresponding residual functions are related by $\varPhi(x,\chi)\mapsto\varPhi^{*}(-x,\chi)$.
\end{corollary}

\begin{proof}
Just compose Propositions \ref{prop_conj} and \ref{prop_refl}.\hfill$\square$
\end{proof}

Thus, the spectrum is symmetric about the imaginary axis when expressed
in the throat-frame frequency $\varpi$. By Proposition \ref{prop_conj}, it suffices to consider $m\geq 0$. The mirror symmetry established in Corollary \ref{cor_mirror} further reduces the reported spectrum to the member of each pair satisfying $\operatorname{Re}\varpi\geq 0$. In the static limit, $\varpi=\widehat{\omega}$ and \eqref{mirror_symmetry} reduces to
$\widehat{\omega}\mapsto-\widehat{\omega}^*$. A further simplification follows from the linear dependence of $\widehat{\Omega}$ on $j$. Let us write $m\widehat{\Omega}(\eta)=mj\widehat{\Omega}_1(\eta)$, where
$\widehat{\Omega}_1$ is the profile at unit spin. Angular separation gives
$\mathcal{L}_m P_L^m=-L(L+1)P_L^m$ for $|m|\leq L$. At fixed $L$, the resulting radial equation therefore depends on $m$ and $j$ only through their product as it can be seen in the next result.

\begin{proposition}\label{prop_mj}
For a fixed multipole $L$, the quasinormal frequencies of the solutions
$\Psi=R(\eta)P_L^m(\chi)$ with $|m|\leq L$ depend on $m$ and $j$
only through $\mu=mj$, i.e.
\begin{equation}\label{mj_scaling}
\widehat{\omega}_{Ln}(m,j)=\mathfrak{W}_{Ln}(mj),\qquad
|m|\leq L,
\end{equation}
where $n$ labels the overtones.
\end{proposition}

\begin{proof}
The separated radial equation is $\partial_\eta[(1+\eta^2)\partial_\eta R]+(1+\eta^2)[\widehat{\omega}-mj\widehat{\Omega}_1(\eta)]^2 R-L(L+1)R=0$. The outgoing conditions, \eqref{PDE_outgoing_boundary_condition} and 
\eqref{PDE_outgoing_boundary_condition_minus} use the frequencies
$\widehat{\omega}$ and $\widehat{\omega}_-=\widehat{\omega}-3\pi mj$. Thus,  neither the radial equation nor its boundary conditions distinguish
$m$ and $j$ separately when $L$ and $mj$ are fixed.\hfill$\square$
\end{proof}

Consequently, two calculations with $mj=m'j'$ give identical spectra in
any multipole family common to both sectors, provided $|m|,|m'|\leq L$. For $m'\neq 0$, this correspondence can be written as $j'=mj/m'$. Notice that it does not equate sectors with different allowed multipoles. The mirror line $\operatorname{Re}\widehat{\omega}=3\pi\mu/2$ and the real-frequency window derived below also depend only on $\mu$. In particular, the entire axisymmetric spectrum is independent of $j$. Conservation of the radial flux provides a further constraint on the spectrum, giving a necessary condition for a real frequency to satisfy the outgoing boundary conditions at both asymptotic ends.

\begin{proposition}\label{prop_window}
Let us suppose that $\widehat{\omega}\in\mathbb{R}$ is a QNM in sector $m$  with $\widehat{\omega}\neq 0$ and $\widehat{\omega}_-\neq 0$. Then,  $\widehat{\omega}\widehat{\omega}_-<0$. In particular, $mj\neq 0$ and
\begin{equation}\label{superradiant_window}
0<\frac{\widehat{\omega}}{m\widehat{\Omega}_{-}}<1,\qquad
\text{i.e.}\qquad
0<\frac{\omega}{m\Omega_{-}}<1,\qquad
|\varpi|<|m\widehat{\Omega}_{0}|.
\end{equation}
\end{proposition}

\begin{proof}
For real $\widehat{\omega}$, we multiply \eqref{divergence_form_PDE} by
$\Psi^{*}$, integrate over $\chi\in(-1,1)$ and take the imaginary part. Since $\int_{-1}^{1}\Psi^{*}\mathcal{L}_{m}\Psi~d\chi=-\int_{-1}^{1}[(1-\chi^{2})|\partial_{\chi}\Psi|^{2}+m^{2}|\Psi|^{2}/(1-\chi^{2})]~d\chi$ is real for axis-regular $\Psi$, the flux
\begin{equation}\label{flux}
\mathcal{F}(\eta)=\operatorname{Im}\int_{-1}^{1}(1+\eta^{2})\Psi^{*}\partial_\eta\Psi~d\chi
\end{equation}
is independent of $\eta$. Evaluating it with \eqref{PDE_outgoing_boundary_condition} and
\eqref{PDE_outgoing_boundary_condition_minus} gives
$mathcal{F}(+\infty)=\widehat{\omega}\|a_{+}\|^{2}$ and $\mathcal{F}(-\infty)=-\widehat{\omega}_{-}\|a_{-}\|^{2}$, where $\|a\|^{2}=\int_{-1}^{1}|a|^{2}~d\chi$. Hence, we obtain
\begin{equation}\label{flux_balance}
\widehat{\omega}\|a_{+}\|^{2}+\widehat{\omega}_{-}\|a_{-}\|^{2}=0.
\end{equation}
If $\widehat{\omega}\widehat{\omega}_->0$, \eqref{flux_balance} forces
both outgoing amplitudes to vanish. To see why this excludes a nontrivial
mode, we project $\Psi$ onto associated Legendre functions. Each radial
component has a unique outgoing solution up to normalisation at the first
infinity and that normalisation is its leading asymptotic coefficient.
A vanishing coefficient therefore makes every component zero, contradicting
the definition of a QNM.\hfill$\square$
\end{proof}

The preceding result also constrains where a continuous branch of QNMs can cross the real axis as the spin varies, provided the crossing occurs away from the two asymptotic thresholds.

\begin{corollary}\label{cor_cross}
Let $\widehat{\omega}(j)$ be a continuous branch of QNMs with fixed $m$. If $\operatorname{Im}\widehat{\omega}(j_{c})=0$ and $\widehat{\omega}(j_{c})\widehat{\omega}_{-}(j_{c})\neq 0$, then $\operatorname{Re}\widehat{\omega}(j_{c})$ lies in the window \eqref{superradiant_window}. For $m=0$ or $j=0$ the window is empty and
the spectrum contains no real nonzero frequencies.
\end{corollary}

\begin{proof}
At $j=j_c$, the frequency is real and differs from both thresholds. Proposition \ref{prop_window} therefore places it in the window \eqref{superradiant_window}. For $m=0$ or $j=0$, we have 
$\widehat{\omega}_-=\widehat{\omega}$, so the necessary condition
$\widehat{\omega}\widehat{\omega}_-<0$ cannot hold for a real nonzero
frequency.\hfill$\square$
\end{proof}

Inside this window, the two asymptotic frequencies have opposite signs.
For a real QNM, the outgoing energy fluxes at the two infinities would therefore have opposite signs when both are defined with respect to the time translation of the first asymptotic frame. The flux balance is analogous to the black-hole superradiance condition $0<\omega<m\Omega_{\rm{H}}$ \cite{Brito2021} with the second asymptotic end providing the additional energy-flux channel. The requirement of unequal asymptotic angular velocities for wormhole superradiance was identified in \cite{Konoplya2016JCAP}. The thresholds $\widehat{\omega}=0$ and $\widehat{\omega}_-=0$ correspond to perturbations that are time-independent in one of the asymptotic rest frames. At the corresponding infinity, the radial equation no longer has the irregular-singular asymptotics used in the preceding argument. Determining whether a mode can cross the real axis at either threshold therefore requires a separate zero-frequency analysis. Finally, the static and axisymmetric limits admit a stronger conclusion,
obtained from the positivity of the separated radial operator.

\begin{proposition}
The static geometry admits no growing scalar QNMs. The same conclusion holds in the axisymmetric sector of the rotating geometry for every $j$.
\end{proposition}

\begin{proof}
When $m=0$, frame dragging drops out of both \eqref{dimensionless_scalar_PDE} and the outgoing boundary conditions, since $\widehat{\omega}_-=\widehat{\omega}$. The axisymmetric spectrum therefore coincides with the $m=0$ spectrum of the static geometry for every $j$. It consequently suffices to establish the absence of growing modes in the static problem. Setting $j=0$ and substituting $\Psi=(1+\eta^2)^{-1/2}u(\eta)P_L^m(\chi)$ into \eqref{divergence_form_PDE} gives
\begin{equation}\label{static_radial}
u''+\left[\widehat{\omega}^{2}-V_{L}(\eta)\right]u=0,\qquad
V_{L}(\eta)=\frac{L(L+1)}{1+\eta^{2}}+\frac{1}{(1+\eta^{2})^{2}}>0,
\end{equation}
where prime denotes differentiation with respect to $\eta$. Let us suppose that a nontrivial outgoing solution exists with $\operatorname{Im}\widehat{\omega}>0$. The outgoing asymptotics $u\sim e^{i\widehat{\omega}\eta}$ as $\eta\to+\infty$ and $u\sim e^{-i\widehat{\omega}\eta}$ as $\eta\to-\infty$ imply exponential decay of both $u$ and $u'$ at the two ends. Multiplying \eqref{static_radial} by $u^*$, integrating over $\eta\in\mathbb{R}$ and integrating by parts therefore yields
\begin{equation}
\widehat{\omega}^{2}=\frac{\displaystyle
\int_{-\infty}^{+\infty}\left(|u'|^2+V_L|u|^2\right)~d\eta}
{\displaystyle\int_{-\infty}^{+\infty}|u|^2~d\eta}>0.
\end{equation}
The boundary terms vanish because of the exponential decay. Since $V_L>0$ and $u$ is nontrivial, the quotient is strictly positive. This requires $\widehat{\omega}$ to be real, contradicting $\operatorname{Im}\widehat{\omega}>0$. Thus, the static geometry admits
no growing scalar QNMs and the same conclusion holds for the axisymmetric sector at every spin.\hfill$\square$
\end{proof}

Any growing scalar mode must therefore belong to a nonaxisymmetric
sector of a rotating geometry. If a continuous branch changes from damping to growth through a finite, nonthreshold real frequency, the preceding corollary places the crossing in the window \eqref{superradiant_window}.

\section{Chebyshev discretisation and eigenvalue selection}\label{spec_meth}

The factorisation \eqref{scalar_asymptotic_factorisation} removes the leading
outgoing factors at both infinities and the regular axis behaviour. The
remaining function $\varPhi(x,\chi)$ is smooth on the compact domain $(x,\chi)\in[-1,1]\times[-1,1]$. We recall that the throat is an ordinary interior point and requires no boundary condition. At the two infinities, the outgoing solution is defined by analytic continuation, as explained after \eqref{excluded_branches}. Smoothness of $\varPhi$ alone does not impose this condition for damped modes. We therefore use the factorised equation to construct the spectral pencil and test the outgoing character of the computed low-lying modes by independent direct integration. The equation takes the quadratic form \eqref{quadratic_differential_eigenproblem}. We approximate $\varPhi$ by a tensor-product Chebyshev expansion. Here, $N_x$ and $N_\chi$ count the radial and angular basis functions, while $n$, $L$ and $m$ denote the overtone, multipole and azimuthal indices, respectively.  We use Chebyshev polynomials of the first kind as basis functions in both coordinates. For a generic argument $y\in[-1,1]$, the polynomial
of degree $k$ is defined by $T_k(y)=\cos{\left(k\arccos{y}\right)} with  k\in\mathbb{N}_0$. Evaluating these polynomials separately in the radial coordinate $x$ and the angular coordinate $\chi$, we write the truncated expansion
\begin{equation}\label{two_dimensional_Chebyshev_expansion}
\varPhi(x,\chi)=\sum_{p=0}^{N_x-1}\sum_{q=0}^{N_\chi-1}a_{pq}T_p(x)T_q(\chi).
\end{equation}
The expansion coefficients $a_{pq}$ are the unknowns. Substituting \eqref{two_dimensional_Chebyshev_expansion} into
\eqref{quadratic_differential_eigenproblem} and evaluating it on the
Chebyshev--Gauss grid yield
\begin{equation}\label{Gauss_nodes}
x_{i}=\cos{\left[\frac{(2i-1)\pi}{2N_{x}}\right]},\quad
i\in\{1,\dots,N_{x}\},\qquad
\chi_{k}=\cos{\left[\frac{(2k-1)\pi}{2N_{\chi}}\right]},\quad
k\in\{1,\dots,N_{\chi}\}.
\end{equation}
This procedure converts the differential problem into a finite-dimensional quadratic matrix pencil of the form
\begin{equation}\label{quadratic_matrix_pencil}
\left(\mathbf{A}_{0}+\widehat{\omega}\mathbf{A}_{1}+\widehat{\omega}^{2}\mathbf{A}_{2}\right)\mathbf{a}=\mathbf{0},
\end{equation}
where $\mathbf{a}$ collects the coefficients $a_{pq}$ while the matrices
$\mathbf{A}_0$, $\mathbf{A}_1$ and $\mathbf{A}_2$ represent the three differential operators. The Gauss grid excludes $x=\pm 1$, so the expressions
\eqref{L100} and \eqref{L000} are not evaluated at their removable singularities. Furthermore, we do not impose the endpoint relations \eqref{endpoint_relations} as additional rows. These relations hold for the
smooth outgoing solutions, but do not by themselves distinguish them from
smooth incoming admixtures. The matrices are assembled symbolically in \textsc{Maple}. For each $(m,j)$, we substitute the truncated expansion and its derivatives into the factorised equation, evaluate the result at the collocation nodes and collect the coefficients of $a_{pq}$ at each power of $\widehat{\omega}$. The resulting matrices are exported in ASCII format to \textsc{Julia}, where we solve the eigenvalue problem and analyse the roots at high precision. The quadratic eigenvalue problem \eqref{quadratic_matrix_pencil} is converted into an equivalent generalised linear eigenvalue problem by introducing the augmented vector
\begin{equation}\label{augmented_eigenvector}
\mathbf{u}=
\begin{pmatrix}
\mathbf{a}\\
\widehat{\omega}\mathbf{a}
\end{pmatrix}.
\end{equation}
The corresponding companion linearisation is
\begin{equation}\label{linearized_pencil_wormhole}
\begin{pmatrix}
\mathbf{0} & \mathbf{I}\\
-\mathbf{A}_{0} & -\mathbf{A}_{1}
\end{pmatrix}\mathbf{u}
=\widehat{\omega}
\begin{pmatrix}
\mathbf{I} & \mathbf{0}\\
\mathbf{0} & \mathbf{A}_{2}
\end{pmatrix}\mathbf{u}.
\end{equation}
Since there are $N_x N_\chi$ expansion coefficients, the linearised pencil has dimension $2N_x N_\chi$. Its generalised eigenvalues form the raw spectrum at the chosen $(m,j)$ and resolution. The convergence tests below determine which of these roots are reported as QNMs. Let us recall that with the time dependence $e^{-i\omega t}$, damping corresponds to $\operatorname{Im}\widehat{\omega}<0$, whereas a mode with
$\operatorname{Im}\widehat{\omega}>0$ would represent exponential growth.

\subsection{Exact angular reduction}\label{Subsandexact}

Because the frame dragging depends only on the radial coordinate, the angular derivatives enter \eqref{quadratic_differential_eigenproblem} only
through the $x$-independent operator $\mathcal{A}_{m}$ of \eqref{angular_operator_Am}, whereas all other coefficients are independent of $\chi$.  The collocation matrices therefore have the Kronecker
structure
\begin{equation}\label{Kronecker_structure}
\mathbf{A}_{0}=\mathbf{R}_{0}\otimes\mathbf{B}_{\chi}+\mathbf{B}_{x}\otimes\mathbf{G}_{\chi},\qquad
\mathbf{A}_{1}=\mathbf{R}_{1}\otimes\mathbf{B}_{\chi},\qquad
\mathbf{A}_{2}=\mathbf{R}_{2}\otimes\mathbf{B}_{\chi},
\end{equation}
where $(\mathbf{B}_{x})_{ip}=T_{p}(x_{i})$, $(\mathbf{B}_{\chi})_{kq}=T_{q}(\chi_{k})$, $(\mathbf{G}_{\chi})_{kq}=(\mathcal{A}_{m}T_{q})(\chi_{k})$ and
$\mathbf{R}_{n}$ are the $N_{x}\times N_{x}$ collocation matrices of the
radial parts of $\mathcal{L}_{n}$ with the constant $|m|(|m|+1)$ assigned
to $\mathcal{A}_{m}$. Moreover, the polynomial structure of $\mathcal{A}_m$ yields an exact decomposition of the two-dimensional matrix pencil into independent radial pencils, each associated with a definite multipole index $L$.

\begin{proposition}\label{prop_winkel}
The space of polynomials in $\chi$ of degree at most $N_\chi-1$ is invariant under $\mathcal{A}_m$. On this space, the Gegenbauer polynomials $C_\nu^{(|m|+1/2)}(\chi)$ with $\nu\in\{0,\ldots,N_\chi-1\}$ form an eigenbasis. The eigenvalue associated with the polynomial of degree $\nu$ is $L(L+1)$, where $L=|m|+\nu$ \cite{NIST2010}. Hence, the discrete pencil \eqref{quadratic_matrix_pencil} is equivalent to the direct sum of $N_\chi$ independent radial pencils, one for each multipole $L\in\{|m|,\ldots,|m|+N_\chi-1\}$, given by
\begin{equation}\label{radial_pencils}
\mathbf{R}_{0}+L(L+1)\mathbf{B}_{x}+\widehat{\omega}\mathbf{R}_{1}
+\widehat{\omega}^{2}\mathbf{R}_{2},\qquad
L\in\{|m|,|m|+1,\dots,|m|+N_{\chi}-1\}.
\end{equation}
\end{proposition}

\begin{proof}
The operator $\mathcal{A}_m$ preserves polynomial degree. The Gegenbauer equation
\begin{equation}
(1-\chi^2)y''-2(|m|+1)\chi y'+\nu(\nu+2|m|+1)y=0
\end{equation}
therefore gives
\begin{equation}
\mathcal{A}_m C_\nu^{(|m|+1/2)}=(|m|+\nu)(|m|+\nu+1)C_\nu^{(|m|+1/2)}.
\end{equation}
Since these polynomials have distinct degrees $\nu\in\{0,\ldots,N_\chi-1\}$, they form an eigenbasis for the polynomial space used in the angular discretisation. Let $\mathbf{D}$ denote the matrix of $\mathcal{A}_m$ in the
Chebyshev basis and let the $\nu$-th column of $\mathbf{V}$ contain
the Chebyshev coefficients of $C_\nu^{(|m|+1/2)}$. The matrix $\mathbf{V}$ is invertible because it relates two bases of the same polynomial space. The eigenvalue relation above implies $\mathbf{D}=\mathbf{V}\boldsymbol{\Lambda}\mathbf{V}^{-1}$ with 
$\boldsymbol{\Lambda}=\operatorname{diag}_{\nu\in\{0,\ldots,N_\chi-1\}}
\left[(|m|+\nu)(|m|+\nu+1)\right]$. Applying $\mathcal{A}_m$ in coefficient space and then evaluating at the collocation nodes gives $\mathbf{G}_\chi=\mathbf{B}_\chi\mathbf{D}$. This identity is exact because $\mathcal{A}_m$ maps the chosen polynomial space into itself. Moreover, $\mathbf{B}_\chi$ is invertible, since a polynomial of degree at most $N_\chi-1$ that vanishes at $N_\chi$ distinct nodes must vanish identically. Using the Kronecker structure \eqref{Kronecker_structure} yields
\begin{equation}\label{pencil_factorisation}
\left(\mathbf{I}_{N_x}\otimes(\mathbf{B}_\chi\mathbf{V})^{-1}\right)
\left(\mathbf{A}_0+\widehat{\omega}\mathbf{A}_1+\widehat{\omega}^{2}\mathbf{A}_2\right)\left(\mathbf{I}_{N_x}\otimes\mathbf{V}\right)=
\left(\mathbf{R}_0+\widehat{\omega}\mathbf{R}_1+\widehat{\omega}^{2}\mathbf{R}_2\right)\otimes\mathbf{I}_{N_\chi}+\mathbf{B}_x\otimes\boldsymbol{\Lambda}.
\end{equation}
Both transformation matrices are independent of
$\widehat{\omega}$ and invertible. This implies that the transformed pencil is equivalent to the original one. Since $\boldsymbol{\Lambda}$ is diagonal, a simultaneous permutation of rows and columns groups the unknowns by angular eigenfunction and produces a block diagonal pencil. Its blocks are precisely the radial pencils \eqref{radial_pencils} with $L=|m|+\nu$. \hfill$\square$
\end{proof}

Since $(1-\chi^{2})^{|m|/2}C_{\nu}^{(|m|+1/2)}(\chi)\propto
P_{|m|+\nu}^{|m|}(\chi)$, choosing $N_\chi=3$ includes exactly the three
multipole families $L\in\{|m|,|m|+1,|m|+2\}$ without angular truncation error. Their accuracy is therefore controlled by the radial resolution $N_x$. We determine the multipole label $L$ of each eigenvalue by matching it
to the spectra of the radial pencils \eqref{radial_pencils}. The same label can be obtained by projecting the angular part of the associated eigenvector onto the Gegenbauer basis. Within each $(L,m)$ family, the
overtone label $n$ orders the modes by increasing damping rate. In exact arithmetic, the two-dimensional spectrum is the union of these radial spectra. Agreement between the two calculations thus tests the assembly of the two-dimensional operator. Notice that this simplification relies on the angular structure of the present metric. If latitude-dependent metric
coefficients couple the multipoles, the Kronecker decomposition
\eqref{Kronecker_structure} generally fails and $N_\chi$ must also be
increased to establish convergence. The same spectral construction can then
be used with the coefficients and boundary factors appropriate to that
background.

\subsection{Convergence criteria and mode selection} \label{subsroot_sel}

We assemble the matrices in \textsc{Maple} at $200$ decimal digits and solve
the generalised eigenvalue problem in \textsc{Julia} at $220$ decimal digits.
The subsequent matching and filtering use the same precision. The metric
\eqref{global_metric} is regular for every $J$, so the rotation parameter
has no extremal upper bound. To sample both small and large spins, we use
$j_*=j/(1+j)\in[0,1)$ with $j_*\in\{0,0.1,\ldots,0.9,0.95\}$. These values correspond to $j\in\{0, 1/9, 1/4,3/7,2/3,1,3/2,7/3,4,9,19\}$. At each spin, we compare the spectra at three radial resolutions $N_{x}\in\{90,95,100\}$ and $N_{\chi}=3$ with the same three angular basis functions. Proposition~\ref{prop_winkel} ensures that radial refinement is sufficient for the included multipoles. For each root $\widehat{\omega}_{90}$, we find the nearest root $\widehat{\omega}_{95}$ at $N_x=95$, then match that root to the nearest $\widehat{\omega}_{100}$ at $N_x=100$. A triplet passes the first test when its largest pairwise separation satisfies
\begin{equation}\label{stability_criterion_wormhole}
\max\left\{\left|\widehat{\omega}_{90}-\widehat{\omega}_{95}\right|,\left|
\widehat{\omega}_{95}-\widehat{\omega}_{100}\right|,\left|\widehat{\omega}_{90}-\widehat{\omega}_{100}\right|\right\}\leq
\max\left[\varepsilon_{\mathrm{abs}},\varepsilon_{\mathrm{rel}}\max\{
1,|\widehat{\omega}_{90}|,|\widehat{\omega}_{95}|,|\widehat{\omega}_{100}|\}
\right].
\end{equation}
Let $\varepsilon_{\mathrm{abs}}$ and $\varepsilon_{\mathrm{rel}}$ be the absolute and relative tolerances used to match eigenvalues across the three radial resolutions. The relative tolerance is scaled by the largest frequency magnitude in the triplet with a minimum scale of unity. We set
$\varepsilon_{\mathrm{abs}}=\varepsilon_{\mathrm{rel}}=10^{-3}$. For each matched triplet, we define the arithmetic mean
\begin{equation}\label{frequency_average_wormhole}
\widehat{\omega}_{\mathrm{av}}=\frac{\widehat{\omega}_{90}+\widehat{\omega}_{95}+\widehat{\omega}_{100}}{3}.
\end{equation}
Moreover, we store all three pairwise differences separately and denote their maximum by $d_{\max}$ which measures the change under the chosen refinements.  A second test is needed near the two thresholds $\widehat{\omega}=0$ and $\widehat{\omega}_-=0$. There the raw spectra contain sequences of roots whose distances from a threshold decrease approximately as $N_x^{-2}$ under refinement (see Section \ref{suscheck}). The fixed absolute tolerance cannot reliably reject such roots near $\widehat{\omega}=0$, while near $\widehat{\omega}_-=0$ the relative tolerance grows to approximately $\varepsilon_{\rm{rel}}|m|\widehat{\Omega}_-$. Moreover, the first test is
not invariant under the mirror map \eqref{mirror_symmetry}. We therefore
also compare $d_{\max}$ with the distance from the nearer threshold, i.e.
\begin{equation}\label{refined_criterion}
d_{\max}\leq\varepsilon_{q}\min\left\{|\widehat{\omega}_{\mathrm{av}}|,|\widehat{\omega}_{\mathrm{av}}-m\widehat{\Omega}_{-}|\right\},\qquad
\varepsilon_{q}=10^{-3}.
\end{equation}
Here $\varepsilon_q$ is the dimensionless relative tolerance for the
second convergence test. The choice $\varepsilon_q=10^{-3}$ requires
the largest pairwise frequency separation $d_{\max}$ to be at most
$0.1\%$ of the distance from $\widehat{\omega}_{\mathrm{av}}$ to
the nearer threshold. since the mirror map exchanges the two distances while leaving $d_{\max}$ unchanged, \eqref{refined_criterion} treats both members of a pair in the same way. We also exclude roots on $\operatorname{Re}\widehat{\omega}=0$ and $\operatorname{Re}\widehat{\omega}_-=0$, where the non-convergent sequences
accumulate. This exclusion limits the part of the spectrum that the present
calculation can identify. In what follows, \emph{converged} means that a root
passes both \eqref{stability_criterion_wormhole} and
\eqref{refined_criterion} and is not on either excluded line. It is worth mentioning that the exact properties discussed in Section \ref{spec_symm} provide additional, resolution-independent diagnostics.  The Gauss grid
\eqref{Gauss_nodes} is symmetric under $x\mapsto-x$ and
$T_{p}(-x)=(-1)^{p}T_{p}(x)$. The map $a_{pq}\mapsto(-1)^{p}a_{pq}^{*}$, which is the discrete counterpart of $\varPhi(x,\chi)\mapsto\varPhi^{*}(-x,\chi)$, therefore carries eigenvectors of the pencil at $\widehat{\omega}$ into eigenvectors at $m\widehat{\Omega}_{-}-\widehat{\omega}^{*}$, so that
Corollary \ref{cor_mirror} holds exactly for the discrete problem and the
computed mirror pairs must agree to working precision, up to the conditioning
of the individual eigenvalues. A violation of this symmetry signals an error
in the assembly of the operator rather than a discretisation error and only
the member of each pair with $\operatorname{Re}\varpi\geq 0$ needs to be
reported. Similarly, by Proposition \ref{prop_winkel} the three multipole
families must be recovered separately from the radial pencils \eqref{radial_pencils}.  Since the radial matrices $\mathbf{R}_n$ depend on $m$ and $j$ only through $mj$, Proposition~\ref{prop_mj} also holds exactly for the discrete problem. Calculations with different azimuthal numbers but
the same product $mj$ must therefore yield identical spectra in their
common multipole families, providing a consistency check on
independently assembled pencils. For instance, the
families $L\in\{2,3\}$ obtained with $m=1$ at spin $j$ must coincide with those obtained with $m=2$ at spin $j/2$. We also inspect the complete unfiltered spectrum at each resolution for roots with $\operatorname{Im}\widehat\omega>10^{-18}$. These candidates are recorded separately and subjected to the three-resolution matching test
\eqref{stability_criterion_wormhole}. Corollary \ref{cor_cross} supplies
an additional consistency condition for a continuous branch crossing the
real axis away from the thresholds. Such a crossing must occur inside the
window \eqref{superradiant_window}, whereas the corollary does not decide
what happens at its edges. We use this distinction when examining candidates
whose imaginary parts are smaller than the matching tolerance. A candidate
that persists under refinement still requires an independent outgoing-mode
test before it can be identified as a growing pole of the continuum problem.

\section{Scalar spectra and their physical interpretation}
\label{results}

We compute the sectors $m\in\{1,2,3\}$ at the eleven spins $j\in\{0, 1/9, 1/4,3/7,2/3,1,3/2,7/3,4,9,19\}$ using $N_x\in\{90,95,100\}$ and $N_\chi=3$. Each of the $99$ raw spectra contains $2N_xN_\chi=6N_x$ eigenvalues
and decomposes into three multipole families, labelled by $L\in\{m,m+1,m+2\}$. We report only the member of each mirror pair
with $\operatorname{Re}\varpi\geq 0$. For comparing sectors, we use $\mu=mj$ since this is the only rotation parameter entering a fixed multipole family. We also use the frequency in the asymptotic frame of the second exterior , i.e. $\widehat{\omega}_-=\widehat{\omega}-3\pi\mu$. In the complex
$\widehat{\omega}_-$-plane, the real-frequency window \eqref{superradiant_window} extends to the vertical strip
$-3\pi\mu<\operatorname{Re}\widehat{\omega}_-<0$. Its right-hand boundary
passes through the threshold $\widehat{\omega}_-=0$.
\begin{figure}[H]
\centering
\includegraphics[width=0.98\linewidth]{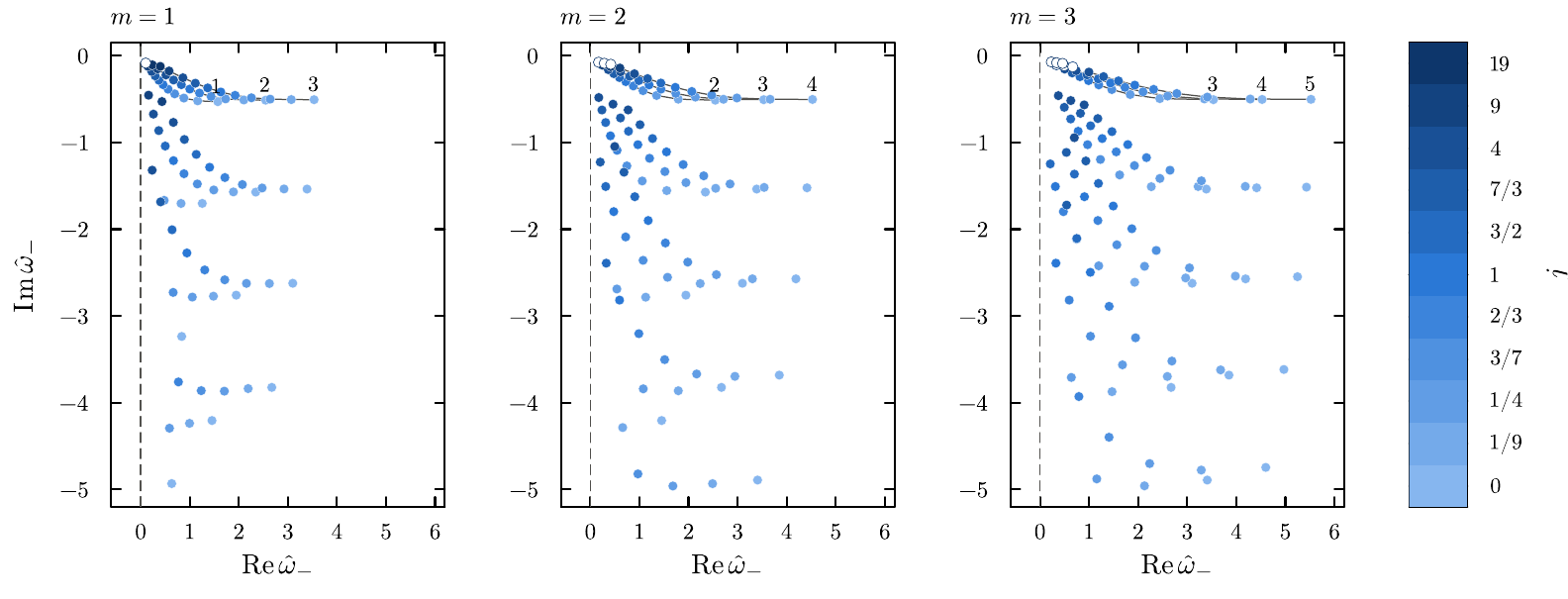}
\caption{\textit{Converged scalar QNMs at all sampled spins for
$m\in\{1,2,3\}$ (left to right), expressed in the second asymptotic frame
as $\widehat{\omega}_-=\widehat{\omega}-3\pi mj$. Only the member of each mirror pair satisfying $\operatorname{Re}\varpi\geq 0$ is shown.
The superradiant window \eqref{superradiant_window} corresponds to
the strip $-3\pi mj<\operatorname{Re}\widehat{\omega}_-<0$, whose right-hand boundary is marked by the dashed line. In each panel, the three angular basis functions represent exactly the multipoles $L\in\{m,m+1,m+2\}$ (see Proposition~\ref{prop_winkel}). Grey lines trace the fundamental branches $(n=0)$ across the sampled spins, with the multipole index $L$ labelled at $j=0$. Open circles mark fundamental frequencies obtained by direct
integration where the spectral solver does not resolve them at
$N_x\leq 100$ (see Table \ref{tab_directm}). Tables \ref{tab_QNMm1}-\ref{tab_QNMm3} also list more strongly damped modes beyond the plotted range.}}\label{fig_overv}
\end{figure}
Figure \ref{fig_overv} shows how the resolved spectra evolve with spin. At $j=0$, each multipole has a sequence of increasingly damped overtones. Along the computed fundamental branches, increasing $j$ reduces both $\operatorname{Re}\widehat{\omega}_-$ and the damping rate. All converged
roots shown remain to the right of the window boundary and below the real
axis. We first examine the numerical checks and the rejected roots, then
use the static limit to identify the mode families. Their evolution with
rotation leads to the light-ring interpretation and the large-spin
limit, which also explain where the fixed spectral resolution becomes
insufficient.

\subsection{Numerical validation and rejected roots}\label{suscheck}

The identities established in Sections \ref{spec_symm} and \ref{Subsandexact} apply to each finite pencil and can be tested before any root is selected. In all $99$ raw spectra, every eigenvalue has the mirror partner \eqref{mirror_symmetry} to within $2.2\times 10^{-163}$. This very small discrepancy tests the discrete symmetry. It does not measure the error relative to the continuum spectrum, for which the much larger resolution differences and independent integration provide the relevant diagnostics. The runs $(m,j)=(2,1)$ and $(3,2/3)$ have $\mu=2$, while
$(2,3/2)$ and $(3,1)$ have $\mu=3$. In each comparison, their common families $L\in\{3,4\}$ coincide as required by Proposition~\ref{prop_mj}. There are $2N_x$ eigenvalues per common family or $4N_x$ in total, giving $360$ at $N_x=90$ and $400$ at $N_x=100$. Their discrepancies are at most $1.5\times 10^{-163}$. Within this comparison, runs with different $\mu$ share no eigenvalues. The static runs likewise agree on their common families. We also assemble the radial companion pencils independently in
midpoint-radius (ball) arithmetic at $256$-bit precision
\cite{Johansson2017Arb}. Their union reproduces all $17820$ eigenvalues
of the $N_x=90$ spectra to a relative accuracy of $5\times 10^{-40}$,
as required by Proposition~\ref{prop_winkel}. In addition to checking the matrix assembly, matching each root to the spectrum of a radial pencil identifies its angular multipole index $L$. An independent test of the continuum formulation requires imposing the outgoing conditions without using the compactification or interpolating phase. For this purpose, we integrate the separated radial equation directly. The substitution $\Psi=(1+\eta^2)^{-1/2}u(\eta)P_L^m(\chi)$ in \eqref{divergence_form_PDE} gives
\begin{equation}\label{radial_rotating}
u''+\left\{\left[\widehat{\omega}-m\widehat{\Omega}(\eta)\right]^{2}-V_{L}(\eta)\right\}u=0
\end{equation}
with $V_L$ defined in \eqref{static_radial}. The outgoing branches
are $u\sim e^{i\widehat{\omega}\eta}$ at positive infinity and $u\sim e^{-i\widehat{\omega}_-\eta}$ at negative infinity. We initialise them with Riccati-Hankel functions on complex rays along which they are subdominant, then integrate inward with an adaptive eighth-order Runge-Kutta method. The quasinormal condition is equality of their logarithmic derivatives at an interior matching point. Choosing subdominant rays suppresses sensitivity to an unwanted dominant solution during inward integration, although the truncated far-field initial data still require numerical checks. This calculation reproduces all $166$ converged fundamental and first-overtone modes ($n\in\{0,1\}$) within their respective $d_{\max}$. The median frequency difference is $9\times 10^{-13}$. When the frequency difference exceeds the double-precision round-off level, it is between $10\%$ and $30\%$ of the corresponding $d_{\max}$. Thus, $d_{\max}$ is a conservative error estimate for the averaged frequencies \eqref{frequency_average_wormhole} in these comparisons. Notice that the check concerns the low-lying modes tested here, rather than every root of the pencil.
\begin{figure}[H]
\centering
\includegraphics[width=0.98\linewidth]{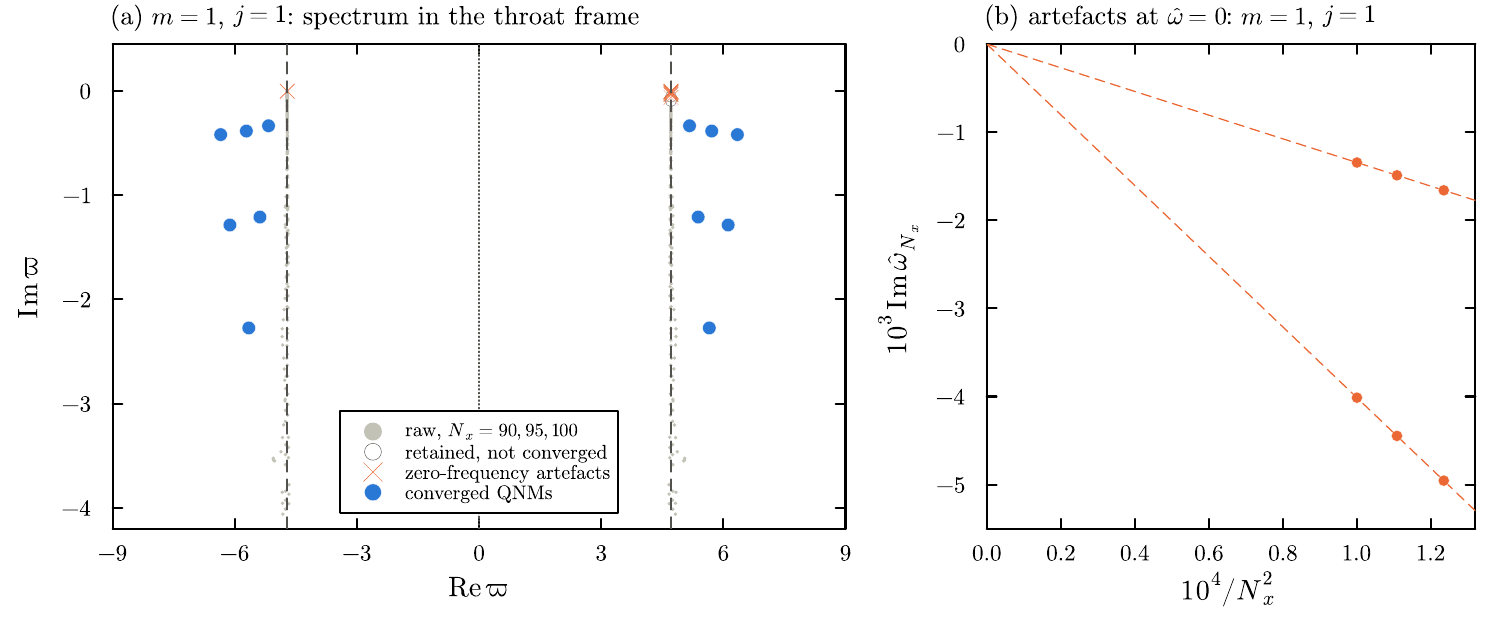}
\caption{\textit{(a) Raw spectra for $m=1$, $j=1$ and $N_x\in\{90,95,100\}$ (grey dots), in the throat frame $\varpi=\widehat{\omega}-3\pi mj/2$. Among the roots passing the first matching criterion \eqref{stability_criterion_wormhole}, filled circles mark converged modes, crosses mark non-convergent roots near $\widehat{\omega}=0$  or $\widehat{\omega}_-=0$ and the open circle marks another non-convergent root. The converged modes form mirror pairs about
$\operatorname{Re}\varpi=0$. Dashed vertical lines mark the two window
edges, near which raw roots accumulate. (b) Imaginary parts of the two roots
closest to $\widehat{\omega}=0$, plotted against $N_x^{-2}$ for the
three resolutions. The dashed fits pass through the origin.}}
\label{fig_anatomy}
\end{figure}
The first-stage criterion \eqref{stability_criterion_wormhole} accepts
$1350$ roots across the $33$ runs. Of these, $674$, forming
$337$ mirror pairs, also pass \eqref{refined_criterion}. Most do so
by a wide margin, with median
$d_{\max}/\min\{|\widehat{\omega}|,|\widehat{\omega}_-|\}=8\times 10^{-12}$. 
Among the $676$ roots rejected by the second test, $395$ move towards
one of the two thresholds. At $N_x=90$, the distance of each root from the nearer threshold is between $1.10$ and $1.40$ times its value at $N_x=100$. This decrease is consistent with the approximate $N_x^{-2}$ scaling
shown in Figure \ref{fig_anatomy}, which predicts a ratio of $(100/90)^2\simeq 1.23$ between the two resolutions. For $\mu\leq 12$, these roots lie below the thresholds along $\operatorname{Re}\widehat{\omega}=0$ and $\operatorname{Re}\widehat{\omega}_-=0$, within wedges of half-angle
$8^\circ$. The roots closest to the thresholds coincide in runs with the
same $L$. At $\mu\geq 18$, they spread over a neighbourhood of $\widehat{\omega}_-=0$. We interpret this behaviour as a discrete
representation of the branch cuts of the frequency-domain Green's function. 
For a massless field in an asymptotically flat exterior, such a cut originates at zero frequency and produces late-time tails \cite{Leaver1986PRD}. Here, the two asymptotic frames give two candidate cut origins. The remaining $281$ rejected roots also fail to converge and are predominantly strongly damped overtone candidates or roots from the large-$mu$ runs discussed in Section \ref{subs_resol}. Among the first-stage roots with $\operatorname{Re}\varpi\geq 0$, only one lies in the upper half-plane. It occurs at $(m,j)=(3,19)$ with $\widehat{\omega}=537.137+i0.094$ or
$\widehat{\omega}_-=-0.075+i0.094$. It is within $0.12$ of the second threshold, but moves by $0.38$ between $N_x=90$ and $N_x=100$ and crosses into the lower half-plane. It passes the first test only because the relative tolerance $\varepsilon_{\rm{rel}}|\widehat{\omega}|\approx
\varepsilon_{\rm{rel}}3\pi\mu$ is $0.54$ at $\mu=57$. It fails \eqref{refined_criterion} by more than three orders of magnitude. This example shows why the threshold-relative test is needed at large spin. The accumulation of discrete roots near the candidate cuts also limits
spectral resolution. At the resolutions used here, modes within a few tenths
of $\operatorname{Re}\widehat{\omega}_-=0$ are no longer recovered as
converged roots. Some strongly damped overtones reach this region even at
small $j$, leaving gaps in the resolved sequences. For example, only
$n\in\{0,1\}$ of the $L=1$ family are resolved at $j=1/4$. Four converged mirror pairs with $\operatorname{Im}\widehat{\omega}<-7.5$ are separated from the rest of their families by such gaps. We leave their overtone labels unassigned and omit them from the frequency tables. Section \ref{subs_resol} discusses the related loss of resolution at large $\mu$.

\subsection{Static limit and parity}

The static limit provides a reference for identifying the rotating families.
At $j=0$, the asymptotic frames coincide, the window is empty and the
spectrum at fixed $L$ is independent of $m$. The three static runs
agree on their common families. Tables \ref{tab_QNMm1}-\ref{tab_QNMm3}
list overtones through $n=3$ for $L=1$, $n=5$ for $L=2$, $n=7$ for $L=3$ and $n=8$ for $L\in\{4,5\}$. The five fundamentals, in order of increasing $L$, are $1.5727-i0.5297$, $2.5467-i0.5127$, $3.5343-i0.5069$, $4.5271-i0.5043$ and $5.5223-i0.5030$. For $L\in\{1,3,5\}$, these agree in every quoted digit with the corresponding massless Ellis-Bronnikov values \cite{Gonzalez2022PRD}. Reflection symmetry makes each static mode even or odd in $\eta$. Direct integration gives $u(-\eta)=(-1)^n u(\eta)$ for all five families and $n\leq 3$. Thus, $u'(0)=0$ for $n\in\{0,2\}$, whereas
$u(0)=0$ for $n\in\{1,3\}$. As $L$ increases, the fundamentals approach
$\widehat{\omega}\simeq L+\frac{1}{2}-\frac{i}{2}$, that is the eikonal value of the throat light ring at the maximum of $V_L$.

\subsection{Evolution of the $m=1$ spectrum}

\begin{figure}[H]
\centering
\includegraphics[width=0.98\linewidth]{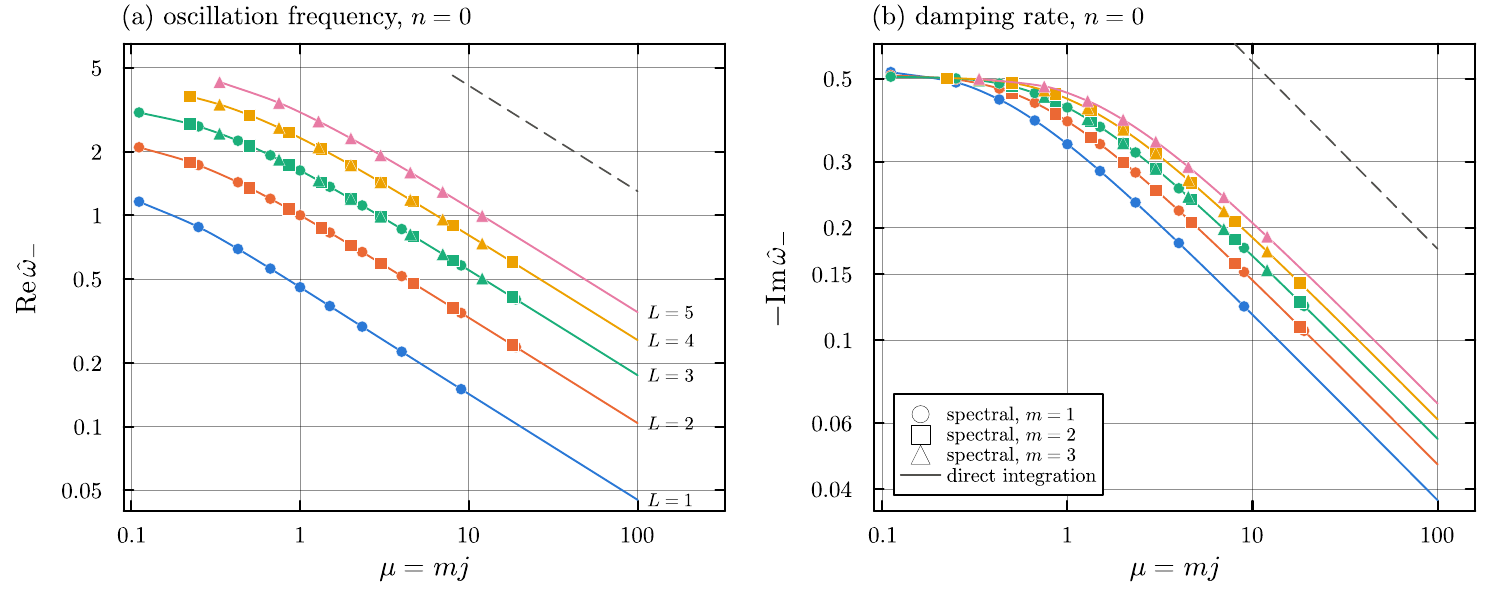}
\caption{\textit{Real part (a) and damping rate (b) of the fundamental modes
($n=0$) in the frame of the second exterior as functions of $\mu=mj$. Symbols are converged roots of the spectral solver with the shape indicating
the sector $m$ and the colour the multipole $L$. Solid lines are the
direct-integration continuation of each family up to $\mu=100$. Families
computed in different sectors coincide, as required by Proposition \ref{prop_mj}. The dashed line indicates the slope $\mu^{-1/2}$ of \eqref{large_mu}.}}\label{fig_mjscal}
\end{figure}

For $m=1$, the $L=1$ fundamental mode evolves from $\widehat{\omega}=1.5727-i0.5297$ at $j=0$ to $\widehat{\omega}_-=0.4572-i0.3339$ or $\widehat{\omega}=9.8819-i0.3339$ at $j=1$. At $j=9$, it reaches $\widehat{\omega}_-=0.1505-i0.1231$ or $\widehat{\omega}=84.9735-i0.1231$. The $L\in\{2,3\}$ fundamentals follow the same trend, with larger shifted real parts and, for $j\geq 1/4$, slightly larger damping rates. At $j=19$, only the $L\in\{2,3\}$ fundamentals pass the spectral convergence test, at $\widehat{\omega}=179.3090-i0.1060$ and $179.4710-i0.1230$. The frequency shift explains the contrasting appearance of this evolution in the two frames. At large spin, $\operatorname{Re}\widehat{\omega}=m\widehat{\Omega}_-+\operatorname{Re}\widehat{\omega}_-$ is dominated by $m\widehat{\Omega}_-=3\pi mj$. The mode therefore oscillates near $m\Omega_-$ as measured in the first asymptotic frame, while its damping rate is identical in both frames. Every converged member with $\operatorname{Re}\varpi\geq 0$ has $\operatorname{Re}\widehat{\omega}_->0$. These members lie beyond the upper-frequency edge of the window and their mirror partners have $\operatorname{Re}\widehat{\omega}=-\operatorname{Re}\widehat{\omega}_-<0$ 
beyond the lower edge. The fundamental damping rates decrease monotonically across the sampled spins, from approximately $0.5$ to approximately $0.1$ (see Figure \ref{fig_mjscal}). Some overtones move towards the vertical line
$\operatorname{Re}\widehat{\omega}_-=0$ more rapidly. For example, the
$(L.n)=(1,1)$ mode changes from $\widehat{\omega}_-=0.8233-i1.7030$ at $j=1/9$ to $0.4797-i1.6675$ at $j=1/4$, after which it is not resolved by the fixed spectral grids.

\subsection{Higher azimuthal sectors and $mj$ scaling}

The sectors $m=2$ and $m=3$ include $L\in\{2,3,4\}$ and $L\in\{3,4,5\}$, respectively, and show the same qualitative evolution (see Figure \ref{fig_overv}). For any multipole common to two sectors, a fixed
$\mu$ is reached at a smaller $j$ when $m$ is larger. Comparisons between different $L$ also involve the multipole dependence. For example, the $L=m$ fundamentals at $j=4$ are $\widehat{\omega}_-=0.2265-i0.1818$, $0.3663-i0.1609$ and $0.5032-i0.1540$ for $m\in\{1,2,3\}$, corresponding to $\mu\in\{4,8,12\}$. The exact $mj$ reduction is more directly tested by common multipoles at equal $\mu$. Both $(m,j)=(2,1)$ and $(3,2/3)$ give $\widehat{\omega}_-=1.1976-i0.3363$ for $(L,n)=(3,0)$ and $1.7318-i0.3654$. Likewise, $(2,3/2)$ and $(3,1)$ both give $0.9894-i0.2873$ and $1.4378-i0.3158$ for these two families. The sector data therefore lie on common fixed-$L$ curves when plotted against $\mu$, as required by $\mathfrak{W}_{L0}(\mu)$ in \eqref{mj_scaling} (see Figure \ref{fig_mjscal}). As it can be evinced from Table \ref{tab_counts}, the number of converged mirror pairs decreases with $j$ in each sector. Moreover, no root passes the convergence tests for $(m,j)=(3,9)$, $(2,19)$ or $(3,19)$.

\subsection{Light rings and the large-spin limit}\label{subs_lightring}

\begin{figure}[H]
\centering
\includegraphics[width=0.98\linewidth]{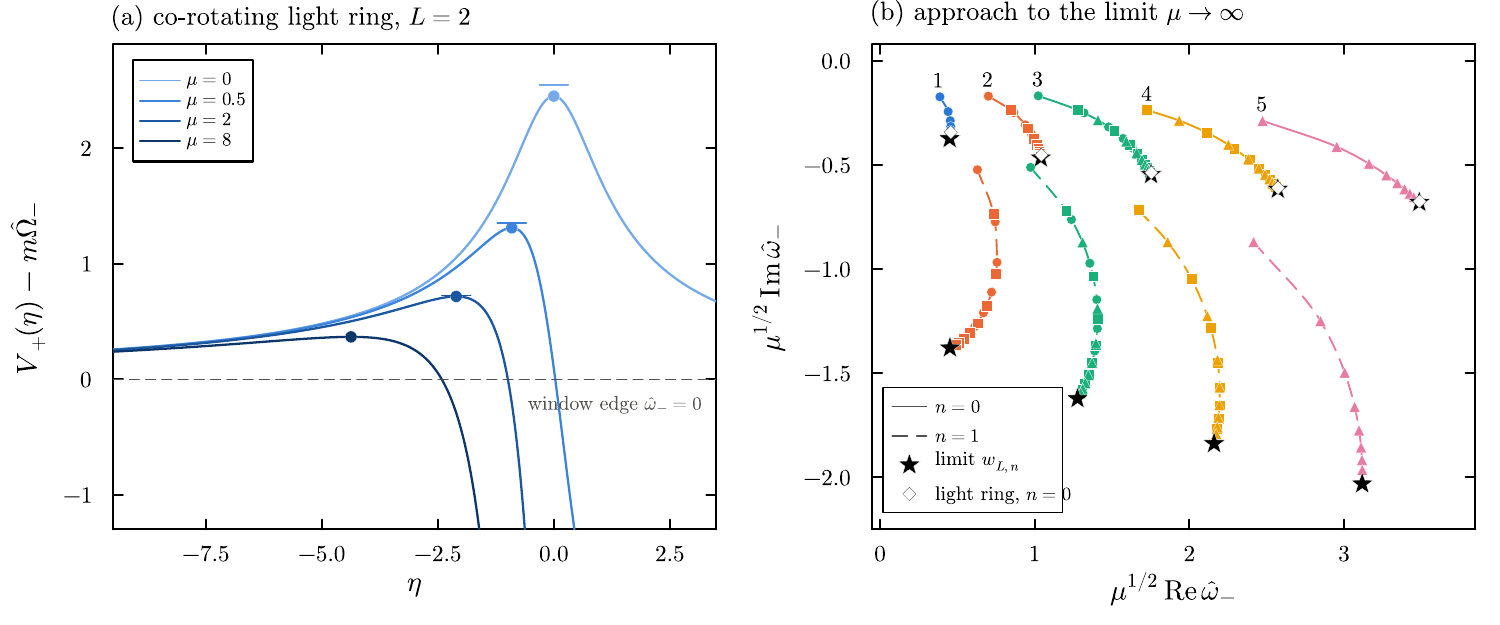}
\caption{\textit{(a) Co-rotating eikonal potential $V_{+}(\eta)$ of \eqref{eikonal_potentials} for $L=2$, measured from the window edge
$m\widehat{\Omega}_{-}$ for $\mu\in\{0,0.5,2,8\}$. Dots indicate the light-ring location given by \eqref{light_ring}, while short horizontal bars show $\operatorname{Re}\widehat{\omega}_-$ for the fundamental mode. (b)
Rescaled frequencies $\mu^{1/2}\widehat{\omega}_{-}$ of the families
$L\in\{1,\dots,5\}$ (labels) for $n=0$ (solid) and $n=1$ (dashed), from the
direct-integration continuation for $0.1\leq\mu\leq 100$, with the converged
spectral roots superimposed (symbols as in Figure \ref{fig_mjscal}). Stars
are the eigenvalues $w_{L,n}$ of the limit problem \eqref{limit_equation}, and diamonds the light-ring estimate \eqref{limit_eikonal}.}}
\label{Fig_lightring}
\end{figure}

The trends of Figures \ref{fig_overv} and \ref{fig_mjscal} have a simple
geometric-optics interpretation. For large $L$, the term $(1+\eta^{2})^{-2}$
in $V_{L}$ is negligible and the bracket in \eqref{radial_rotating} factorises as
\begin{equation}\label{eikonal_potentials}
\left[\widehat{\omega}-V_{+}(\eta)\right]\left[\widehat{\omega}-V_{-}(\eta)\right],\qquad
V_{\pm}(\eta)=m\widehat{\Omega}(\eta)\pm\frac{\ell_{L}}{(1+\eta^{2})^{1/2}},
\qquad
\ell_{L}=\sqrt{L(L+1)} .
\end{equation}
The co-rotating branch $V_+$ tends to $m\widehat{\Omega}_-$ at the second infinity and to zero at the first. Between these limits, it has one
maximum. Indeed, \eqref{dimensionless_frame_dragging} gives $V_+'(\eta)=-(1+\eta^2)^{-2}[6\mu+\ell_L\eta\sqrt{1+\eta^2}]$, which vanishes only at
\begin{equation}\label{light_ring}
\eta_{*}\sqrt{1+\eta_{*}^{2}}=-\frac{6\mu}{\ell_{L}} ,
\end{equation}
where $V_{+}$ has its maximum. This co-rotating light ring lies at the throat
for $\mu=0$ and moves into the second exterior as $\mu$ increases (see Figure \ref{Fig_lightring}a). Since $V_{+}$ approaches $m\widehat{\Omega}_{-}$ from above as $\eta\to-\infty$, its maximum exceeds
$m\widehat{\Omega}_{-}$ for every $\mu>0$. The first-order WKB condition
\cite{SchutzWill1985ApJL,IyerWill1987PRD} applied to the product \eqref{eikonal_potentials} gives
\begin{equation}\label{eikonal_qnm}
\widehat{\omega}\simeq V_{+}(\eta_{*})-i\left(n+\frac{1}{2}\right)
\sqrt{\frac{-2V_{+}''(\eta_{*})}{V_{+}(\eta_{*})-V_{-}(\eta_{*})}},
\end{equation}
which reduces to $\widehat{\omega}\simeq\ell_L-i(n+1/2)$ at zero
spin. For every converged fundamental in the sampled range, \eqref{eikonal_qnm} reproduces both $\operatorname{Re}\widehat{\omega}_-$ and
$\operatorname{Im}\widehat{\omega}_-$ to within $2.2\%$ for $L\geq 3$, $4.2\%$ for $L=2$ and $11\%$ for $L=1$. The extrema also clarify the radial stability of the associated photon orbits. For $\mu>0$, $V_+$ has a single maximum in the second exterior, whereas $V_-$ has a single minimum at the reflected point in the first exterior, because $V_-(\eta)=m\widehat{\Omega}_- -V_+(-\eta)$. At either circular orbit, the radial ray function in \eqref{eikonal_potentials} has a double zero and positive second derivative. Both orbits are therefore radially unstable despite the different types of extremum. The stable-light-ring trapping mechanism associated with some horizonless ergoregion instabilities
\cite{Cardoso2008PRD,Cardoso2014PRD,Cunha2017PRL} is absent in this
geometric-optics description. However, this observation does not replace the
finite-$L$ spectral calculation. Moreover, it also follows that the co-rotating eikonal modes lie beyond the upper-frequency window edge, with their mirror partners beyond the lower edge, at every spin.
The light ring also accounts for the approach to the window edge at large
$\mu$. For $\mu\gg 1$, \eqref{light_ring} gives $|\eta_{*}|\simeq\sqrt{6\mu/\ell_{L}}\gg 1$, where $m\widehat{\Omega}(\eta)\simeq m\widehat{\Omega}_{-}-2\mu|\eta|^{-3}$ by \eqref{Omega_hat_asymptotics}. With $\eta=\sqrt{\mu}y$ and $\widehat{\omega}_{-}=w/\sqrt{\mu}$, \eqref{radial_rotating} becomes, to
leading order in $\mu^{-1}$
\begin{equation}\label{limit_equation}
\frac{d^{2}u}{dy^{2}}+\left[\left(w+\frac{2}{|y|^{3}}\right)^{2}-\frac{L(L+1)}{y^{2}}\right]u=0
\end{equation}
with $y<0$. The inner end $y\to 0^-$ corresponds to the overlap region
$1\ll|\eta|\ll\sqrt{\mu}$. There, the local wavenumber $w+2|y|^{-3}$ grows without bound. The transmitted branch describes waves travelling towards the throat and into the first exterior. The WKB approximation suggests that reflection in this region becomes small at large $\mu$. Neglecting that reflection gives the limiting boundary conditions
\begin{equation}\label{limit_bc}
u\sim\left(w+\frac{2}{|y|^{3}}\right)^{-1/2}e^{i(wy+y^{-2})}\quad (y\to 0^{-}),
\qquad
u\sim e^{-iwy}\quad (y\to-\infty).
\end{equation}
For continuation to complex $y$, the real-axis expression $|y|^{-3}$ in \eqref{limit_equation}-\eqref{limit_bc} is understood as $-y^{-3}$, since $y<0$ on the original domain. The amplitude branch in \eqref{limit_bc} is continued consistently with the chosen transmitted solution. We solve the limiting problem by the complex-contour integration of Section \ref{suscheck}, starting at $\arg {y}=3\pi/4$, where the transmitted solution is subdominant as $y\to 0$. Table \ref{Tab_lim} lists the eigenvalues, which vary by less than $10^{-11}$ under changes of contour. This limiting calculation gives the leading asymptotic relation
\begin{equation}\label{large_mu}
\widehat{\omega}_{-}\simeq\frac{w_{L,n}}{\sqrt{\mu}},\qquad
\mu=mj\to\infty.
\end{equation}
The same analysis applied to \eqref{limit_equation} gives the light-ring
estimate
\begin{equation}\label{limit_eikonal}
w_{L,n}\simeq\frac{2}{3\sqrt6}\ell_{L}^{3/2}
-i\left(n+\frac{1}{2}\right)\sqrt{\frac{\ell_{L}}{3}}.
\end{equation}
In dimensional variables, \eqref{large_mu} becomes $\omega-m\Omega_-\simeq w_{L,n}/\sqrt{mJ}$. The leading shifted frequency is therefore independent of $r_0$, although the unshifted frequency still contains $m\Omega_-$. At large $mj$, the relevant radial scale is set by the far-field dragging $2J/|\ell|^3$ in the second exterior with $|\ell_*|\simeq\sqrt{6mJ/\ell_L}\gg r_0$. Figures \ref{fig_mjscal} and \ref{Fig_lightring}b show the approach to this limit. In the computed families, the relative deviation $|\sqrt{\mu}\widehat{\omega}_- -w_{L,n}|/|w_{L,n}|$ decreases as $1/\mu$. At $\mu=100$, it is at most $1.5\times 10^{-3}$ for $n=0$ and $2.2\times 10^{-3}$ for $n=1$. These observations are
consistent with a correction of order $\mu^{-3/2}$ to \eqref{large_mu}, but do not constitute a proof of that remainder. The approximation is already useful at $\mu=9$, where the spectral $L=1$ fundamental mode is $3\widehat{\omega}_-=0.4516-i0.3692$, i.e. within $1\%$ of $w_{1,0}$.
\begin{table}[H]
\centering
\caption{\textit{Eigenvalues $w_{L,n}$ of the large-$\mu$ limit problem \eqref{limit_equation}-\eqref{limit_bc}, compared with $\sqrt{\mu}\widehat{\omega}_{-}$ of the fundamental mode at $\mu=100$ from
the direct integration of \eqref{radial_rotating} and with the light-ring
estimate \eqref{limit_eikonal} for $n=0$. The $n=1$ mode of $L=1$ approaches the branch cut of the second exterior (see Section \ref{subs_resol}) and is not listed.}}
\label{Tab_lim}
\begingroup
\small
\setlength{\tabcolsep}{5pt}
\renewcommand{\arraystretch}{1.15}
\begin{tabular}{@{}ccccc@{}}
\toprule
 & \multicolumn{3}{c}{Fundamental mode ($n=0$)}
 & First overtone ($n=1$) \\
\cmidrule(lr){2-4}\cmidrule(l){5-5}
$L$
& $w_{L,0}$
& $10\widehat{\omega}_{-}$
& Light-ring estimate
& $w_{L,1}$ \\
\midrule
$1$
& $0.4504-i0.3738$
& $0.4505-i0.3734$
& $0.4580-i0.3430$
& --- \\
$2$
& $1.0399-i0.4664$
& $1.0396-i0.4655$
& $1.0430-i0.4520$
& $0.4522-i1.3787$ \\
$3$
& $1.7525-i0.5462$
& $1.7514-i0.5447$
& $1.7550-i0.5370$
& $1.2770-i1.6224$ \\
$4$
& $2.5722-i0.6166$
& $2.5698-i0.6144$
& $2.5740-i0.6100$
& $2.1603-i1.8379$ \\
$5$
& $3.4873-i0.6802$
& $3.4830-i0.6771$
& $3.4890-i0.6760$
& $3.1181-i2.0314$ \\
\bottomrule
\end{tabular}
\endgroup
\end{table}
For the limiting modes in Table~\ref{Tab_lim}, $\operatorname{Re}w_{L,n}>0$ and $\operatorname{Im}w_{L,n}<0$. Hence, \eqref{large_mu} describes an approach to the upper window edge from outside, with damping of order $1/\sqrt{\mu}$, rather
than a crossing. In the first asymptotic frame, these modes oscillate near
$m\Omega_-$ and their quality factor grows as $\operatorname{Re}\widehat{\omega}/(2|\operatorname{Im}\widehat{\omega}|)
\simeq 3\pi\mu^{3/2}/(2|\operatorname{Im}w_{L,n}|)$. For example, the $L=3$ fundamental mode at $\mu=57$ has a quality factor of approximately $3733$ when evaluated using the frequency $\widehat{\omega}$ in the first asymptotic frame. Using the shifted frequency $\widehat{\omega}_{-}$ in the second asymptotic frame
gives a quality factor of approximately $1.61$ for the same mode. The damping rate is the same in both frames, while the real frequency changes. Whether these modes dominate a waveform depends on their excitation amplitudes, which are not calculated here.

\subsection{Stability of resolved modes and resolution limits}\label{subs_resol}

The stability assessment must account for how the range of resolved modes changes as the spin increases. The decreasing number of converged roots at large spin therefore calls for a closer examination of the limitations imposed by the fixed radial grids. Direct integration provides an independent way to follow the fundamental branches when the spectral calculation loses resolution.
\begin{table}[H]
\centering
\caption{\textit{Number $N_{\mathrm{c}}$ of converged mirror pairs and
largest imaginary part among the converged roots for each sector and spin. No converged root has $\operatorname{Im}\widehat{\omega}\geq 0$ and none lies
inside the window \eqref{superradiant_window}. For $(m,j)=(3,9)$, $(2,19)$
and $(3,19)$ no root converges at $N_{x}\leq 100$. The fundamental modes of
these runs are listed in Table~\ref{tab_directm}. $N_{\mathrm{c}}$ includes four converged mirror pairs omitted from Tables~\ref{tab_QNMm1}-\ref{tab_QNMm3} because gaps in the resolved overtone sequences leave their overtone labels ambiguous. These comprise two pairs at $(m,j)=(1,1/4)$ and one pair
at each of $(m,j)=(1,3/7)$ and $(2,3/7)$.}}\label{tab_counts}
\begingroup
\small
\setlength{\tabcolsep}{6pt}
\renewcommand{\arraystretch}{1.2}
\begin{tabular}{@{}crr@{\hspace{1.5em}}rr@{\hspace{1.5em}}rr@{}}
\toprule
 & \multicolumn{2}{c}{$m=1$}
 & \multicolumn{2}{c}{$m=2$}
 & \multicolumn{2}{c}{$m=3$} \\
\cmidrule(lr){2-3}
\cmidrule(lr){4-5}
\cmidrule(l){6-7}
$j$
& $N_{\mathrm{c}}$ & $\max\operatorname{Im}\widehat{\omega}$
& $N_{\mathrm{c}}$ & $\max\operatorname{Im}\widehat{\omega}$
& $N_{\mathrm{c}}$ & $\max\operatorname{Im}\widehat{\omega}$ \\
\midrule
$0$
& $18$ & $-0.5069$
& $23$ & $-0.5043$
& $26$ & $-0.5030$ \\
$1/9$
& $15$ & $-0.5060$
& $19$ & $-0.5010$
& $22$ & $-0.4945$ \\
$1/4$
& $15$ & $-0.4884$
& $14$ & $-0.4583$
& $19$ & $-0.4473$ \\
$3/7$
& $10$ & $-0.4398$
& $11$ & $-0.4029$
& $13$ & $-0.3902$ \\
$2/3$
& $7$ & $-0.3864$
& $9$ & $-0.3489$
& $12$ & $-0.3363$ \\
$1$
& $6$ & $-0.3339$
& $9$ & $-0.2989$
& $11$ & $-0.2873$ \\
$3/2$
& $6$ & $-0.2832$
& $9$ & $-0.2521$
& $10$ & $-0.2419$ \\
$7/3$
& $6$ & $-0.2333$
& $8$ & $-0.2069$
& $9$ & $-0.1983$ \\
$4$
& $6$ & $-0.1818$
& $7$ & $-0.1609$
& $7$ & $-0.1540$ \\
$9$
& $5$ & $-0.1231$
& $3$ & $-0.1087$
& $0$ & \multicolumn{1}{c}{---} \\
$19$
& $2$ & $-0.1059$
& $0$ & \multicolumn{1}{c}{---}
& $0$ & \multicolumn{1}{c}{---} \\
\bottomrule
\end{tabular}
\endgroup
\end{table}
We observe that Table \ref{tab_counts} contains no converged root with
$\operatorname{Im}\widehat{\omega}\geq 0$ and no root inside the window. The least damped spectral root is the $L=2$ fundamental mode at $(m,j)=(1,19)$ with
$\widehat{\omega}=179.3090-i0.1060$. The decreasing number of converged pairs must nevertheless be taken into account when interpreting this absence. At $\mu\in\{27,38,57\}$, corresponding to $(m,j)=(3,9)$, $(2,19)$ and $(3,19)$, no root converges on the grids with $N_x\leq 100$. Two effects make these runs difficult. The local wavenumber $\widehat{\omega}-m\widehat{\Omega}(\eta)$ is of order $3\pi\mu/2$ at the throat and approaches $\widehat{\omega}\approx3\pi\mu$ in the first exterior. The resulting radial oscillations have a scale proportional
to $1/mu$, so progressively more radial basis functions are needed. At the same time, \eqref{large_mu} brings the frequencies towards the second threshold, where the discrete roots associated with the candidate branch cut accumulate. Direct integration allows us to follow the fundamental modes beyond this fixed-grid
limit. Table \ref{tab_directm} gives the modes from the unresolved runs and
the missing $L=1$ fundamental mode at $(m,j)=(1,19)$. Their frequencies change by less than $10^{-12}$ under the tested variations of contour, matching point and integration tolerance. All are damped and lie beyond the upper-frequency window edge in agreement with \eqref{large_mu}.
\begin{table}[H]
\centering
\caption{\textit{Fundamental modes ($n=0$) unresolved by the spectral solver at
$N_x\leq 100$, obtained by direct integration of \eqref{radial_rotating}. The real parts are given in the second and first asymptotic frames, related by $\widehat{\omega}=\widehat{\omega}_{-}+3\pi\mu$, where $\mu=mj$. The imaginary part is common to both frames.}}
\label{tab_directm}
\begingroup
\setlength{\tabcolsep}{8pt}
\renewcommand{\arraystretch}{1.15}
\begin{tabular}{@{}cccrrr@{}}
\toprule
$(m,j)$
& $\mu$
& $L$
& $\operatorname{Re}\widehat{\omega}_{-}$
& $\operatorname{Re}\widehat{\omega}$
& $\operatorname{Im}\widehat{\omega}$ \\
\midrule
$(1,19)$ & $19$ & $1$
& $0.1035$ & $179.1742$ & $-0.0853$ \\
\addlinespace
$(3,9)$ & $27$ & $3$
& $0.3365$ & $254.8055$ & $-0.1040$\\
 & & $4$
& $0.4933$ & $254.9623$ & $-0.1171$ \\
 & & $5$
& $0.6681$ & $255.1371$ & $-0.1287$ \\
\addlinespace
$(2,19)$ & $38$ & $2$
& $0.1686$ & $358.3101$ & $-0.0753$ \\
 & & $3$
& $0.2838$ & $358.4254$ & $-0.0879$ \\
 & & $4$
& $0.4162$ & $358.5578$ & $-0.0991$ \\
\addlinespace
$(3,19)$ & $57$ & $3$
& $0.2319$ & $537.4442$ & $-0.0720$ \\
 & & $4$
& $0.3401$ & $537.5525$ & $-0.0811$ \\
 & & $5$
& $0.4609$ & $537.6733$ & $-0.0894$ \\
\bottomrule
\end{tabular}
\endgroup
\end{table}
The resolved calculations thus provide no evidence of a growing scalar mode. Every converged spectral root and every fundamental mode obtained by direct
integration is damped and lies outside the window. Direct integration follows
the fundamental modes of $L\leq 5$ to $\mu=100$ with the same behaviour. The large-$\mu$ limit, for which $\operatorname{Im}w_{L,n}<0$ in the computed families, supports their continued damping at larger spins. We underline that this is evidence for those branches, rather than an exclusion of all possible instabilities. The thresholds, the neighbourhoods of the two cuts and multipoles
$L\geq |m|+3$ at finite $\mu$ remain outside the resolved spectral calculation, apart from the geometric-optics information presented in Section \ref{subs_lightring}. Cut-approaching overtones require particular
care. For $L=1$ and $n=1$, direct integration gives $\widehat{\omega}_-=0.1419-i1.5039$ at $\mu=0.5$ and $0.0638-i1.4311$ at $\mu=0.6$, before the branch
reaches the second-exterior cut at approximately $\mu=0.7$. Following it beyond this point requires an explicit treatment of the appropriate Riemann sheet. Future work will investigate this continuation using Jost-function
or continued-fraction methods and examine the associated late-time response through time-domain evolution.

\section{Kerr benchmark and numerical validation}\label{Kerr_valid}

The exact rotating wormhole considered here has no independently tabulated
quasinormal spectrum against which to benchmark the calculation. We therefore
apply the same two-dimensional spectral method to massless scalar
perturbations of Kerr, whose frequencies are accurately known from Leaver's
continued-fraction construction \cite{refLeaver1985_Kerr}. Our principal
reference is the open-source \textsc{qnm} implementation \cite{Stein2019},
which assigns mode labels by continuation in spin from Schwarzschild. We also
compare with the published tabulations summarised in Table \ref{Tab_KerrLit}. Although the Kerr equation separates, we solve it directly in the radial and
polar coordinates. The comparison thus tests the assembly and collocation of
the two-dimensional operator, the asymptotic factorisation and the extraction
and filtering of its eigenvalues. It also tests angular convergence, which is
absent from the present wormhole calculation because its angular subspace is
exactly represented (see Proposition \ref{prop_winkel}). In the Kerr formulation,
the term $a^2\omega^2\chi^2$ couples spherical multipoles. Both radial and
angular resolution must therefore be varied, as would be necessary for a rotating wormhole with latitude-dependent coefficients. We use Boyer-Lindquist coordinates $(t,r,\vartheta,\varphi)$ and set $\chi=\cos{\vartheta}\in[-1,1]$. Moreover, we adopt units $c=G=2M=1$, where $M$ is the Kerr mass. Thus, $r$ and $a$ are measured in units of $2M$ and $\omega$ in units of $1/(2M)$. The tables instead use the conventional combinations $a/M$ and $M\omega$, obtained by doubling the
rescaled $a$ and halving the rescaled $\omega$, respectively. The horizon function is
\begin{equation}
\Delta(r)=r^2-r+a^2=(r-r_-)(r-r_+),\qquad
r_\pm=\frac{1\pm\sqrt{1-4a^2}}{2},
\end{equation}
where the radii $r_+$ and $r_-$ locate the event and Cauchy horizons. We restrict the calculation to $0\leq a<1/2$ because these horizons are distinct for that choice. we recall that at $a=1/2$, the two regular singular points coalesce, so the extremal problem requires a different asymptotic treatment. Following \cite{refLeaver1985_Kerr}, we write the equation for the Teukolsky master
variable $\psi_s$ with spin weight $s\in\{0,-1,-2\}$, corresponding to scalar,
electromagnetic and gravitational perturbations. All numerical results below
use $s=0$, for which the master equation is the Klein--Gordon equation $\nabla_\mu\nabla^\mu\psi_0=0$. Stationarity and axisymmetry allow us
to extract the dependence on $t$ and $\varphi$ while keeping the radial and
polar dependence together. For a fixed azimuthal number $m$, we write the ansatz
\begin{equation}
\psi_s(t,r,\chi,\varphi)=e^{-i\omega t}e^{im\varphi}\Psi^{(s)}_{\omega m}(r,\chi).
\end{equation}
Here, $\omega\in\mathbb{C}$ is the frequency and $\Psi^{(s)}_{\omega m}(r,\chi)$ is the unknown function of both spatial coordinates. Substitution into the Teukolsky equation gives
\begin{equation}\label{Kerr2DTeuk}
\Delta\partial_{rr}\Psi^{(s)}_{\omega m}+(s+1)(2r-1)\partial_r\Psi^{(s)}_{\omega m}+\partial_\chi\left[(1-\chi^2)\partial_\chi\Psi^{(s)}_{\omega m}\right]+\mathcal{V}^{(s)}_{\omega m}(r,\chi)\Psi^{(s)}_{\omega m}(r,\chi)=0
\end{equation}
with $r>r_+$, $-1<\chi<1$ and
\begin{eqnarray}
\mathcal{V}^{(s)}_{\omega m}(r,\chi)&=&
\frac{1}{\Delta}\left\{(r^2+a^2)^2\omega^2-2am\omega r+a^2m^2+is\left[
        am(2r-1)-\omega(r^2-a^2)
    \right]
\right\}\nonumber\\
&+&2is\omega r-a^2\omega^2(1-\chi^2)-2as\omega\chi+s-\frac{(m+s\chi)^2}{1-\chi^2}.\label{Kerr2Dpot}
\end{eqnarray}
Equations \eqref{Kerr2DTeuk}-\eqref{Kerr2Dpot} contain the same information as the separated radial and angular Teukolsky equations. Here, however, the numerical unknowns are determined from the full partial differential equation, without introducing an angular eigenvalue $A_{\ell m}$ or a separately computed angular eigenfunction. With the convention $e^{-i\omega t}$, the quasinormal boundary conditions are purely ingoing radiation at the event horizon and purely outgoing radiation at spatial infinity. Their leading radial factors are
\begin{equation}\label{Kerrhasymp}
\Psi^{(s)}_{\omega m}(r,\chi)\underset{r\longrightarrow r_+^+}{\sim}(r-r_+)^{-s-i\sigma_+},\qquad
\sigma_+=\frac{\omega r_+-am}{r_+-r_-}=\frac{\omega r_+-am}{\sqrt{1-4a^2}}
\end{equation}
and
\begin{equation}\label{KerrInfasympt}
\Psi^{(s)}_{\omega m}(r,\chi)\underset{r\longrightarrow\infty}{\sim}
r^{-1-2s+i\omega}e^{i\omega r}.
\end{equation}
Regularity on the two parts of the rotation axis requires the leading factors
\begin{equation}\label{KerrNaxisasymp}
\Psi^{(s)}_{\omega m}(r,\chi)\underset{\chi\longrightarrow-1}{\sim}(1+\chi)^{\frac{|m-s|}{2}},\qquad
\Psi^{(s)}_{\omega m}(r,\chi)\underset{\chi\longrightarrow+1}{\sim}
(1-\chi)^{\frac{|m+s|}{2}}.
\end{equation}
The expressions in \eqref{Kerrhasymp}-\eqref{KerrNaxisasymp} specify only the singular or oscillatory factors. The dependence on the other coordinate is regular. We extract these factors before discretisation by writing
\begin{equation} \label{Kerrfullfact}
\Psi^{(s)}_{\omega m}(r,\chi)=e^{i\omega r}(r-r_-)^{-1-s+i\omega+i\sigma_+}
(r-r_+)^{-s-i\sigma_+}e^{a\omega\chi}(1+\chi)^{\frac{|m-s|}{2}}(1-\chi)^{\frac{|m+s|}{2}}\Phi^{(s)}_{\omega m}(r,\chi),
\end{equation}
where the residual function $\Phi^{(s)}_{\omega m}(r,\chi)$ is required to remain regular at $r=r_+$, $r\to\infty$ and $\chi=\pm 1$. At the event horizon, the factors in \eqref{Kerrfullfact} other than $(r-r_+)^{-s-i\sigma_+}$ remain finite. At infinity, the radial powers add to $(-1-s+i\omega+i\sigma_+)+(-s-i\sigma_+)=-1-2s+i\omega$, which reproduces \eqref{KerrInfasympt}. The polar powers similarly give the required axis behaviour. The additional factor
$e^{a\omega\chi}$ is the exponential prefactor in Leaver's angular series
\cite{refLeaver1985_Kerr}. Since it is entire and nonzero, it leaves the axis
regularity conditions unchanged. Moreover, its derivatives introduce terms proportional to $a\omega$ in the residual angular operator and reduces to unity in the Schwarzschild limit. We map the exterior radial domain to $x\in[-1,1]$ using
\begin{equation}
x=-1+\frac{2(r-r_+)}{r-r_-}.
\end{equation}
Notice that the horizon and spatial infinity are mapped to $x=-1$ and $x=1$, respectively. We then expand the residual field as
\begin{equation}
\Phi^{(s)}_{\omega m}(x,\chi)\simeq\sum_{p=0}^{N_r-1}\sum_{q=0}^{N_\chi-1}
c_{pq}T_p(x)T_q(\chi).
\end{equation}
All coefficients $c_{pq}$ are solved for simultaneously. The tensor-product
basis therefore provides a two-dimensional discretisation without requiring
separation of the differential equation. Collocation on the Chebyshev-Gauss grid \eqref{Gauss_nodes} with $N_r$ radial and $N_\chi$ angular nodes, gives a quadratic pencil of the form \eqref{quadratic_matrix_pencil}. We apply the companion linearisation given in \eqref{linearized_pencil_wormhole}. The asymptotic factorisation identifies the desired solution branches and comparison with an independently computed spectrum tests whether the numerical scheme
recovers the corresponding modes. We compute the scalar sector $s=0$, $m=1$ at
$a/M\in\{0,0.1,\ldots,0.9,0.99\}$. The matrices are assembled in
\textsc{Maple} with $40$ significant digits and the generalised eigenproblem
is solved in \textsc{Julia} using double-double arithmetic, which provides
about $32$ significant digits. To test both sources of discretisation error,
we increase the radial and angular basis sizes together through $(N_r,N_\chi)\in\{(60,8),\,(70,10),\,(80,12)\}$. A root is accepted when it satisfies the three-resolution criterion \eqref{stability_criterion_wormhole} with
$\varepsilon_{\mathrm{abs}}=\varepsilon_{\mathrm{rel}}=10^{-6}$. We report the
three-resolution mean \eqref{frequency_average_wormhole} and identify its
$(\ell,n)$ label by comparison with the QNM branch continued from
$a=0$. This continuation is particularly important near extremality, where overtone sequences reorganise \cite{YangEtAl2013}. At $a/M=0.99$, for example, the
branch labelled $(\ell,n)=(2,2)$ has $M\omega=0.5189-i0.1888$ with damping between that of the $n=1$ and $n=3$ branches. Its label is fixed by continuation and should not be reassigned by extending a damping-based ordering from smaller spin.

\subsection{Accuracy and convergence}

At each spin, the computed spectrum includes the $\ell\in\{1,2\}$ families, higher
multipoles and the frequencies $-\omega^*$ corresponding to the
counter-rotating $m=-1$ sector. Across the eleven spins, $418$ roots pass the
three-resolution test. Of these, $409$ match QNM frequencies with
$\ell\leq 9$ and $n\leq 11$, with a largest relative deviation of $3.1\times 10^{-7}$. The other nine occur at $a/M=0.99$ close to the negative
imaginary axis and have no counterpart in this reference set. Their location
is consistent with a discrete representation of the Green-function branch
cut along that axis \cite{Leaver1986PRD}. We exclude them from the quantitative
comparison, without treating the resolution test alone as evidence that they
are quasinormal poles.
\begin{table}[H]
\caption{\textit{Fundamental scalar Kerr frequencies $M\omega$ for $m=1$, $\ell\in\{1,2\}$ and $n=0$. The spectral-method (SM) values are three-resolution means. The relative deviation $\delta=|\omega_{\mathrm{SM}}-\omega_{\mathrm{qnm}}|/|\omega_{\mathrm{qnm}}|$ is measured against the continued-fraction values of the \textup{\textsc{qnm}} package \cite{Stein2019}. Ten decimal places are displayed to facilitate comparison, but only the precision supported by $\delta$ is significant.}}
\label{Tab_Kerrfund}
\centering
\begingroup
\footnotesize
\setlength{\tabcolsep}{4pt}
\begin{tabular}{ccccc}
\toprule
 & \multicolumn{2}{c}{$\ell=1$, $n=0$} & \multicolumn{2}{c}{$\ell=2$, $n=0$} \\
$a/M$ & $M\omega_{\mathrm{SM}}$ & $\delta$ & $M\omega_{\mathrm{SM}}$ & $\delta$ \\
\midrule
$0$ & $0.2929361333-i0.0976599889$ & $1.4\times 10^{-15}$ & $0.4836438722-i0.0967587760$ & $9.2\times 10^{-16}$ \\
$0.1$ & $0.3010445314-i0.0975471619$ & $7.7\times 10^{-16}$ & $0.4916103420-i0.0966737568$ & $2.0\times 10^{-15}$ \\
$0.2$ & $0.3100426350-i0.0972450176$ & $3.1\times 10^{-15}$ & $0.5005547304-i0.0964111930$ & $4.5\times 10^{-15}$ \\
$0.3$ & $0.3201264576-i0.0966913169$ & $2.2\times 10^{-15}$ & $0.5106524802-i0.0959256915$ & $1.3\times 10^{-13}$ \\
$0.4$ & $0.3315664990-i0.0957914724$ & $5.7\times 10^{-15}$ & $0.5221421624-i0.0951472846$ & $1.5\times 10^{-12}$ \\
$0.5$ & $0.3447531808-i0.0943945193$ & $3.8\times 10^{-14}$ & $0.5353594169-i0.0939647149$ & $1.1\times 10^{-11}$ \\
$0.6$ & $0.3602846874-i0.0922424822$ & $2.2\times 10^{-13}$ & $0.5507987600-i0.0921910818$ & $5.8\times 10^{-11}$ \\
$0.7$ & $0.3791585307-i0.0888481922$ & $1.1\times 10^{-12}$ & $0.5692373150-i0.0894842286$ & $2.6\times 10^{-10}$ \\
$0.8$ & $0.4032726486-i0.0831323404$ & $4.7\times 10^{-12}$ & $0.5920175599-i0.0851252497$ & $1.0\times 10^{-9}$ \\
$0.9$ & $0.4372338080-i0.0718481321$ & $2.0 \times10^{-11}$ & $0.6218286117-i0.0771749434$ & $3.8\times 10^{-9}$ \\
$0.99$ & $0.4934232838-i0.0367119850$ & $7.1\times 10^{-11}$ & $0.6595134656-i0.0597467382$ & $1.3\times 10^{-8}$ \\
\bottomrule
\end{tabular}
\endgroup
\end{table}
Table \ref{Tab_Kerrfund} compares the fundamental modes of
$\ell\in\{1,2\}$ with the continued-fraction values. The relative deviations range
from about $10^{-15}$ at small spin, where the reference comparison reaches
the round-off level, to $1.3\times 10^{-8}$ at $a/M=0.99$. Table \ref{Tab_KerrDev} extends this comparison to $n\leq 3$. Of these $88$ target modes, $74$ pass the three-resolution test. Their largest relative deviation is $2.6\times 10^{-7}$ and it remains below $2\times 10^{-8}$ for $n\leq 1$. The median deviation is $6\times 10^{-13}$ for $n=0$ and $2\times 10^{-11}$ for $n=1$.
Except for four fundamental modes at the round-off level, the observed
relative deviations are smaller than $d_{\max}/|\omega|$. Thus $d_{\max}$
provides a conservative empirical error estimate for these benchmark modes, as it does for the independently checked wormhole modes. Failure of the resolution criterion does not necessarily mean that a mode is absent from the discrete spectrum. Of the $14$ target modes that fail the test, $13$ are present at all three resolutions and approach the \textsc{qnm} values monotonically. These comprise the $\ell=1$ mode with $n=2$ at $a=0$, the $\ell=1$, $n=3$ modes for $a/M\leq 0.7$ and four modes at $a/M=0.99$. For the first two groups, the finest-grid relative deviations range from $6\times 10^{-8}$ to $6\times 10^{-6}$. At $a/M=0.99$, the deviations for $(\ell,n)\in\{(1,2),(2,1),(2,2)\}$ are
$1.3\times 10^{-6}$, $1.5\times 10^{-7}$ and $1.2\times 10^{-6}$,
respectively, while that for $(1,3)$ is $1.1\times 10^{-3}$. For example, the $(2,2)$ deviation falls from $6.8\times 10^{-5}$ to $9.0\times 10^{-6}$ and then $1.2\times 10^{-6}$ over $(N_r,N_\chi)\in\{(60,8),\,(70,10),\,(80,12)\}$. This mode is identifiable and improving under refinement, although it has not passed the convergence test. Only $(2,3)$ at $a/M=0.99$ is not resolved on these grids.
\begin{table}[H]
\caption{\textit{Base-ten logarithm of the relative deviation $\delta$ from the \textup{\textsc{qnm}} frequencies \cite{Stein2019} for scalar Kerr modes with $m=1$, $\ell=1,2$ and $n\leq 3$. Spectral frequencies are computed on the three grids $(N_r,N_\chi)\in\{(60,8),\,(70,10),\,(80,12)\}$. A dash indicates failure of the three-resolution test at tolerance $10^{-6}$, not necessarily absence of the mode from the raw spectra.}}
\label{Tab_KerrDev}
\centering
\begingroup
\footnotesize
\setlength{\tabcolsep}{5pt}
\begin{tabular}{c|cccc|cccc}
\toprule
 & \multicolumn{4}{c|}{$\ell=1$} & \multicolumn{4}{c}{$\ell=2$} \\
$a/M$ & $n=0$ & $n=1$ & $n=2$ & $n=3$ & $n=0$ & $n=1$ & $n=2$ & $n=3$ \\
\midrule
$0$ & $-14.9$ & $-9.7$ & --- & --- & $-15.0$ & $-14.2$ & $-10.6$ & $-7.9$ \\
$0.1$ & $-15.1$ & $-9.9$ & $-6.6$ & --- & $-14.7$ & $-14.3$ & $-10.7$ & $-8.0$ \\
$0.2$ & $-14.5$ & $-10.2$ & $-6.8$ & --- & $-14.3$ & $-14.2$ & $-10.9$ & $-8.2$ \\
$0.3$ & $-14.7$ & $-10.4$ & $-7.1$ & --- & $-12.9$ & $-12.6$ & $-11.0$ & $-8.3$ \\
$0.4$ & $-14.2$ & $-10.7$ & $-7.3$ & --- & $-11.8$ & $-11.5$ & $-11.0$ & $-8.5$ \\
$0.5$ & $-13.4$ & $-10.9$ & $-7.6$ & --- & $-11.0$ & $-10.7$ & $-10.2$ & $-8.6$ \\
$0.6$ & $-12.7$ & $-11.1$ & $-7.9$ & --- & $-10.2$ & $-10.0$ & $-9.6$ & $-9.0$ \\
$0.7$ & $-12.0$ & $-11.4$ & $-8.3$ & --- & $-9.6$ & $-9.4$ & $-9.0$ & $-8.4$ \\
$0.8$ & $-11.3$ & $-11.1$ & $-8.6$ & $-6.7$ & $-9.0$ & $-8.8$ & $-8.5$ & $-8.1$ \\
$0.9$ & $-10.7$ & $-10.5$ & $-9.1$ & $-6.6$ & $-8.4$ & $-8.3$ & $-8.1$ & $-7.7$ \\
$0.99$ & $-10.1$ & $-7.7$ & --- & --- & $-7.9$ & --- & --- & --- \\
\bottomrule
\end{tabular}
\endgroup
\end{table}
The growth of the fundamental-mode deviations from the round-off level at
$a=0$ to $10^{-11}$-$10^{-8}$ at $a/M=0.99$ reflects errors in both spatial
directions. In Schwarzschild, the residual angular functions of the
$\ell\in\{1,2\}$ families are polynomials of degree $0$ and $1$, respectively.
Thus, $N_\chi=2$ represents them exactly, just as the corresponding finite angular subspace is exact for the wormhole. For $a\neq 0$, this polynomial
closure is lost. To isolate the angular error, Table~\ref{Tab_Kerr_wink} varies $N_\chi$ at fixed $N_r=60$ and $a/M=0.9$, using the same operator and grids in double precision. With $N_\chi=2$, relative errors reach $3.6\times 10^{-2}$ and would be missed by varying $N_r$ alone. Beyond $N_\chi=4$, adding two angular nodes reduces the error by roughly three orders of magnitude until radial or
round-off errors dominate at $N_\chi=10$-$12$. This separate convergence test
shows why angular refinement is essential when rotation couples the
multipoles.
\begin{table}[H]
\caption{\textit{Angular convergence of four scalar Kerr modes at $a/M=0.9$ and $m=1$. Entries are relative deviations from the \textup{\textsc{qnm}} values \cite{Stein2019}, with radial resolution fixed at $N_r=60$ and the number of angular nodes $N_\chi$ varied.}}
\label{Tab_Kerr_wink}
\centering
\begingroup
\footnotesize
\setlength{\tabcolsep}{6pt}
\begin{tabular}{ccccc}
\toprule
$N_\chi$ & $(\ell,n)=(1,0)$ & $(1,1)$ & $(2,0)$ & $(2,1)$ \\
\midrule
$2$  & $3.5\times 10^{-2}$ & $3.6\times 10^{-2}$ & $7.2\times 10^{-3}$ & $7.8\times 10^{-3}$ \\
$4$  & $8.7\times 10^{-5}$ & $1.1\times 10^{-4}$ & $2.9\times 10^{-3}$ & $3.1\times 10^{-3}$ \\
$6$  & $8.0\times 10^{-8}$ & $1.1\times 10^{-7}$ & $9.4\times 10^{-6}$ & $1.2\times 10^{-5}$ \\
$8$  & $6.0\times 10^{-11}$ & $8.7\times 10^{-11}$ & $1.1\times 10^{-8}$ & $1.5\times 10^{-8}$ \\
$10$ & $5.6\times 10^{-14}$ & $8.9\times 10^{-12}$ & $1.4\times 10^{-11}$ & $1.8\times 10^{-11}$ \\
$12$ & $5.3\times 10^{-14}$ & $8.7\times 10^{-12}$ & $1.5\times 10^{-13}$ & $2.6\times 10^{-12}$ \\
\bottomrule
\end{tabular}
\endgroup
\end{table}

\subsection{Comparison with published spectra}

The \textsc{qnm} package supplies reference frequencies at every spin used
here, but several earlier calculations use the same continued-fraction
construction. Comparisons with published tabulations therefore provide
additional checks against independent implementations and other methods.
For rotating Kerr, we use the continued-fraction results of \cite{KonoplyaZhidenko2006,BertiCardoso2006,LinEtAl2017} and the matrix
method of \cite{LinEtAl2017}. In the Schwarzschild limit, we also use
Leaver's continued-fraction values \cite{refLeaver1985_Kerr}, the
pseudospectral and asymptotic-iteration results of \cite{Mamani2022EPJC,ChoEtAl2012} and WKB results at sixth
\cite{Konoplya2003PRD} and third
\cite{SchutzWill1985ApJL,IyerWill1987PRD,Iyer1987PRD} order. For each reference, we include every tabulated massless-scalar frequency whose mode and spin match a converged root of our calculation. The common parameter range is $|m|=1$ and $a/M\in\{0,0.1,\dots,0.9,0.99\}$. Published $m=-1$ values are compared with the corresponding $-\omega^*$ images in our $m=1$ spectrum. Table~\ref{Tab_KerrLit} summarises the comparison and Appendix~\ref{AppKerrLit} gives the individual
frequencies.
\begin{table}[H]
\caption{\textit{Comparison of published scalar quasinormal frequencies with the two-dimensional spectral method (SM). The listed modes and spins are those common to both calculations, with $m=1$ unless stated. `\emph{Decimals} gives the quoted precision of $M\omega$, $N$ the number of frequencies compared and $N_{\mathrm{d}}$ the number agreeing in every digit after the SM value is rounded to that precision. The largest relative difference is $\delta_{\max}=\max|\omega_{\mathrm{SM}}-\omega_{\mathrm{ref}}|/|\omega_{\mathrm{SM}}|$. For \textup{\textsc{qnm}}, $N$ counts identified converged roots and $\delta_{\max}$ is their largest deviation. CF denotes continued fractions, PS pseudospectral collocation and AIM asymptotic iteration. $^{a}$Six significant digits in units $2M=1$. $^{b}$Values tabulated in \cite{Konoplya2003PRD}. Tables~\ref{TabKerrLitSpin} and \ref{TabKerrLitSchw} give the individual comparisons.}}
\label{Tab_KerrLit}
\centering
\begingroup
\footnotesize
\setlength{\tabcolsep}{4pt}
\begin{tabular}{llllcccc}
\toprule
Ref. & Method & $(\ell,n)$ & $a/M$ & Decimals & $N$ & $N_{\mathrm{d}}$ & $\delta_{\max}$ \\
\midrule
\cite{Stein2019} & CF (\textsc{qnm}) & $\ell\leqslant9$, $n\leqslant11$; $m=\pm1$ & $0$--$0.99$ & -- & 409 & -- & $3.1\times 10^{-7}$ \\
\cite{KonoplyaZhidenko2006} & CF & $(1,0)$ & $0$--$0.9$, $0.99$ & $6$ & 11 & 9 & $2.1\times 10^{-6}$ \\
\cite{BertiCardoso2006} & CF & $(2,0)$, $(2,1)$; $m=\pm1$ & $0$, $0.2$, $0.4$--$0.6$, $0.8$, $0.9$ & $5$ & 28 & 24 & $1.2\times 10^{-5}$ \\
\cite{LinEtAl2017} & CF & $(1,0)$, $(2,0)$ & $0$, $0.4$, $0.8$ & $6^{a}$ & 6 & 5 & $4.1\times 10^{-6}$ \\
\cite{LinEtAl2017} & matrix method & $(1,0)$, $(2,0)$ & $0$, $0.4$, $0.8$ & $6^{a}$ & 6 & 2 & $1.1\times 10^{-5}$ \\
\cite{refLeaver1985_Kerr}$^{b}$ & CF & $\ell=1,2$; $n\leqslant\ell$ & $0$ & $4$ & 5 & 4 & $1.7\times 10^{-4}$ \\
\cite{Mamani2022EPJC} & PS, AIM & $1\leqslant\ell\leqslant4$; $n\leqslant\ell$ & $0$ & $6$ & 14 & 14 & $1.3\times 10^{-6}$ \\
\cite{ChoEtAl2012} & AIM & $\ell=1,2$; $n\leqslant\ell$ & $0$ & $4$ & 5 & 3 & $1.7\times 10^{-4}$ \\
\cite{Konoplya2003PRD} & WKB, 6th order & $\ell=1,2$; $n\leqslant\ell$ & $0$ & $4$ & 5 & 1 & $6.1\times 10^{-4}$ \\
\cite{IyerWill1987PRD,Iyer1987PRD}$^{b}$ & WKB, 3rd order & $\ell=1,2$; $n\leqslant\ell$ & $0$ & $4$ & 5 & 0 & $7.9\times 10^{-3}$ \\
\bottomrule
\end{tabular}
\endgroup
\end{table}
Of the $69$ published continued-fraction, pseudospectral and asymptotic-iteration values, $59$ agree with our spectral frequencies in every quoted digit. These include the near-extremal frequency at $a/M=0.99$ in \cite{KonoplyaZhidenko2006} and the counter-rotating modes in \cite{BertiCardoso2006}. The other ten differ by one unit in the last quoted digit. In each case, the accurate frequency lies within $0.08$ of a unit in that digit of a rounding boundary, so the last-digit mismatch is consistent with rounding sensitivity at the quoted precision. The matrix-method results of \cite{LinEtAl2017}, computed with $18$ interpolation points, differ by at most $1.1\times 10^{-5}$, consistent with the accuracy reported there. The largest differences from WKB are about $6\times 10^{-4}$ at sixth order and $8\times 10^{-3}$ at third order, as expected from truncating the WKB expansion at low multipole number \cite{Konoplya2003PRD}. The comparisons therefore support agreement at the precision of the tabulated continued-fraction, pseudospectral and asymptotic-iteration values and within the stated accuracy of the matrix and WKB approximations. Across $0\leq a/M\leq 0.99$, the benchmark reproduces fundamental frequencies with relative deviations of order $10^{-15}$-$10^{-8}$ and the identified overtones to approximately $3\times 10^{-7}$ or better. Agreement with the independent calculations in Table~\ref{Tab_KerrLit} supports both the two-dimensional implementation and the use of $d_{\max}$ as an empirical diagnostic for the modes tested here. The comparison exercises the asymptotic factorisation, radial compactification, angular coupling, collocation and companion linearisation without using Kerr separation in the solve. In the wormhole application, the horizon factor in \eqref{Kerrfullfact} is replaced by the outgoing factor appropriate to the second asymptotic frame in \eqref{scalar_asymptotic_factorisation}. Its three-multipole angular subspace is exact at $N_\chi=3$, while the collocation, linearisation and
resolution-matching procedure follow the same construction.

\section{Conclusions and outlook}

We have formulated and solved the scalar quasinormal-mode problem on the
complete two-ended geometry of the exact spinning Morris-Thorne wormhole
\cite{Batic2026EPJC}. The formulation follows the global treatment of
rotating Ellis-Bronnikov wormholes in \cite{Khoo2024PRD}, with an
analytic background that permits exact spectral identities and an
independent radial reduction. The signed proper distance makes the
throat an ordinary interior point. At the two infinities, the relative
rotation of the asymptotic rest frames requires outgoing conditions with
frequencies $\omega$ and $\omega-m\Omega_-$. Extracting these asymptotic factors and compactifying the radial coordinate gives an operator with analytic coefficients on a closed square and a quadratic Chebyshev matrix pencil for $\widehat{\omega}=r_0\omega$. The analytic structure provides checks that are independent of numerical resolution. Each azimuthal spectrum is mirror symmetric about $\operatorname{Re}\widehat{\omega}=3\pi jm/2$ and each fixed multipole
family depends on $m$ and $j$ only through their product. The angular
discretisation is exact for the included multipoles. With three
Chebyshev polynomials, the two-dimensional pencil decomposes into the
three radial pencils for $L\in\{|m|,|m|+1,|m|+2\}$. These identities label the
mode families and test the assembled matrices, although they do not by
themselves measure the error relative to the continuum spectrum. The static scalar problem is modally stable because its separated radial potential is positive. The same conclusion holds for the axisymmetric sector at every spin, since rotation drops out of that sector exactly. For nonaxisymmetric modes, the conserved flux confines nonthreshold real QNMs to $0<\omega/(m\Omega_-)<1$. It therefore restricts crossings of continuous mode branches away from the thresholds, but leaves the threshold cases open. The global ergoregion also requires careful interpretation. The value $j_{\rm{ergo}}=2/(3\pi)$ marks its
arrival at the throat in a chosen asymptotic frame. For every $j>0$, the ergoregion of that frame already extends to the opposite infinity, outside the hypotheses of the usual compact-ergoregion instability argument. We examine the sectors $m\in\{1,2,3\}$, with three multipole families in each sector, at eleven spins up to $j=19$. Every converged frequency in these calculations is damped and lies outside the superradiant window. The fundamental modes recovered by direct
integration where the fixed spectral grids lose resolution have the same
properties. The least damped converged spectral root has
$\operatorname{Im}\widehat{\omega}=-0.106$ at $j=19$, while the least
damped directly integrated mode among these runs has $\operatorname{Im}\widehat{\omega}=-0.072$ at $(m,j)=(3,19)$. These observations concern the resolved poles. However, they do not establish the absence of modes near the two thresholds, in unresolved overtone sequences or in the additional multipoles $L\geq |m|+3$. Moreover, the static frequencies reproduce the Ellis-Bronnikov benchmarks of \cite{Gonzalez2022PRD}, with alternating parity in the checked overtone sequences. The rotating fundamental modes move towards the window edge as $mj$ increases. Their geometric-optics description is governed by an unstable circular null orbit that moves from the throat into the second exterior. There is no accompanying stable orbit in the radial eikonal branches
analysed here, although this observation is not a general proof of
finite-multipole stability. At large $mj$, rescaling the radial equation
gives a limiting problem controlled by the far-field frame dragging and the leading relation $\widehat{\omega}_-\simeq w_{L,n}(mj)^{-1/2}$. Its inner condition neglects reflection of waves transmitted through the throat into the
first exterior. In that exterior, the computed modes oscillate close to
multiples of $\Omega_-$ with a large quality factor. Their excitation in an observable waveform requires information beyond the pole frequencies. The main numerical limitation occurs near the zero-frequency points of
the two asymptotic frames. Nonconvergent roots accumulate there with
behaviour consistent with discretised branch cuts and some physical
mode sequences approach the same regions. Measuring convergence relative
to the distance from the nearest threshold is necessary to reject
candidates that pass the initial frequency-relative tolerance. Direct
integration supplies an independent outgoing-boundary check for the
low-lying modes and recovers fundamental modes beyond the range of the fixed
grids. The Kerr benchmark further tests the two-dimensional scheme when
angular convergence is required. For the identified converged modes,
agreement with the continued-fraction reference values ranges from
approximately $10^{-15}$ to $3\times 10^{-7}$ over $0\leq a/M\leq 0.99$, consistent with the reported convergence diagnostics and the published comparisons in Section \ref{Kerr_valid}.

Further progress requires resolving the spectrum near the two cuts and
their endpoints. A Jost-function calculation on specified Riemann sheets
would allow the cut-approaching overtones to be followed beyond the
present continuation, while time-domain evolution would test their
excitation and the associated late-time tails. A separate threshold
analysis is needed when $\omega=0$ or $\omega=m\Omega_-$, where one
asymptotic end changes singularity type. The large-spin approximation
also motivates a quantitative treatment of reflection and matching
errors. Beyond this background, the same spectral framework can be
adapted to latitude-dependent frame dragging, nontrivial $K(r,\chi)$ or other perturbing fields after deriving the appropriate equations and
boundary factors. Ultimately, perturbations of the supporting source
and the metric must be treated together to assess stability beyond the
scalar test-field sector.

\section*{Code and data availability}

The code for assembling the collocation matrices, solving and analysing
the quadratic eigenvalue problems, integrating the radial equation and
performing the Kerr benchmark is available at
\begin{center}
\url{https://github.com/MarkSukaiti/wormhole-qnm-chebyshev-2d}
\end{center}
The repository also provides the converged frequencies, the rejected
roots with their classifications, and the data used to generate the
tables and figures.

\bibliography{QNMS}

\appendix
\section{Scalar QNMs tables}

Tables \ref{tab_QNMm1}--\ref{tab_QNMm3} collect the converged scalar
frequencies for $m\in\{1,2,3\}$ with multipole and overtone labels assigned as in
Section \ref{Subsandexact}. Each entry satisfies the two convergence
criteria of Section~\ref{subsroot_sel}. We omit four converged mirror pairs from the tables because unresolved intervening modes prevent an unambiguous assignment of their overtone indices $n$, as explained in Sectio \ref{suscheck}. The repository listed under Code and Data Availability supplies the tabulated data, their frequencies in the second asymptotic frame, the convergence diagnostics  and the classification of rejected roots.

\begin{table}[H]
\centering
\caption{Converged QNMs for $m=1$, arranged by spin $j$ and overtone index $n$  with separate columns for the multipoles $L\in\{1,2,3\}$. Only the member of each mirror pair with $\operatorname{Re}\varpi\geq 0$ is listed. Its partner is $3\pi j-\widehat{\omega}^{*}$ (see Corollary \ref{cor_mirror}). Values are the three-resolution means in \eqref{frequency_average_wormhole}, quoted with at most six decimal places according to $d_{\max}$.}
\label{tab_QNMm1}
\begingroup
\small
\setlength{\tabcolsep}{5pt}
\renewcommand{\arraystretch}{1.12}
\setlength{\heavyrulewidth}{0.6pt}
\setlength{\lightrulewidth}{0.4pt}
\setlength{\cmidrulewidth}{0.3pt}
\setlength{\aboverulesep}{2pt}
\setlength{\belowrulesep}{3pt}
\begin{tabular*}{\linewidth}{@{\extracolsep{\fill}}cclll@{}}
\toprule
 & & \multicolumn{3}{c}{QNMN $\widehat{\omega}$} \\
\cmidrule(l){3-5}
$j$ & $n$ & \multicolumn{1}{c}{$L=1$} & \multicolumn{1}{c}{$L=2$} & \multicolumn{1}{c}{$L=3$} \\
\midrule
$0$ & $0$ & $1.572708-i0.529701$ & $2.546656-i0.512669$ & $3.534287-i0.506922$ \\
 & $1$ & $1.255828-i1.702451$ & $2.344977-i1.572498$ & $3.390100-i1.537211$ \\
 & $2$ & $0.8368-i3.2361$ & $1.947768-i2.760388$ & $3.099862-i2.623226$ \\
 & $3$ & $0.633-i4.932$ & $1.45331-i4.20521$ & $2.672197-i3.823518$ \\
 & $4$ & \multicolumn{1}{c}{---} & $1.079-i5.878$ & $2.163009-i5.211938$ \\
 & $5$ & \multicolumn{1}{c}{---} & $0.865-i7.602$ & $1.69686-i6.80488$ \\
 & $6$ & \multicolumn{1}{c}{---} & \multicolumn{1}{c}{---} & $1.360-i8.510$ \\
 & $7$ & \multicolumn{1}{c}{---} & \multicolumn{1}{c}{---} & $1.137-i10.233$ \\
\addlinespace[2pt]
$1/9$ & $0$ & $2.209126-i0.520483$ & $3.148802-i0.510104$ & $4.117084-i0.505996$ \\
 & $1$ & $1.870525-i1.702957$ & $2.937505-i1.570190$ & $3.968562-i1.535976$ \\
 & $2$ & \multicolumn{1}{c}{---} & $2.530650-i2.772266$ & $3.671291-i2.626159$ \\
 & $3$ & \multicolumn{1}{c}{---} & $2.0434-i4.2378$ & $3.237706-i3.836997$ \\
 & $4$ & \multicolumn{1}{c}{---} & $1.689-i5.918$ & $2.729206-i5.240430$ \\
 & $5$ & \multicolumn{1}{c}{---} & $1.478-i7.660$ & $2.2717-i6.8459$ \\
 & $6$ & \multicolumn{1}{c}{---} & \multicolumn{1}{c}{---} & $1.948-i8.557$ \\
\addlinespace[2pt]
$1/4$ & $0$ & $3.237246-i0.488407$ & $4.086443-i0.497724$ & $4.994640-i0.500631$ \\
 & $1$ & $2.8359-i1.6674$ & $3.846099-i1.547782$ & $4.832100-i1.524842$ \\
 & $2$ & \multicolumn{1}{c}{---} & $3.409498-i2.782525$ & $4.512047-i2.624547$ \\
 & $3$ & \multicolumn{1}{c}{---} & $2.945-i4.294$ & $4.061910-i3.866926$ \\
 & $4$ & \multicolumn{1}{c}{---} & \multicolumn{1}{c}{---} & $3.5636-i5.3124$ \\
 & $5$ & \multicolumn{1}{c}{---} & \multicolumn{1}{c}{---} & $3.138-i6.942$ \\
 & $6$ & \multicolumn{1}{c}{---} & \multicolumn{1}{c}{---} & $2.866-i8.639$ \\
\addlinespace[2pt]
$3/7$ & $0$ & $4.732958-i0.439767$ & $5.474889-i0.470381$ & $6.296495-i0.485288$ \\
 & $1$ & \multicolumn{1}{c}{---} & $5.194517-i1.478682$ & $6.111976-i1.485111$ \\
 & $2$ & \multicolumn{1}{c}{---} & $4.7019-i2.7287$ & $5.753131-i2.583367$ \\
 & $3$ & \multicolumn{1}{c}{---} & \multicolumn{1}{c}{---} & $5.273655-i3.860785$ \\
 & $4$ & \multicolumn{1}{c}{---} & \multicolumn{1}{c}{---} & $4.789-i5.352$ \\
\addlinespace[2pt]
$2/3$ & $0$ & $6.843273-i0.386440$ & $7.482362-i0.430919$ & $8.208628-i0.457413$ \\
 & $1$ & \multicolumn{1}{c}{---} & $7.166696-i1.360705$ & $7.999841-i1.404661$ \\
 & $2$ & \multicolumn{1}{c}{---} & \multicolumn{1}{c}{---} & $7.588816-i2.469644$ \\
 & $3$ & \multicolumn{1}{c}{---} & \multicolumn{1}{c}{---} & $7.053-i3.760$ \\
\addlinespace[2pt]
$1$ & $0$ & $9.881934-i0.333924$ & $10.426608-i0.384532$ & $11.057251-i0.418717$ \\
 & $1$ & \multicolumn{1}{c}{---} & $10.093508-i1.209814$ & $10.830184-i1.286051$ \\
 & $2$ & \multicolumn{1}{c}{---} & \multicolumn{1}{c}{---} & $10.367494-i2.273982$ \\
\addlinespace[2pt]
$3/2$ & $0$ & $14.509452-i0.283197$ & $14.966707-i0.334211$ & $15.503427-i0.371677$ \\
 & $1$ & \multicolumn{1}{c}{---} & $14.638437-i1.041474$ & $15.270587-i1.137733$ \\
 & $2$ & \multicolumn{1}{c}{---} & \multicolumn{1}{c}{---} & $14.77783-i2.00648$ \\
\addlinespace[2pt]
$7/3$ & $0$ & $22.288673-i0.233277$ & $22.662376-i0.280562$ & $23.105548-i0.317388$ \\
 & $1$ & \multicolumn{1}{c}{---} & $22.36024-i0.86328$ & $22.882601-i0.965724$ \\
 & $2$ & \multicolumn{1}{c}{---} & \multicolumn{1}{c}{---} & $22.3976-i1.6850$ \\
\addlinespace[2pt]
$4$ & $0$ & $37.925611-i0.181785$ & $38.215004-i0.221920$ & $38.560911-i0.254551$ \\
 & $1$ & \multicolumn{1}{c}{---} & $37.95927-i0.67354$ & $38.365389-i0.768698$ \\
 & $2$ & \multicolumn{1}{c}{---} & \multicolumn{1}{c}{---} & $37.934-i1.320$ \\
\addlinespace[2pt]
$9$ & $0$ & $84.973541-i0.123083$ & $85.168481-i0.152072$ & $85.402940-i0.176422$ \\
 & $1$ & \multicolumn{1}{c}{---} & $84.98414-i0.45532$ & $85.257618-i0.528394$ \\
\addlinespace[2pt]
$19$ & $0$ & \multicolumn{1}{c}{---} & $179.309-i0.106$ & $179.471-i0.123$ \\
\bottomrule
\end{tabular*}
\par\smallskip
\begin{minipage}{\linewidth}
\footnotesize\raggedright
\textit{Note.} A dash indicates that no frequency is tabulated for that combination of $j$, $L$ and $n$. Roots failing \eqref{refined_criterion} are excluded. Converged roots with $\operatorname{Im}\widehat{\omega}<-7.5$ whose overtone indices remain ambiguous because of unresolved intervening modes, are also omitted (see Section \ref{suscheck}).
\end{minipage}
\endgroup
\end{table}

\begin{table}[H]
\centering
\caption{\textit{Converged QNMs for $m=2$, arranged by spin $j$ and overtone index $n$ with separate columns for the multipoles $L\in\{2,3,4\}$. Only the member of each mirror pair with $\operatorname{Re}\varpi\geq 0$ is listed. Its partner is $6\pi j-\widehat{\omega}^{*}$ (see Corollary \ref{cor_mirror}). Values are the three-resolution means in \eqref{frequency_average_wormhole}, quoted with at most six decimal places according to $d_{\max}$.}}
\label{tab:qnm-m2}
\begingroup
\small
\setlength{\tabcolsep}{5pt}
\renewcommand{\arraystretch}{1.06}
\setlength{\heavyrulewidth}{0.6pt}
\setlength{\lightrulewidth}{0.4pt}
\setlength{\cmidrulewidth}{0.3pt}
\setlength{\aboverulesep}{2pt}
\setlength{\belowrulesep}{3pt}
\begin{tabular*}{\linewidth}{@{\extracolsep{\fill}}cclll@{}}
\toprule
 & & \multicolumn{3}{c}{QNM $\widehat{\omega}$} \\
\cmidrule(l){3-5}
$j$ & $n$ & \multicolumn{1}{c}{$L=2$} & \multicolumn{1}{c}{$L=3$} & \multicolumn{1}{c}{$L=4$} \\
\midrule
$0$ & $0$ & $2.546656-i0.512669$ & $3.534287-i0.506922$ & $4.527050-i0.504334$ \\
 & $1$ & $2.344977-i1.572498$ & $3.390100-i1.537211$ & $4.415134-i1.522662$ \\
 & $2$ & $1.947768-i2.760388$ & $3.099862-i2.623226$ & $4.189863-i2.572245$ \\
 & $3$ & $1.45331-i4.20521$ & $2.672197-i3.823518$ & $3.851094-i3.681826$ \\
 & $4$ & $1.079-i5.878$ & $2.163009-i5.211938$ & $3.408396-i4.891569$ \\
 & $5$ & $0.865-i7.602$ & $1.69686-i6.80488$ & $2.898498-i6.248085$ \\
 & $6$ & \multicolumn{1}{c}{---} & $1.360-i8.510$ & $2.396258-i7.772664$ \\
 & $7$ & \multicolumn{1}{c}{---} & $1.137-i10.233$ & $1.9765-i9.4260$ \\
 & $8$ & \multicolumn{1}{c}{---} & \multicolumn{1}{c}{---} & $1.660-i11.138$ \\
\addlinespace[2pt]
$1/9$ & $0$ & $3.886073-i0.501043$ & $4.807784-i0.502190$ & $5.757444-i0.502127$ \\
 & $1$ & $3.652206-i1.554762$ & $4.648486-i1.528406$ & $5.637379-i1.518193$ \\
 & $2$ & $3.222672-i2.783693$ & $4.333759-i2.626546$ & $5.397395-i2.571978$ \\
 & $3$ & $2.753-i4.286$ & $3.887297-i3.862171$ & $5.041163-i3.696189$ \\
 & $4$ & \multicolumn{1}{c}{---} & $3.38610-i5.29900$ & $4.585048-i4.932980$ \\
 & $5$ & \multicolumn{1}{c}{---} & $2.954-i6.925$ & $4.074353-i6.325530$ \\
 & $6$ & \multicolumn{1}{c}{---} & $2.669-i8.630$ & $3.5880-i7.8848$ \\
 & $7$ & \multicolumn{1}{c}{---} & \multicolumn{1}{c}{---} & $3.195-i9.562$ \\
\addlinespace[2pt]
$1/4$ & $0$ & $6.063830-i0.458291$ & $6.854023-i0.477360$ & $7.699780-i0.487341$ \\
 & $1$ & $5.770354-i1.444148$ & $6.661275-i1.462937$ & $7.559293-i1.478682$ \\
 & $2$ & $5.253-i2.689$ & $6.286110-i2.554480$ & $7.281034-i2.523881$ \\
 & $3$ & \multicolumn{1}{c}{---} & $5.79150-i3.83999$ & $6.879393-i3.668778$ \\
 & $4$ & \multicolumn{1}{c}{---} & \multicolumn{1}{c}{---} & $6.395430-i4.960003$ \\
 & $5$ & \multicolumn{1}{c}{---} & \multicolumn{1}{c}{---} & $5.8999-i6.4185$ \\
 & $6$ & \multicolumn{1}{c}{---} & \multicolumn{1}{c}{---} & $5.470-i8.015$ \\
\addlinespace[2pt]
$3/7$ & $0$ & $9.152470-i0.402944$ & $9.819705-i0.434704$ & $10.551934-i0.455305$ \\
 & $1$ & $8.823502-i1.270616$ & $9.598464-i1.335659$ & $10.389476-i1.384806$ \\
 & $2$ & \multicolumn{1}{c}{---} & $9.154247-i2.358610$ & $10.063942-i2.379071$ \\
 & $3$ & \multicolumn{1}{c}{---} & \multicolumn{1}{c}{---} & $9.593205-i3.503506$ \\
 & $4$ & \multicolumn{1}{c}{---} & \multicolumn{1}{c}{---} & $9.048-i4.821$ \\
\addlinespace[2pt]
$2/3$ & $0$ & $13.443257-i0.348915$ & $14.006670-i0.385869$ & $14.634618-i0.412706$ \\
 & $1$ & $13.111349-i1.090742$ & $13.773967-i1.182690$ & $14.458392-i1.254895$ \\
 & $2$ & \multicolumn{1}{c}{---} & $13.28647-i2.08919$ & $14.096940-i2.158717$ \\
 & $3$ & \multicolumn{1}{c}{---} & \multicolumn{1}{c}{---} & $13.55493-i3.20373$ \\
\addlinespace[2pt]
$1$ & $0$ & $19.572638-i0.298864$ & $20.047136-i0.336307$ & $20.581378-i0.365393$ \\
 & $1$ & $19.259536-i0.923674$ & $19.819045-i1.025571$ & $20.402780-i1.108440$ \\
 & $2$ & \multicolumn{1}{c}{---} & $19.3266-i1.7972$ & $20.029300-i1.899658$ \\
 & $3$ & \multicolumn{1}{c}{---} & \multicolumn{1}{c}{---} & $19.445-i2.817$ \\
\addlinespace[2pt]
$3/2$ & $0$ & $28.868329-i0.252084$ & $29.263741-i0.287257$ & $29.712084-i0.315804$ \\
 & $1$ & $28.58680-i0.77039$ & $29.052153-i0.870860$ & $29.541529-i0.954777$ \\
 & $2$ & \multicolumn{1}{c}{---} & $28.588-i1.508$ & $29.180800-i1.625278$ \\
 & $3$ & \multicolumn{1}{c}{---} & \multicolumn{1}{c}{---} & $28.599-i2.391$ \\
\addlinespace[2pt]
$7/3$ & $0$ & $44.460490-i0.206912$ & $44.782115-i0.238019$ & $45.148669-i0.264051$ \\
 & $1$ & $44.21888-i0.62596$ & $44.595971-i0.717408$ & $44.995003-i0.795337$ \\
 & $2$ & \multicolumn{1}{c}{---} & $44.185-i1.226$ & $44.668436-i1.343080$ \\
\addlinespace[2pt]
$4$ & $0$ & $75.764501-i0.160852$ & $76.012775-i0.186393$ & $76.296830-i0.208253$ \\
 & $1$ & $75.57048-i0.48233$ & $75.860305-i0.558776$ & $76.168501-i0.624943$ \\
 & $2$ & \multicolumn{1}{c}{---} & \multicolumn{1}{c}{---} & $75.895633-i1.046749$ \\
\addlinespace[2pt]
$9$ & $0$ & $169.8907-i0.1087$ & $170.0576-i0.1267$ & $170.249-i0.142$ \\
\bottomrule
\end{tabular*}
\par\smallskip
\begin{minipage}{\linewidth}
\footnotesize\raggedright
\textit{Note.} A dash indicates that no frequency is tabulated for that combination of $j$, $L$ and $n$. Roots failing \eqref{refined_criterion} are excluded. Converged roots with $\operatorname{Im}\widehat{\omega}<-7.5$ whose overtone indices remain ambiguous because of unresolved intervening modes are also omitted (see Section \ref{suscheck}).
\end{minipage}
\endgroup
\end{table}

\begin{table}[H]
\centering
\caption{\textit{Converged scalar QNMs for $m=3$, arranged by spin $j$ and overtone index $n$ with separate columns for the multipoles $L\in\{3,4,5\}$. Only the member of each mirror pair with $\operatorname{Re}\varpi\geq 0$ is listed. Its partner is $9\pi j-\widehat{\omega}^{*}$ (see Corollary \ref{cor_mirror}). Values are the three-resolution means in \eqref{frequency_average_wormhole}, quoted with at most six decimal places according to $d_{\max}$.}}
\label{tab_QNMm3}
\begingroup
\small
\setlength{\tabcolsep}{5pt}
\renewcommand{\arraystretch}{1.00}
\setlength{\heavyrulewidth}{0.6pt}
\setlength{\lightrulewidth}{0.4pt}
\setlength{\cmidrulewidth}{0.3pt}
\setlength{\aboverulesep}{2pt}
\setlength{\belowrulesep}{3pt}
\begin{tabular*}{\linewidth}{@{\extracolsep{\fill}}cclll@{}}
\toprule
 & & \multicolumn{3}{c}{Quasinormal frequency $\widehat{\omega}$} \\
\cmidrule(l){3-5}
$j$ & $n$ & \multicolumn{1}{c}{$L=3$} & \multicolumn{1}{c}{$L=4$} & \multicolumn{1}{c}{$L=5$} \\
\midrule
$0$ & $0$ & $3.534287-i0.506922$ & $4.527050-i0.504334$ & $5.522309-i0.502958$ \\
 & $1$ & $3.390100-i1.537211$ & $4.415134-i1.522662$ & $5.430901-i1.515247$ \\
 & $2$ & $3.099862-i2.623226$ & $4.189863-i2.572245$ & $5.247167-i2.547654$ \\
 & $3$ & $2.672197-i3.823518$ & $3.851094-i3.681826$ & $4.970140-i3.617053$ \\
 & $4$ & $2.163009-i5.211938$ & $3.408396-i4.891569$ & $4.601248-i4.745744$ \\
 & $5$ & $1.69686-i6.80488$ & $2.898498-i6.248085$ & $4.149772-i5.962761$ \\
 & $6$ & $1.360-i8.510$ & $2.396258-i7.772664$ & $3.641923-i7.299879$ \\
 & $7$ & $1.137-i10.233$ & $1.9765-i9.4260$ & $3.126489-i8.775709$ \\
 & $8$ & \multicolumn{1}{c}{---} & $1.660-i11.138$ & $2.6598-i10.3753$ \\
\addlinespace[2pt]
$1/9$ & $0$ & $5.582886-i0.494495$ & $6.491923-i0.498004$ & $7.426817-i0.499473$ \\
 & $1$ & $5.410095-i1.509689$ & $6.364048-i1.507873$ & $7.325648-i1.506853$ \\
 & $2$ & $5.072659-i2.611453$ & $6.109828-i2.561937$ & $7.123524-i2.540605$ \\
 & $3$ & $4.610267-i3.872541$ & $5.737177-i3.697447$ & $6.822207-i3.621834$ \\
 & $4$ & $4.120-i5.341$ & $5.271071-i4.958215$ & $6.428200-i4.776407$ \\
 & $5$ & \multicolumn{1}{c}{---} & $4.76663-i6.38060$ & $5.958572-i6.034419$ \\
 & $6$ & \multicolumn{1}{c}{---} & $4.303-i7.961$ & $5.448430-i7.423377$ \\
 & $7$ & \multicolumn{1}{c}{---} & $3.931-i9.641$ & $4.95144-i8.95137$ \\
 & $8$ & \multicolumn{1}{c}{---} & \multicolumn{1}{c}{---} & $4.520-i10.592$ \\
\addlinespace[2pt]
$1/4$ & $0$ & $8.906881-i0.447339$ & $9.669439-i0.465422$ & $10.479121-i0.476837$ \\
 & $1$ & $8.691908-i1.374316$ & $9.512466-i1.415011$ & $10.357404-i1.442366$ \\
 & $2$ & $8.265171-i2.421946$ & $9.199550-i2.427682$ & $10.114639-i2.445878$ \\
 & $3$ & $7.708-i3.709$ & $8.749264-i3.563378$ & $9.756664-i3.519795$ \\
 & $4$ & \multicolumn{1}{c}{---} & $8.2253-i4.8782$ & $9.303747-i4.701678$ \\
 & $5$ & \multicolumn{1}{c}{---} & $7.719-i6.367$ & $8.798475-i6.021663$ \\
 & $6$ & \multicolumn{1}{c}{---} & $7.432-i7.994$ & $8.2977-i7.4834$ \\
 & $7$ & \multicolumn{1}{c}{---} & \multicolumn{1}{c}{---} & $7.839-i9.058$ \\
\addlinespace[2pt]
$3/7$ & $0$ & $13.581297-i0.390184$ & $14.217867-i0.416655$ & $14.906279-i0.436054$ \\
 & $1$ & $13.348895-i1.196337$ & $14.042389-i1.267057$ & $14.767530-i1.320183$ \\
 & $2$ & $12.86364-i2.11408$ & $13.683195-i2.179880$ & $14.486802-i2.244784$ \\
 & $3$ & \multicolumn{1}{c}{---} & $13.14660-i3.23402$ & $14.064540-i3.251593$ \\
 & $4$ & \multicolumn{1}{c}{---} & \multicolumn{1}{c}{---} & $13.524550-i4.397817$ \\
 & $5$ & \multicolumn{1}{c}{---} & \multicolumn{1}{c}{---} & $12.937-i5.728$ \\
\addlinespace[2pt]
$2/3$ & $0$ & $20.047136-i0.336307$ & $20.581378-i0.365393$ & $21.165356-i0.388410$ \\
 & $1$ & $19.819045-i1.025571$ & $20.402780-i1.108440$ & $21.019895-i1.174716$ \\
 & $2$ & $19.3266-i1.7972$ & $20.029300-i1.899658$ & $20.721143-i1.994466$ \\
 & $3$ & \multicolumn{1}{c}{---} & $19.445-i2.817$ & $20.258647-i2.888926$ \\
 & $4$ & \multicolumn{1}{c}{---} & \multicolumn{1}{c}{---} & $19.6422-i3.9288$ \\
\addlinespace[2pt]
$1$ & $0$ & $29.263741-i0.287257$ & $29.712084-i0.315804$ & $30.205525-i0.339443$ \\
 & $1$ & $29.052153-i0.870860$ & $29.541529-i0.954777$ & $30.062927-i1.024543$ \\
 & $2$ & $28.588-i1.508$ & $29.180800-i1.625278$ & $29.767362-i1.732640$ \\
 & $3$ & \multicolumn{1}{c}{---} & $28.599-i2.391$ & $29.300251-i2.497109$ \\
\addlinespace[2pt]
$3/2$ & $0$ & $43.225556-i0.241852$ & $43.598353-i0.268137$ & $44.010569-i0.290555$ \\
 & $1$ & $43.037169-i0.729277$ & $43.443094-i0.807878$ & $43.878016-i0.874939$ \\
 & $2$ & $42.621-i1.248$ & $43.113205-i1.365120$ & $43.602125-i1.472478$ \\
 & $3$ & \multicolumn{1}{c}{---} & \multicolumn{1}{c}{---} & $43.161220-i2.106580$ \\
\addlinespace[2pt]
$7/3$ & $0$ & $66.629646-i0.198264$ & $66.932436-i0.221197$ & $67.268376-i0.241162$ \\
 & $1$ & $66.468914-i0.595052$ & $66.797689-i0.664327$ & $67.151423-i0.724525$ \\
 & $2$ & \multicolumn{1}{c}{---} & $66.511130-i1.114717$ & $66.907746-i1.213352$ \\
 & $3$ & \multicolumn{1}{c}{---} & \multicolumn{1}{c}{---} & $66.516536-i1.722098$ \\
\addlinespace[2pt]
$4$ & $0$ & $113.600500-i0.153989$ & $113.833997-i0.172635$ & $114.093720-i0.189115$ \\
 & $1$ & $113.47188-i0.46032$ & $113.72472-i0.51700$ & $113.99762-i0.56693$ \\
 & $2$ & \multicolumn{1}{c}{---} & \multicolumn{1}{c}{---} & $113.798-i0.945$ \\
\bottomrule
\end{tabular*}
\par\smallskip
\begin{minipage}{\linewidth}
\footnotesize\raggedright
\textit{Note.} A dash indicates that no frequency is tabulated for that combination of $j$, $L$ and $n$. Roots failing \eqref{refined_criterion} are excluded. Converged roots with $\operatorname{Im}\widehat{\omega}<-7.5$ whose overtone indices remain ambiguous because of unresolved intervening modes are also omitted (see Section~\ref{suscheck}).
\end{minipage}
\endgroup
\end{table}

\section{Comparison with Kerr and Schwarzschild reference spectra}\label{AppKerrLit}

Tables \ref{TabKerrLitSchw} and \ref{TabKerrLitSpin} give the individual
comparisons summarised in Table~\ref{Tab_KerrLit}. The spectral-method
values are the three-resolution means defined in Section \ref{Kerr_valid}. For every entry here, their relative deviation from the \textsc{qnm} reference \cite{Stein2019} is below $10^{-8}$. All frequencies are expressed as $M\omega$ so that the different source normalisations can be compared directly.

\begin{table}[H]
\centering
\caption{\textit{Scalar Schwarzschild QNMs $M\omega$ at $a=0$. Spectral-method (SM) values are rounded to eight decimal places. Panel (a) compares SM with the pseudospectral and asymptotic-iteration (PS/AIM) results of \cite{Mamani2022EPJC}. Panel (b) gives comparisons with Leaver's continued-fraction (CF) results \cite{refLeaver1985_Kerr}, as tabulated in \cite{Konoplya2003PRD} and the $15$-iteration AIM results of \cite{ChoEtAl2012}. Panel (c) contrasts SM with sixth-order WKB \cite{Konoplya2003PRD} and third-order WKB \cite{IyerWill1987PRD,Iyer1987PRD}, where the latter values are taken from \cite{Konoplya2003PRD}.}}
\label{TabKerrLitSchw}
\begingroup
\small
\setlength{\tabcolsep}{5pt}
\renewcommand{\arraystretch}{1.12}
\setlength{\heavyrulewidth}{0.6pt}
\setlength{\lightrulewidth}{0.4pt}
\setlength{\cmidrulewidth}{0.3pt}
\setlength{\aboverulesep}{2pt}
\setlength{\belowrulesep}{3pt}
\begin{tabular*}{\linewidth}{@{\extracolsep{\fill}}cll@{}}
\toprule
\multicolumn{3}{l}{(a) Comparison with PS/AIM} \\
\addlinespace[2pt]
$(\ell,n)$ & \multicolumn{1}{c}{SM} & \multicolumn{1}{c}{PS/AIM} \\
\midrule
$(1,0)$ & $0.29293613-i0.09765999$ & $0.292936-i0.097660$ \\
$(1,1)$ & $0.26444865-i0.30625739$ & $0.264449-i0.306257$ \\
\addlinespace[3pt]
$(2,0)$ & $0.48364387-i0.09675878$ & $0.483644-i0.096759$ \\
$(2,1)$ & $0.46385058-i0.29560394$ & $0.463851-i0.295604$ \\
$(2,2)$ & $0.43054405-i0.50855840$ & $0.430544-i0.508558$ \\
\addlinespace[3pt]
$(3,0)$ & $0.67536623-i0.09649963$ & $0.675366-i0.096500$ \\
$(3,1)$ & $0.66067150-i0.29228479$ & $0.660671-i0.292285$ \\
$(3,2)$ & $0.63362581-i0.49600823$ & $0.633626-i0.496008$ \\
$(3,3)$ & $0.59877325-i0.71122121$ & $0.598773-i0.711221$ \\
\addlinespace[3pt]
$(4,0)$ & $0.86741564-i0.09639169$ & $0.867416-i0.096392$ \\
$(4,1)$ & $0.85580804-i0.29087602$ & $0.855808-i0.290876$ \\
$(4,2)$ & $0.83369213-i0.49032489$ & $0.833692-i0.490325$ \\
$(4,3)$ & $0.80328811-i0.69748155$ & $0.803288-i0.697482$ \\
$(4,4)$ & $0.76773262-i0.91401943$ & $0.767733-i0.914019$ \\
\bottomrule
\end{tabular*}
\par\medskip
\begin{tabular*}{\linewidth}{@{\extracolsep{\fill}}clll@{}}
\toprule
\multicolumn{4}{l}{(b) Comparison with CF and AIM} \\
\addlinespace[2pt]
$(\ell,n)$ & \multicolumn{1}{c}{SM} & \multicolumn{1}{c}{CF} & \multicolumn{1}{c}{AIM} \\
\midrule
$(1,0)$ & $0.29293613-i0.09765999$ & $0.2929-i0.0977$ & $0.2929-i0.0977$ \\
$(1,1)$ & $0.26444865-i0.30625739$ & $0.2645-i0.3063$ & $0.2645-i0.3063$ \\
\addlinespace[3pt]
$(2,0)$ & $0.48364387-i0.09675878$ & $0.4836-i0.0968$ & $0.4836-i0.0968$ \\
$(2,1)$ & $0.46385058-i0.29560394$ & $0.4639-i0.2956$ & $0.4639-i0.2956$ \\
$(2,2)$ & $0.43054405-i0.50855840$ & $0.4305-i0.5086$ & $0.4306-i0.5086$ \\
\bottomrule
\end{tabular*}
\par\medskip
\begin{tabular*}{\linewidth}{@{\extracolsep{\fill}}clll@{}}
\toprule
\multicolumn{4}{l}{(c) Comparison with WKB approximations} \\
\addlinespace[2pt]
$(\ell,n)$ & \multicolumn{1}{c}{SM} & \multicolumn{1}{c}{WKB$_6$} & \multicolumn{1}{c}{WKB$_3$} \\
\midrule
$(1,0)$ & $0.29293613-i0.09765999$ & $0.2929-i0.0978$ & $0.2911-i0.0980$ \\
$(1,1)$ & $0.26444865-i0.30625739$ & $0.2645-i0.3065$ & $0.2622-i0.3074$ \\
\addlinespace[3pt]
$(2,0)$ & $0.48364387-i0.09675878$ & $0.4836-i0.0968$ & $0.4832-i0.0968$ \\
$(2,1)$ & $0.46385058-i0.29560394$ & $0.4638-i0.2956$ & $0.4632-i0.2958$ \\
$(2,2)$ & $0.43054405-i0.50855840$ & $0.4304-i0.5087$ & $0.4317-i0.5034$ \\
\bottomrule
\end{tabular*}
\endgroup
\end{table}

\begin{table}[H]
\centering
\caption{\textit{Scalar Kerr QNMs $M\omega$ for $m=\pm 1$. The spectral-method (SM) entries are three-resolution means rounded to eight decimal places. In panel (a), the continued-fraction (CF) values are taken from \cite{KonoplyaZhidenko2006} for $(\ell,m,n)=(1,1,0)$ and \cite{BertiCardoso2006} for $\ell=2$. Panel (b) compares SM with the CF and matrix-method results of \cite{LinEtAl2017}.}}
\label{TabKerrLitSpin}
\begingroup
\small
\setlength{\tabcolsep}{4pt}
\renewcommand{\arraystretch}{1.03}
\setlength{\heavyrulewidth}{0.6pt}
\setlength{\lightrulewidth}{0.4pt}
\setlength{\cmidrulewidth}{0.3pt}
\setlength{\aboverulesep}{2pt}
\setlength{\belowrulesep}{3pt}
\begin{tabular*}{\linewidth}{@{\extracolsep{\fill}}ccll@{}}
\toprule
\multicolumn{4}{l}{(a) Comparison with published continued-fraction results} \\
\addlinespace[2pt]
$(\ell,m,n)$ & $a/M$ & \multicolumn{1}{c}{SM} & \multicolumn{1}{c}{CF} \\
\midrule
$(1,1,0)$ & $0$ & $0.29293613-i0.09765999$ & $0.292936-i0.097660$ \\
 & $0.1$ & $0.30104453-i0.09754716$ & $0.301045-i0.097547$ \\
 & $0.2$ & $0.31004263-i0.09724502$ & $0.310043-i0.097245$ \\
 & $0.3$ & $0.32012646-i0.09669132$ & $0.320126-i0.096691$ \\
 & $0.4$ & $0.33156650-i0.09579147$ & $0.331567-i0.095792$ \\
 & $0.5$ & $0.34475318-i0.09439452$ & $0.344753-i0.094395$ \\
 & $0.6$ & $0.36028469-i0.09224248$ & $0.360285-i0.092243$ \\
 & $0.7$ & $0.37915853-i0.08884819$ & $0.379159-i0.088848$ \\
 & $0.8$ & $0.40327265-i0.08313234$ & $0.403273-i0.083132$ \\
 & $0.9$ & $0.43723381-i0.07184813$ & $0.437234-i0.071848$ \\
 & $0.99$ & $0.49342328-i0.03671199$ & $0.493423-i0.036712$ \\
\addlinespace[3pt]
$(2,1,0)$ & $0$ & $0.48364387-i0.09675878$ & $0.48364-i0.09676$ \\
 & $0.2$ & $0.50055473-i0.09641119$ & $0.50055-i0.09641$ \\
 & $0.4$ & $0.52214216-i0.09514728$ & $0.52214-i0.09515$ \\
 & $0.5$ & $0.53535942-i0.09396471$ & $0.53536-i0.09397$ \\
 & $0.6$ & $0.55079876-i0.09219108$ & $0.55080-i0.09219$ \\
 & $0.8$ & $0.59201756-i0.08512525$ & $0.59202-i0.08513$ \\
 & $0.9$ & $0.62182861-i0.07717494$ & $0.62183-i0.07718$ \\
\addlinespace[3pt]
$(2,1,1)$ & $0$ & $0.46385058-i0.29560394$ & $0.46385-i0.29560$ \\
 & $0.2$ & $0.48220572-i0.29405282$ & $0.48221-i0.29405$ \\
 & $0.4$ & $0.50577967-i0.28955215$ & $0.50578-i0.28955$ \\
 & $0.5$ & $0.52020102-i0.28555128$ & $0.52020-i0.28555$ \\
 & $0.6$ & $0.53696577-i0.27968490$ & $0.53697-i0.27969$ \\
 & $0.8$ & $0.58067221-i0.25688281$ & $0.58067-i0.25688$ \\
 & $0.9$ & $0.60984573-i0.23150305$ & $0.60984-i0.23150$ \\
\addlinespace[3pt]
$(2,-1,0)$ & $0$ & $0.48364387-i0.09675878$ & $0.48364-i0.09676$ \\
 & $0.2$ & $0.47014608-i0.09650394$ & $0.47015-i0.09650$ \\
 & $0.4$ & $0.45932466-i0.09576922$ & $0.45932-i0.09577$ \\
 & $0.5$ & $0.45476916-i0.09522894$ & $0.45477-i0.09523$ \\
 & $0.6$ & $0.45072867-i0.09456518$ & $0.45073-i0.09457$ \\
 & $0.8$ & $0.44407271-i0.09280356$ & $0.44407-i0.09280$ \\
 & $0.9$ & $0.44140634-i0.09165429$ & $0.44141-i0.09165$ \\
\addlinespace[3pt]
$(2,-1,1)$ & $0$ & $0.46385058-i0.29560394$ & $0.46385-i0.29560$ \\
 & $0.2$ & $0.44939775-i0.29517801$ & $0.44940-i0.29518$ \\
 & $0.4$ & $0.43805094-i0.29314251$ & $0.43805-i0.29314$ \\
 & $0.5$ & $0.43336782-i0.29153475$ & $0.43337-i0.29153$ \\
 & $0.6$ & $0.42926541-i0.28950182$ & $0.42927-i0.28950$ \\
 & $0.8$ & $0.42252661-i0.28394916$ & $0.42253-i0.28395$ \\
 & $0.9$ & $0.41969039-i0.28028727$ & $0.41969-i0.28029$ \\
\bottomrule
\end{tabular*}
\par\medskip
\begin{tabular*}{\linewidth}{@{\extracolsep{\fill}}cclll@{}}
\toprule
\multicolumn{5}{l}{(b) Comparison with Ref.~\cite{LinEtAl2017}} \\
\addlinespace[2pt]
$(\ell,m,n)$ & $a/M$ & \multicolumn{1}{c}{SM} & \multicolumn{1}{c}{CF} & \multicolumn{1}{c}{Matrix method} \\
\midrule
$(1,1,0)$ & $0$ & $0.29293613-i0.09765999$ & $0.292936-i0.09766$ & $0.2929395-i0.09766$ \\
 & $0.4$ & $0.33156650-i0.09579147$ & $0.3315665-i0.0957915$ & $0.331569-i0.0957935$ \\
 & $0.8$ & $0.40327265-i0.08313234$ & $0.4032725-i0.0831325$ & $0.4032735-i0.0831355$ \\
\addlinespace[3pt]
$(2,1,0)$ & $0$ & $0.48364387-i0.09675878$ & $0.483644-i0.096759$ & $0.483644-i0.096759$ \\
 & $0.4$ & $0.52214216-i0.09514728$ & $0.52214-i0.0951475$ & $0.52214-i0.0951475$ \\
 & $0.8$ & $0.59201756-i0.08512525$ & $0.59202-i0.0851255$ & $0.59202-i0.0851255$ \\
\bottomrule
\end{tabular*}
\par\smallskip
\begin{minipage}{\linewidth}
\footnotesize\raggedright
\textit{Notes.} For $m=-1$, the SM values follow from $\omega\mapsto-\omega^*$ applied to the corresponding roots of the computed $m=1$ spectrum. Frequencies from \cite{LinEtAl2017}, originally quoted in units $2M=1$, have been divided by two to obtain $M\omega$.
\end{minipage}
\endgroup
\end{table}

\end{document}